\documentclass[11pt]{article}          
\usepackage{latexsym}
\usepackage{bbm}
\makeatletter
\def\@begintheorem#1#2{\trivlist%
 \item[\hskip \labelsep{\sffamily\bfseries #2\ #1}]\itshape}
\newtheorem{teo}{Theorem}[section]
\newtheorem{defi}[teo]{Definition}
\newtheorem{cor}[teo]{Corollary}
\newtheorem{lem}[teo]{Lemma}
\newtheorem{pro}[teo]{Proposition}
\newtheorem{_rem}[teo]{Remark}
\newtheorem{_eje}[teo]{Example}
\newenvironment{rem}{\def\@begintheorem##1##2{\trivlist%
 \item[\hskip\labelsep{\sffamily\bfseries ##2\ ##1}]}\begin{_rem}}{\end{_rem}}

\makeatother

\newenvironment{beweis}{{\em Proof:}}{\hfill $\rule{2mm}{2mm}$
\vspace{3mm}

}

\DeclareMathAlphabet{\Ma}{U}{msa}{m}{n}
\DeclareMathAlphabet{\Mb}{U}{msb}{m}{n}
\DeclareMathAlphabet{\Meuf}{U}{euf}{m}{n}

\def\Lzwei#1{\mbox{{\rm L$^{2}(#1)$}}}
\def\R{\Mb{R}}
\def\C{\Mb{C}}
\def\P{\Mb{P}}
\def\Z{\Mb{Z}}
\def\z#1{\Mb{#1}}
\def\got#1{\Meuf{#1}}

\def\mr #1.{\mathrm{#1\,}}
\def\mb #1.{\mathbf{#1\,}}

\DeclareSymbolFont{ASMa}{U}{msa}{m}{n}
\DeclareSymbolFont{ASMb}{U}{msb}{m}{n}
\DeclareMathSymbol{\hrist}{\mathord}{ASMa}{"16}
\DeclareMathSymbol{\varkappa}{\mathalpha}{ASMb}{"7B}
\DeclareMathSymbol{\CrPr}{\mathord}{ASMb}{"6F}

\def\1{\mathbbm 1}
\def\EINS{\1}

\def\restriction{{\mathchoice{
 \mbox{\unitlength1cm\begin{picture}(.2,.4)%
  \bezier{5}(.07,.3)(.1,.27)(.13,.24)%
  \put(.07,.35){\line(0,-1){.5}}\end{picture}}}{
 \mbox{\unitlength1cm\begin{picture}(.2,.4)%
  \bezier{5}(.07,.3)(.1,.27)(.13,.24)%
  \put(.07,.35){\line(0,-1){.5}}\end{picture}}}{
  \hrist}{\hrist}}}

  \def\al #1.{{\mathcal{#1}}}
  \def\ot #1.{{\got{#1}}}
  \def\C{\Mb{C}}
  \def\N{\Mb{N}}
  \def\R{\Mb{R}}

\def\DD{\Mb{D}}

\def\B{\al O.}
\def\BR{\al B.(\R^4)}

\def\ed{\end{document}}
\def\be{\begin{equation}}
\def\ee{\end{equation}}
\def\bea{\begin{eqnarray}}
\def\eea{\end{eqnarray}}
\def\beaO{\begin{eqnarray*}}
\def\eeaO{\end{eqnarray*}}

\def\g#1#2{\mbox{$\Wort{#1}(#2)$}}
\def\Wort#1{\mbox{\fontfamily{cmr}\selectfont\mdseries\upshape #1}}
\def\In#1{{\fontsize{8pt}{10pt}\selectfont\Wort{#1}}}
\def\KIn#1{{\fontsize{6pt}{8pt}\selectfont\Wort{#1}}}
\newcommand{\bp}{{\breve{p}}} 

\newcommand{\SPgrr}  {\mbox{$\cal P_{+}^{\uparrow}$}}

\newcommand{\UPgr}   {\widetilde{\SPgrr}}
\newcommand{\mudp}   {\mbox{$\;\mu_{0}(\mbox{{\rm d}}p)$}}
\newcommand{\mumdp}  {\mbox{$\;\mu(\mbox{{\rm d}}p)$}}
\newcommand{\HmPlus}{\mbox{${\got C}^{+}_{m}$}}
\newcommand{\Lk}     {\mbox{${\cal C}_{+}^{\circ}$}}
\newcommand{\Lkf}     {\mbox{${\cal C}_{+}$}}
\newcommand{\kSPgr}{\mbox{$\seins_{\kern-.2em\szwei}^{\kern-.1em\sdrei}$}}

\def\ccr#1{\Wort{CCR}(#1)}
\def\car#1{\Wort{CAR}(#1)}
\def\TestO#1{\Wort{C}_0^\infty(#1)}
\def\Test#1{{\cal S}( #1)}

\DeclareMathSymbol{\hsemi}{\mathord}{ASMb}{"6E}
\newcommand{\semi}[2]{\mbox{$#1\kern.1em\hsemi\kern.1em#2$}}
\def\askp#1#2{\mbox{$\langle#1\mbox{\bf ,}\;#2\rangle$}}

\def\LA{\left\langle\bgroup}
\def\LE{\left[\bgroup}
\def\LG{\left\{\bgroup}
\def\LR{\left(\bgroup}
\def\RA{\egroup^{\rule{0mm}{0mm}}\right\rangle}
\def\RE{\egroup^{\rule{0mm}{0mm}}\right]}
\def\RG{\egroup^{\rule{0mm}{0mm}}\right\}}
\def\RR{\egroup^{\rule{0mm}{0mm}}\right)}
\def\Ldummy{\left.\bgroup}
\def\Rdummy{\egroup^{\rule{0mm}{2mm}}\right.}

\def\Kbegin{\begin{equation} \left. \begin{array}{rcl}}
\def\Kend{\end{array} \right\} \end{equation}}
\newenvironment{Klammer}{\Kbegin}{\Kend}
\newcommand{\rep}     {rep\-resenta\-tion}
                   
\newcommand{\rwe}{relativistic wave equation} 

\title{\bf Massless relativistic wave equations and quantum field theory}
\author{
 {\sc Fernando Lled\'o}\\[2mm]
 {\footnotesize Institute for Pure and Applied Mathematics,}     \\
 {\footnotesize RWTH-Aachen,}
 {\footnotesize Templergraben 55,}             \\ 
 {\footnotesize D-52056 Aachen, Germany.}                        \\
 {\footnotesize lledo@iram.rwth-aachen.de}}

\date{\today{}}

\begin{document}
\maketitle

\begin{center}
{\sl Dedicated to Rudolf Haag on his 80th birthday.}
\end{center}
\vspace{.5cm}

\begin{abstract}
We give a simple and direct construction of a massless quantum field
with arbitrary discrete helicity that satisfies Wightman axioms and the 
corresponding relativistic wave equation in the distributional sense. 
We underline the mathematical differences to massive models. 
The construction is based
on the notion of massless free net (cf.~Section~\ref{AxiomsFreeNets})
and the detailed analysis of covariant and massless canonical (Wigner) 
representations of the Poincar\'e group. A characteristic feature of massless
models with nontrivial helicity is the fact that the fibre degrees of 
freedom of the covariant and canonical representations do not coincide.
We use massless \rwe{s} as constraint equations reducing the fibre degrees
of freedom of the covariant representation. They are characterised by 
invariant (and in contrast with the massive case non reducing) one-dimensional
projections. The definition of one-particle Hilbert space structure
that specifies the quantum field uses distinguished elements of the 
intertwiner space between $\al E.(2)$ (the two-fold cover of the 2-dimensional
Euclidean group) and $\overline{\al E.(2)}$.

We conclude with a brief  comparison between the free nets constructed
in Section~\ref{AxiomsFreeNets} and a recent alternative construction 
that uses the notion of modular localisation.
\end{abstract}

\section{Introduction}\label{Intro}

The transformation character of a quantum field involves typically two
different types of representations of the corresponding 
spacetime symmetry group. The first one is a 
so-called {\em covariant representation} 
which acts reducibly on the test function space of the
quantum field. The second one is a unitary and irreducible representation,
which is called {\em canonical} (and in certain cases 
also {\em Wigner representation}), and which acts
on the one-particle Hilbert space associated to the quantum field
(see e.g.~\cite[Section~2]{WightmanIn81}).
In the context of non-scalar free quantum field theory on Minkowski space,
one chooses as test function space $\al H.$-valued Schwartz functions,
where $\al H.$ is a fixed finite-dimensional Hilbert space with
$\mr dim.\,\al H.\geq 2$. Moreover, in order to describe massless models 
with discrete helicity, the corresponding space carrying the Wigner 
representations is a $\C$-valued $L^2$ space over the positive light cone.
(In contrast with it, massive theories use $\al H.$-valued $L^2$ functions
over the positive mass shell.) The fact that there is a difference
between the dimensions of the
image Hilbert spaces (fibres) of the preceding two function 
spaces ($\al H.$ vs.~$\C$-valued functions), forces one to introduce
some additional set of constraints in the construction of massless models
with non-trivial helicity. These constraints guarantee for example 
that one can embed the test functions into the  
space carrying the corresponding massless Wigner representation. 
Physically they express the fact that for massless particles the 
helicity is parallel to the momentum in {\em all} Lorentz frames.
In previous papers we have shown that this
reduction in the fibre space can be performed in two different ways 
(see Remark~\ref{Rem.3.3.1} below). In this paper we will develop 
a third point of view in order to explain the reduction, where
massless relativistic wave equations will play an essential role. 
Indeed, one may consider massless
relativistic wave equations as constraint equations that reduce the fibre
degrees of freedom. 

Free massless quantum fields with discrete
helicity were constructed by Weinberg in 
\cite{Weinberg64a}. The necessary reduction of degrees of freedom mentioned 
above has been done in this reference as follows: to define a massless quantum 
field of helicity $j$, $2j\in\Z$, one 
usually constructs first a $2j+1$-component free
quantum field. This initial step is a clear
reminiscence of massive theories and
in fact the unnecessary components are ruled out afterwards by imposing
on the quantum field itself a first order constraint equation. This 
construction procedure has been reproduced almost unchanged several 
times in the literature (cf.~\cite{WeinbergIn65,Hislop88,bWeinberg00}).
In the present paper we propose an alternative and direct construction
of a free massless quantum field with arbitrary discrete helicity which 
satisfies the corresponding massless \rwe{} in a distributional 
sense. We will underline the mathematical structures characteristic to
massless theories that appear in 
the group theoretical as well as in the 
quantum field theoretical context. The construction is naturally
suggested by a detailed mathematical analysis of the 
covariant and canonical representations and, in fact,
the reduction of the degrees of freedom can be encoded 
in suitable one-dimensional invariant projections. 
In this way the covariant transformation character of the 
quantum fields becomes completely transparent. They will also
satisfy the corresponding Wightman axioms. 

The main aspects of this paper may be summarised in the following 
three items:
\begin{itemize}
\item[(i)] We analyse from a mathematical point of view the role of classical
   massless relativistic wave equations in the context of 
   induced representations of the Poincar\'e group. 
   We show that these equations are characterised by certain
   invariant (but not reducing) one-dimensional projections.
   This analysis extends a systematic
   study of Niederer and O'Rafeartaigh (cf.~\cite{Niederer74b})
   concerning massive relativistic wave equations.
   We will point out the differences w.r.t.~the massive case
   (see Section~\ref{Indu}).
\item[(ii)] We will give an alternative construction of 
   {\em massless free nets}
   using as the essential element an embedding $\ot I.$ from the test
   function space to the space of solution of the corresponding massless 
   relativistic wave equation. This embedding intertwines the covariant
   and canonical representations mentioned above and therefore we can partly
   interpret the free net construction in the group theoretical context
   (see Section~\ref{AxiomsFreeNets}).
\item[(iii)] We can finally reinterpret the previous embedding
   $\ot I.$ as a one-particle Hilbert 
   space structure and this allows to give a new construction procedure for
   massless quantum fields with nontrivial helicity. These fields will satisfy
   directly the relativistic wave equation in the distributional sense as
   well as the Wightman axioms. We will finally mention some further properties
   of these fields like e.g.~the conformal covariance (see Section~\ref{Mqf}).
\end{itemize}

In order to describe induced representations
we will consider in the following section the ele\-gant 
fibre bundle language. In particular 
the crucial covariant and canonical 
representations of the Poincar\'e group can be described as two different
special cases in this framework. (In this context
we can even describe the representations of the conformal group
that restrict to the massless Wigner
representations with discrete helicity 
(cf.~\cite{Lledo01})). From a mathematical point of view,
the reason for the need of reducing 
the fibre degrees of freedom mentioned at the beginning of the introduction 
lies in the following facts: on the one hand, the canonical or Wigner 
representations
describing massless particles with discrete helicity are induced from
non-faithful, one-dimensional representations of the corresponding little
group $\al E.(2)$ (the two-fold cover of the 2-dimensional Euclidean group).
On the other hand, the covariant representations are induced from 
at least two-dimensional irreducible representation of the little group
$\g{SL}{2,\C}$ which {\em do not} restrict 
(for non-scalar models) to the inducing representation
of the canonical representation when considering $\al E.(2)$ as a subgroup
of $\g{SL}{2,\C}$. In its turn the use of non-faithful representations 
is due to the fact
that the massless little group $\al E.(2)$ is non-compact, 
solvable and has a semi-direct product structure. (Recall that,
in contrast with the previous attributes,
the massive little group  $\g{SU}{2}$ satisfies 
the complementary properties of being compact and simple.)
In the context of massless canonical representations massless relativistic
wave equations will naturally appear as constraints performing the mentioned
reduction and indeed we may associate with them invariant (but in contrast
to massive equations non reducing) projections. It becomes clear that
massless \rwe{s} have a different character than massive ones and are 
in a certain sense unavoidable for nonscalar models, even if one does
not consider discrete transformations of Minkowski space.

We will also use in the present paper spinor fields. They can be
roughly seen as being ``square roots'' of null vector fields
(cf.~\cite{bWald84}) and massless \rwe{s} can be simply
and systematically expressed as differential equations involving 
these type of fields (cf.~\cite{Penrose65,bPenrose86I}). 
The spinorial language is also particularly well adapted to the general
group theoretical framework mentioned before. In the following
section we will in addition work out explicitly the Weyl equation 
as well as Maxwell's equations (in terms of the field strength,
$\ot F.$-eq.~for short). These two concrete examples describing models 
with helicity $\frac12$ and $1$ will be the base for the construction of
massless fermionic and bosonic free nets/fields with nontrivial helicity.
For completeness we will include in our group theoretical context 
the discussion of the classical Maxwell Equations 
in terms of the vector potential field
(${\got A}$-eq.~for short). For a detailed treatment of 
quantum electromagnetism in terms of the vector potential 
(including constraints) we refer to \cite{Lledo00}.
An alternative and systematic analysis of 
relativistic particles in the context of geometric quantisation 
is given in \cite{Diazmiranda94,Diazmiranda96}. 
There are other approaches to classical massless relativistic wave
equations with different degrees of mathematical rigor,
e.g.~\cite[Chapter~II]{WightmanIn78}, \cite[Chapter~9]{bVaradarajan85} or 
\cite{Kruglov01,Gersten00,Navarro98,Lopuszanski78,Bacry76,Duerr70,Dirac36}.

Concerning item (ii) mentioned previously, we define 
and prove in Section~\ref{AxiomsFreeNets} the main properties 
of a massless free net. Recall that free nets, 
as considered in \cite{Lledo95,tLledo98} (see also \cite{bBaumgaertel92}),
are the result of a direct and natural
way of constructing nets of abstract C*-algebras indexed by open
and bounded regions in Minkowski space. They also satisfy 
the axioms of local quantum physics. The construction is based on group
theoretical arguments (in particular on the covariant and canonical
representations of the Poincar\'e group mentioned before) and
standard theory of CAR- and CCR-algebras \cite{ArakiIn87,Manuceau73}.
No representation of
the C*-algebra is used and no quantum fields are initially needed.
This agrees with the point of view in local quantum
physics that the abstract
algebraic structure should be a primary definition of the theory and
the corresponding Hilbert space representation a secondary
\cite[Section~4]{Buchholz00}. Nevertheless, the free nets will 
afterwards canonically suggest the construction procedure for
massless quantum fields. 
Therefore the massless free net construction presented here 
justifies the picture that free nets are the counterpart of 
free fields at the abstract C*-algebraic level. In addition, the 
notion of free nets (cf.~Part~(ii) of Definition~\ref{Tuples})
contains some aspects of Segal's concept of
quantisation for bosonic systems
(cf.~\cite[p.~750]{SegalIn81},\cite[p.~106]{bBaez92}). 
Through the requirements on the
embedding $\got I$ we incorporate to this program the axioms of 
local quantum physics.
Note, however, that since Haag-Kastler axioms are stated in terms of
abstract C*-algebras, we do not initially require (in contrast with
Segal's approach) that the abstract CCR-algebra is represented in
any Hilbert space nor it is necessary to specify any regular state.
This point of view is relevant when constraints are present
(cf.~\cite{Grundling88c}). Furthermore, the construction
presented in Section~\ref{AxiomsFreeNets}, is in a certain sense 
complementary to the construction considered in \cite{Lledo01}, 
but still will produce
isomorphic nets. Instead of using semidefinite sesquilinear forms
and the related factor spaces as in \cite{Lledo01} 
we will use in the present paper explicitly the space
of solutions of the corresponding massless \rwe{} and therefore we need
to introduce a different embedding $\ot I.$ for characterising the net. 
Further in \cite{Lledo01} we focused on the covariance of the 
massless free net under the Poincar\'e and conformal group. Here
we prove the rest of the properties satisfied by the free net
(e.g.~additivity, causality etc.).

The previous construction of a free net and in particular the group
theoretical interpretation of the embedding $\ot I.$  particularly 
pays off in the new construction procedure of massless quantum fields
as well as in the verification of the corresponding axioms. Indeed, for
the construction procedure of the massless quantum field
(cf.~item (iii) above) we need to reinterpret $\ot I.$ as a one-particle
Hilbert space structure. In this context we show the continuity of
$\ot I.$ w.r.t.~the corresponding Schwartz space and Hilbert space
topologies. Moreover, these fields will directly satisfy the 
(constraint) massless \rwe{s}. E.g.~in the helicity $\frac12$ case
the quantum field satisfies in the distributional sense
\[
\partial^{\KIn{$\;CC'$}}\,\phi( f_{\KIn{$C$}})=0\,,
\]
where $f$ is the corresponding vector-valued Schwartz function.
The construction presented here is considerably simpler 
than what is usually done in QFT. In order to define the fields
we will use inducing representations
of the type $D^{\KIn{$(\frac{n}{2},0)$}}$ and 
$D^{\KIn{$(0,\frac{n}{2})$}}$
since these are enough to construct massless quantum fields with 
helicity $\pm\frac{n}{2}$ (see e.g.~\cite{Niederer79}).

We conclude this paper commenting on the relations of the free net
construction in Section~\ref{AxiomsFreeNets} with the construction given
in \cite{Brunetti02} which is based on the notion of modular localisation
(see also \cite{Fassarella02}).

\section{Representations of the Poincar\'e group and relativistic 
         wave equations}
\label{Indu}
In the present section we will summarise some results concerning the theory
of induced \rep{s} in the context of fibre bundles. For details and
further generalisations we refer to 
\cite{Asorey85,bSimms68,bSternberg95} and \cite[Section~5.1]{bWarner72}.
We will specify these structures in the example of the Poincar\'e group
to introduce so-called covariant and massive/massless canonical 
representations. In this group theoretical context we will consider
massive and massless \rwe{s} and analyse their different character.
For a study of the conformal group (in 4-dimensions) in this frame 
and for a proof of the extension of the massless canonical representations
with discrete helicity to a corresponding representation of the
conformal group see \cite{Lledo01} and references cited therein.

Let $\cal G$ be a Lie group that acts 
differentiably and transitively on a
C$^\infty$-manifold $M$. For $u_0\in M$ denote by
$\al K._0:=\{g\in \al G.\mid gu_0=u_0\}$ the corresponding 
isotropy group w.r.t.~this action. Then, from
\cite[Theorem~II.3.2 and Proposition~II.4.3]{bHelgason78} we have that
$g\al K._0\mapsto gu_0$ characterises the diffeomorphism
\[
{\al G.}/{\al K._0}\cong \DD :=\{gu_0\mid g\in \al G.\}\,.
\]
In this context we may consider the following principal 
${\cal K}_0$-bundle,  
\begin{equation}
\label{PBundle1}
  {\cal B}_1:= ( \al G.,\, \Wort{pr}_1,\, \DD ).
\end{equation}
$\Wort{pr}_1\colon\ \al G.\to \DD$ denotes the canonical
projection onto the base space $\DD$. Given a representation
$\tau\colon\ {\cal K}_0\to \g{GL}{{\cal H}}$ on the
finite-dimensional Hilbert space ${\cal H}$, one can construct the 
associated vector bundle 
\begin{equation}
\label{PBundle2}
 {\cal B}_2(\tau):= 
    ( \al G.\times_{\KIn{${\cal K}_0$}}{\cal H},\;\Wort{pr}_2,\;\DD )
\end{equation}
(see \cite[Section~I.5]{bKobayashi63} for further details).
The action of $\cal G$ on $M$
specifies the following further actions on $\DD$ and on 
$\al G.\times_{\KIn{${\cal K}_0$}}{\cal H}$: for 
$g,g_0\in {\cal G}$, $v\in \al H.$, put
\begin{equation}\label{PMActions}
\left. \begin{array}{rcl}
{\cal G}\times \DD &\longrightarrow& \DD , \kern2.5cm 
                            g_0\; \Wort{pr}_1(g):=\Wort{pr}_1(g_0g) \\
{\cal G}\times ( \al G.\times_{\KIn{${\cal K}_0$}}{\cal H} )
     &\longrightarrow& \al G.\times_{\KIn{${\cal K}_0$}}{\cal H}, \kern1.82cm 
           g_0\; [g,v]:=[g_0g,v]\,, \end{array} \right\}
\end{equation}
where $[g,v]=[gk^{-1},\tau(k)v]$, $k\in\al K._0$, denotes 
the equivalence class characterising a point in the 
total space of the associated bundle.
Finally we define the (from $\tau$) induced representation of ${\cal G}$ on
the vector space of sections
of the vector bundle ${\cal B}_2$, which we denote by
$\Gamma (\al G.\times_{\KIn{${\cal K}_0$}}{\cal H})$:
let $\psi$ be such a section and for
$g\in {\cal G}$, $p\in \DD$:
\begin{equation}
\label{FbInd}
 ( T(g)\psi)(p):=g\; \psi( g^{-1} p ) .
\end{equation}

\begin{rem}\label{HFunct}
We will now present a way of rewriting the preceding induced
representation in (for physicists more usual)
terms of vector-valued functions. Indeed, choose a fixed
(not necessarily continuous) section
$s\colon\ \DD\to \al G.$ of the principal ${\cal K}_0$-bundle 
${\cal B}_1$. 
Now for $\psi\in
\Gamma (\al G.\times_{\KIn{${\cal K}_0$}}{\cal H})$ we put
$\psi(p)=[s(p),\varphi(p)]$, $p\in\DD$, for a suitable function
$\varphi\colon\ \DD\to\al H.$ and we may rewrite the induced \rep{} as 
\begin{equation}\label{VfInd}
  ( T(g)\varphi)(p)=\tau
  ( s(p)^{-1}g\, s ( g^{-1}p) ) \varphi ( g^{-1} p),
\end{equation}
where it can be easily seen that 
$s(p)^{-1}g\, s ( g^{-1}p)\in {\cal K}_0$.
\end{rem}

Note that till now we have not completely specified the induced 
representation $T$. In fact, we have to fix 
the structure of the representation space  
$\Gamma (\al G.\times_{\KIn{${\cal K}_0$}}{\cal H})$ (or of the
set of $\al H.$-valued functions).  
This must be done separately in
the three concrete situations considered below: covariant representations 
as well as
massive and massless canonical representations. In these cases
we have to specify completely the structure of the corresponding
representation spaces. We will also give regularity conditions on the 
fixed section $s$ of the principal bundle $\al B._1$ 
considered before. Typically we will work with fixed Borel or continuous 
sections of the corresponding principal bundles.

\subsection{The Poincar\'e group:}\label{thep}
We will apply next the general scheme of 
induced representations presented above
to specify the so-called covariant and canonical
representations of the Poincar\'e group. 
These will play a fundamental 
role in the definition of the free net in the next section. 
For additional results concerning induced representations
and for the physical interpretation of the canonical (or Wigner)
representations we refer to
\cite{Barut72,bBarut80,bMackey76,bSimms68,Wigner39} as well as
\cite[Section~2.1]{Landsman95b} and \cite[Chapters~2 and 3]{bThaller92}.

\paragraph{Covariant representations:}
In the general analysis considered above let 
$\al G.:=\semi{\g{SL}{2,\C}}{\R^4}=\UPgr$ be the universal covering
of the proper orthocronous component of the Poincar\'e group. 
It acts on $M:=\R^4$ in the usual way 
$(A,a)\,x:=\Lambda_{A}x+a$, $(A,a)\in\semi{\g{SL}{2,\C}}{\R^4}$, 
$x\in\R^4$, where $\Lambda_{A}$ is the Lorentz transformation 
associated to $\pm A\in\g{SL}{2,{\C}}$ which
describes the action of $\g{SL}{2,\C}$ on $\R^4$
in the semi-direct product. Putting now $u_0:=0$ gives
$\al K._0=\semi{\g{SL}{2,\C}}{\{0\}}$, 
${{\cal G}}/{(\semi{\g{SL}{2,\C}}{\{0\}})}\cong
\R^4$, and the principal $\g{SL}{2,\C}$-bundle is in this case
${\cal B}_1:=({\cal G},\; \Wort{pr}_1,\; \R^4)$.
As inducing representation we use the finite-dimensional 
irreducible \rep{s} of $\g{SL}{2,{\C}}$ acting on the spinor 
space ${\cal H}^{\KIn{$(\frac{j}{2},\frac{k}{2})$}}
:=\Wort{Sym}\Big(\mathop{\otimes}\limits^j \z{C}^2\Big)
\otimes \Wort{Sym} \Big( \mathop{\otimes}\limits^k \z{C}^2 \Big)$ 
(cf.~\cite{bStreater89}):
i.e.~$\tau^{\KIn{(cov)}}(A,0):=D^{\KIn{$(\frac{j}{2},\frac{k}{2})$}}(A)
=\Big(\mathop{\otimes}\limits^j A\Big)
  \otimes \Big( \mathop{\otimes}\limits^k \overline{A} \Big)$, 
$(A,0)\in \semi{\g{SL}{2,\C}}{\{0\}}$.
From this we have (if no confusion arises we will omit in the 
following the index \In{$(\frac{j}{2},\frac{k}{2})$} in $D(\cdot)$ and 
in $\cal H$),
\begin{equation}
\label{CovB2}
  {\cal B}_2 (\tau^{\KIn{(cov)}}):= 
                   ({\cal G}\times_{\KIn{SL$(2,{\C})$}}{\cal H},\;
                   \Wort{pr}_2,\; \R^4).
\end{equation}
Recalling Remark~\ref{HFunct} we specify a global continuous 
section $s$ of ${\cal B}_1$ (i.e.~${\cal B}_1$ is a trivial bundle):
\[
 s\colon\ \R^4\longrightarrow {\cal G},\qquad s(x):=(\EINS,x)\in
 \semi{\g{SL}{2,{\C}}}{\R^{4}}={\cal G}\,.
\]
Note that since $\tau^{\KIn{(cov)}}$ is {\em not} a unitary representation
and since we want to relate the following so-called
{\em covariant \rep{}} with the irreducible and
unitary canonical ones presented below,  it is enough to define 
$T$ on the space of $\cal H$-valued Schwartz functions
$\Test{\R^{4},{\cal H}}$
\begin{equation}
\label{VfCov}
 ( T(g)f )(x):= D(A)\, f (\Lambda_{A}^{-1}
 (x-a)),\quad f\in \Test{\R^{4},{\cal H}}\,,
\end{equation}
where we have used that $s(x)^{-1}\,(A,a)\,s ((A,a)^{-1}x) =(A,0)$, 
$(A,a)\in {\cal G}$.
$T$ is an algebraically reducible \rep{} even if the
inducing \rep{} $\tau^{\KIn{(cov)}}$ is irreducible.

\begin{rem}
We will show later that the covariant \rep{} above
is related with the covariant transformation character of quantum 
fields. Thus a further reason for considering this \rep{}
space is the fact that in the heuristic picture we want to smear
quantum fields with test functions in $\Test{\R^{4},{\cal H}}$.
\end{rem}

\paragraph{Canonical representations:}
Next we will consider unitary and irreducible canonical representations 
of $\UPgr$ and, in particular, we will specify the massive and
those massless ones with discrete helicity. We will also state
the mathematical differences between these two types of representations.
We will mainly apply here Mackey's theory of 
induced \rep{s} of regular semi-direct products, 
where each subgroup is locally compact and one of them 
abelian \cite{bMackey76,bSimms68,bBarut80}. (However, see 
also Remark~\ref{NonUni}~(b)). 

First note that in the general context of the beginning of this section 
if $\tau$ is a unitary \rep{} of $\al K._0$ on $\al H.$, then
$\Gamma (\al G.\times_{\KIn{${\cal K}_0$}}{\cal H})$
turns naturally into a Hilbert space. Indeed, the fibres
$\Wort{pr}_2^{-1}(p)$, $p\in \DD$, inherit
a unique (modulo unitary equivalence) Hilbert space structure 
from ${\cal H}$. Assume further that $\DD$ allows 
a ${\cal G}$-invariant  measure $\mu$. (The following construction goes also
through with little modifications if we only require the existence on $\DD$
of a quasi-invariant measure w.r.t.~${\cal G}$.) Then
$\Gamma ( \al G.\times_{\KIn{${\cal K}_0$}}{\cal H})$ is the Hilbert
space of all Borel sections $\psi$ of ${\cal B}_2(\tau)$ 
that satisfy,
\[
 \askp{\psi}{\psi}=\int\limits_\DD\askp{\psi(p)}{\psi(p)}_p \mumdp <\infty,
\]
where $\langle\cdot,\cdot\rangle_p$ denotes the scalar product on the Hilbert
space $\Wort{pr}_2^{-1}(p)$, $p\in \DD$, and the induced \rep{}
given in Eq.~(\ref{FbInd}) is unitary on it.

Put now $\al G.:=\semi{{\g{SL}{2,\C}}}{\R^4}$ which acts on $\widehat{\R}^4$
by means of the dual action canonically given by the semi-direct product 
structure of $\UPgr$.
It is defined by 
\begin{equation}\label{DualGroupAction}
 \widetilde{\gamma}\colon\semi{{\g{SL}{2,\C}}}{\R^4}
         \rightarrow \Wort{Aut}\,\widehat{\R}^4
         \quad\mr and.\quad
(\widetilde{\gamma}_{\KIn{$(A,a)$}}\chi)(x)
:=\chi(\Lambda_{A}^{-1} (x)) \,,
\end{equation}
where $\chi\in\widehat{\R}^4$, $A\in \g{SL}{2,\C}$, $a,x\in \R^4$.
For a fixed character $\chi\in\widehat{\R}^4$ consider $M$ as the 
orbit generated by the previous action and
the corresponding isotropy subgroup is given by 
\[
 {\cal G}_\chi:=\LG (A,a)\in\semi{{\g{SL}{2,\C}}}{\R^4}  \mid 
 \widetilde{\gamma}_{\KIn{$(A,a)$}}\chi=\chi \RG 
 \quad\Wort{and recall that}\quad
 {\UPgr}/{{\cal G}_{\chi}} \cong\DD\,.
\]
We have now the principal ${\cal G}_{\chi}$-bundle and the associated
bundle given respectively by
\[
  {\cal B}_1:=\LR\UPgr,\;\Wort{pr}_1,\;\DD\RR\quad\Wort{and}\quad
  {\cal B}_2\!\LR\tau^{\KIn{(can)}}\RR:= 
          \LR\UPgr\times_{\KIn{${\cal G}_{\chi}$}}{\cal H},
          \;\Wort{pr}_2,\;\DD\RR ,
\]
where $\tau^{\KIn{(can)}}$ is a unitary \rep{} of
${\cal G}_{\chi}$ on ${\cal H}$.
If $\tau^{\KIn{(can)}}$ is irreducible, then the corresponding induced
\rep{}, which is called the {\em canonical representation}, is irreducible.
Even more, every irreducible \rep{} of ${\cal G}$ is obtained (modulo unitary
equivalence) in this way. Recall also that the canonical \rep{} is 
unitary iff $\tau^{\KIn{(can)}}$ is unitary.

\begin{rem}\label{AllgRwe}
In the present context \rwe{s} appear if one considers reducible \rep{s}
of the isotropy group ${\cal G}_\chi$. On the subspace of
solutions of these equations (which can be consequently characterised
by projections in $\cal H$) the reducible induced \rep{} will turn 
irreducible. We will
study \rwe{s} in the following for massive and massless \rep{s}. We will 
show that these have a fundamentally different character as a consequence
of the complementary properties of the respective little groups.
(Here we use the name little group to denote the subgroup of
$\g{SL}{2,\C}$ appearing in the isotropy group.)
Indeed, the \rep{s} of the massive little group considered
will be unitary and fully decomposable, 
while the corresponding \rep{s} of the massless little 
group will not satisfy these properties. 
Massive \rwe{s} will be characterised by 
reducing projections, while the massless
ones are associated to invariant (but not reducing) projections.
\end{rem}

\subsection{Massive canonical \rep{s}}\label{MassiveCase}

Choose a character $\chi_{\bp}$, with $\bp:=(m,0,0,0)$, $m>0$, i.e.
$\chi_{\bp}(a)=e^{-i\bp a}$, where $a\in\z{R}^4$ and $\bp\, a$ means the
Minkowski scalar product. In this case we have ${\cal G}_{{\chi}_{\bp}}=
\semi{\g{SU}{2}}{\z{R}^4}$. As unitary \rep{} of the isotropy
subgroup on ${\cal H}^{\KIn{$(\frac{n}{2},0)$}}$ we take 
\begin{equation}
\label{SigmaCan}
\tau^{\KIn{(can)}}(U,a):=e^{-i\bp a}\,D^{\KIn{$(\frac{n}{2},0)$}}(U)\,,\quad 
                         (U,a)\in {\cal G}_{{\chi}_{\bp}}.
\end{equation}
We can now consider (omitting for simplicity the index 
$\In{$(\frac{n}{2},0)$}$) the bundles,
\[
{\cal B}_1^{\KIn{(can)}} := \LR {\UPgr},\; \Wort{pr}_1,\; 
                              \HmPlus \RR \quad\Wort{and}\quad
{\cal B}_2\!\LR\tau^{\KIn{(can)}}\RR
                     := \LR {\UPgr}\times_{\KIn{${\cal G}_{{\chi}_{\bp}}$}}{
                             \cal H},\; \Wort{pr}_2,\; \HmPlus \RR,
\]
where we have used the diffeomorphism 
${{\UPgr}}/{{\cal G}_{{\chi}_{\bp}}}\cong\HmPlus
:=\{p\in\z{R}^4\mid p^2=m^2\;,\;p^0>0\}$
between the factor space and the 
positive mass shell $\HmPlus$.
$\mumdp$ denotes the corresponding invariant measure on $\HmPlus$.

The principal ${\cal G}_{{\chi}_{\bp}}$-bundle ${\cal B}_1^{\KIn{(can)}}$
is trivial \cite{Boya74}. Indeed, a global continuous section $s$ is
\begin{equation}
\label{MassiveSection}
 s\colon\ \HmPlus\longrightarrow \UPgr,\qquad s(p):= ( H_p,\,0 )\in
 \semi{\g{SL}{2,{\z{C}}}}{\z{R}^{4}}=\UPgr \,,
\end{equation}
where $H_{p}:=\frac{1}{\sqrt{2m(m+p_{0})}}(m+P)$,
$P=\sum\limits_\mu p_\mu\sigma_\mu$ 
\cite[Section~A.1]{Lledo95} and
$\sigma_{\mu}$, $\mu=0,1,2,3$, are the unit and the 
Pauli matrices. The assignment $p\to P$ 
defines a vector space isomorphism between
$\R^4$ and the set of self-adjoint elements in $\Wort{Mat}_2(\z{C})$.

So, once the section $s$ is fixed and using $s(p)^{-1}\,(A,a)\,s(q)
= ( H_p^{-1}\,A\,H_q,\; \Lambda_{H_p^{-1}}\,a )\in
{\cal G}_{{\chi}_{\bp}}$, as well as
$q=\Lambda_A^{-1}p\in \HmPlus$, we have on 
$\Lzwei{\HmPlus,{\cal H},\mumdp}$ the massive canonical \rep{} 
(cf.~Eq.~(\ref{VfInd})),
\begin{equation}
\label{VfCan}
 \LR V^{\KIn{(can)}}(g)\varphi\RR\!(p)=e^{-ipa} 
    D( H_p^{-1} A\, H_q) \varphi (q),
\end{equation}
where $g=(A,a)\in \semi{\g{SL}{2,\z{C}}}{\z{R}^4}$ and $\varphi\in
\Lzwei{\HmPlus,{\cal H},\mumdp}$.
The element $H_p^{-1} A\, H_q\in\g{SU}{2}$ is called the Wigner rotation
(e.g. \cite[Section~2.3]{WeinbergIn65}). $V^{\KIn{(can)}}$ is unitary
w.r.t.~usual L$^2$-scalar product,
\[
 \askp{\varphi_1}{\varphi_2}:=\int\limits_{\HmPlus}
                   \askp{\varphi_1(p)}{\varphi_2(p)}_{\cal H} \mumdp.
\]

\begin{rem}
\label{Rem.3.2.1}
The representation
$( \Lzwei{\HmPlus,{\cal H},\mumdp},\;V^{\KIn{(can)}},\;
\langle\cdot,\cdot\rangle)$ is equivalent to the \rep{} $( h_m,\;V_1,\;
\langle\cdot,\cdot\rangle_{\beta})$ used in \cite[Section~A.1]{Lledo95},
where we define first for $\varphi, \psi$ a pair of ${\cal
H}$-valued functions the sesquilinear form
\[
 \askp{\varphi}{\psi}_{\beta}:= \int\limits_{\HmPlus}
 \askp{\varphi(p)}{\beta(p)\ \psi(p)}_{\cal H} \mumdp, 
\]
with
\begin{eqnarray}
 \label{betaPlus}
\beta(p) & := & 
  D^{\KIn{$(\frac{n}{2},0)$}}({P^\KIn{$\dagger$}})\kern2mm=
            \kern2mm\mathop{\otimes}\limits^n P^{\KIn{$\dagger$}}\,,
                               \\ \label{PPlus}
 P^{\KIn{$\dagger$}}\label{Dagger} 
         & := &\frac1m \left( p_{0}\sigma_{0}-\sum\limits_{i=1}^{3} 
                                                 p_{i}\sigma_{i}\right)
 \kern3mm=\kern3mm \LR H_p^{-1}\RR^*H_p^{-1}\,.
\end{eqnarray}
Then we put
\begin{equation}
 h_m:=\LG\varphi\colon\ \HmPlus\longrightarrow {\cal H} \mid
 \varphi\; \Wort{is Borel and}\quad 
 \langle\varphi ,\varphi\rangle_{\beta}< \infty \RG. 
\end{equation}
Finally, the \rep{}
\begin{equation}
  \LR V_1(g)\kern.2em\varphi \RR(p)
  := e^{-ipa}\ D^{\KIn{$(\frac{n}{2},0)$}}(A)\ \varphi(q),
\end{equation}
for $g=(A,a)\in\UPgr=\semi{\g{SL}{2,{\z{C}}}}{\z{R}^{4}}$, $\varphi\in h_m$
and $q:=\Lambda_{A}^{-1}p$, is unitary 
w.r.t.~$\langle\cdot,\cdot\rangle_{\beta}$. 
Roughly, we have removed the $H_p$-matrices from the definition of 
$V_1$ at the price of introducing a positive
operator-valued function on $\HmPlus$,
$p\mapsto\beta(p)$, in the definition of the corresponding scalar product.
This equivalent definition of canonical \rep{} has been
very useful in order to construct massive free nets \cite{Lledo95}.
We will also adapt this idea to the massless case.

The equivalence of the \rep{s} mentioned above is given explicitly by 
the following isometry 
$\Psi\colon\ \Lzwei{\HmPlus,{\cal H},\mumdp}\to h_m$,
\begin{equation}
\label{Isometry}
 \LR\Psi\varphi\RR\!(p):=D(H_p)\varphi(p),
\end{equation}
and it is immediately checked that 
$\Psi\; V^{\KIn{(can)}}(g)=V_1(g)\;\Psi$, $g\in\UPgr$.
\end{rem}

\begin{rem}
\label{Rem.3.2.2}
It is now easy to relate the algebraically reducible \rep{} 
$T^{\KIn{(cov)}}$ in (\ref{VfCov}) with 
the canonical \rep{} $V^{\KIn{(can)}}$ given in Eq.~(\ref{VfCan}). 
Indeed, recalling the definitions
introduced in Remark~\ref{Rem.3.2.1}, we consider the embedding 
\[
 {\got I}\colon\ \Test{\z{R}^{4},{\cal H}}
 \longrightarrow h_m \,,
\] 
defined as Fourier transformation, 
$\hat{f}(p)=\int_{\R^4} e^{-ipx} f(x) d^4x$,
and restriction to $\HmPlus$ (recall \cite[Section~IX.9]{bReedII}). 
Then the equation
\begin{equation}
\label{Intertwiner}
  {\got I}\; T^{\KIn{(cov)}}(g)=V_1(g)\; {\got I},\quad g\in\UPgr\,,
\end{equation}
holds. From the preceding Remark we already know that $V_1$ and 
$V^{\KIn{(can)}}$ are equivalent.
\end{rem}

\subsection{Massive relativistic wave equations}
\label{Subsec2_3}

As already mentioned in Remark~\ref{AllgRwe} \rwe{s} appear when one
considers reducible \rep{s} of the little group $\g{SU}{2}$. 
They serve to reduce the corresponding induced \rep{} and therefore
it is natural to associate with \rwe{s} 
corresponding projections on $\cal H$. We will also illustrate the 
preceding results with two examples that have nontrivial spin, 
namely the Dirac and the Proca Equations.

Niederer and O'Rafeartaigh present Eq.~(20.8) in \cite{Niederer74b} as
``\ldots the most general covariant wave equation corresponding to a 
given (nonzero) mass and spin \ldots" (see Definition~\ref{MassRWEp}
below). It is useful to recognise 
that in \cite{Niederer74b} and \cite[Chapter~21]{bBarut80} the spaces 
$\LR h_m,\;\langle\cdot,\cdot\rangle_{\beta}\RR$ 
of Remark~\ref{Rem.3.2.1} are essentially used. In our context we can
equivalently write the mentioned equation also for the space
$(\Wort{L}^2,\,V^{\KIn{(can)}})$. Using Eq.~(\ref{Isometry})
the following result is a straightforward consequence of the mentioned
equivalence. 
\begin{lem}
\label{Lem.3.2.3}
Let $\pi$ be a reducing orthoprojection w.r.t.~$\tau^{\KIn{(can)}}$, 
i.e. $\pi\, \tau^{\KIn{(can)}}(g)= \tau^{\KIn{(can)}}(g)\,\pi$, 
$g\in\semi{\g{SU}{2}}{\z{R}^{4}}$. Then the following 
equations are equivalent:
\begin{itemize}
\item[{\rm (i)}] $\pi\,\widetilde{\varphi}(p)=\widetilde{\varphi}(p),\;$ 
          for $\;\widetilde{\varphi}\in\Lzwei{\HmPlus,{\cal H},\mumdp}$.
\item[{\rm (ii)}] $\pi(p)\,\varphi(p)=\varphi(p),\;$ 
                  where $\pi(p):= D(H_p)\,\pi\, D(H_p)^{-1}$ and 
                  $\;\varphi(p):=D(H_p)\widetilde{\varphi}(p) \in h_m$, 
                  $p\in\HmPlus$.
\end{itemize}
\end{lem}

Following Niederer and O'Rafeartaigh we introduce the notion
of massive relativistic wave equation
(see \cite[\S~20 and \S~21]{Niederer74b} for further details and motiviation).

\begin{defi}\label{MassRWEp}
Let $\pi$ be a reducing orthoprojection 
w.r.t.~$\tau^{\KIn{(can)}}\restriction (\semi{\g{SU}{2}}{\z{R}^{4}})$.
Then we call 
\[
 \pi(p)\,\varphi(p)=\varphi(p)\;,\quad \varphi \in h_m\,,\;p\in\HmPlus\,,
\]
a massive relativistic wave equation associated with $\pi$ 
(cf.~Lemma~\ref{Lem.3.2.3}~(ii)).
\end{defi}

\begin{rem}
\begin{itemize}
\item
The equation in the previous definition  
coincides with \cite[Eq.~(20.8)]{Niederer74b} or
\cite[Eq.~(17) of Section~21.1]{bBarut80}. Specifying 
$\tau^{\KIn{(can)}}$ and $\pi$ in this context one obtains 
the conventional massive relativistic wave equations
written in momentum space.
For convenience of the reader we mention the examples of Dirac's and
Proca's equation.
More examples of massive \rwe{s} and the corresponding
projections are summarised in \cite[Table~2]{Niederer74b}.

\item
Note further, that the subspace
of $h_m$ characterised by the equation in 
Lemma~\ref{Lem.3.2.3}~(ii) is $V_1$-invariant. Use
for example the relation:
\[
 D(A)^{-1}\,\pi(p)\,D(A)=\pi (\Lambda_A^{-1}p).
\]
Thus we have seen that the reducing subspaces of $\cal H$ are in
correspondence with relativistic wave equations. Recall that
by compactness the unitary representation
$D^{\KIn{$(\frac{j}{2},\frac{k}{2})$}}\restriction\g{SU}{2}$, 
$(j,k)\not= (0,0)$, is fully decomposable, i.e.~it can be decomposed
as a direct sum of irreducible subrepresentations.
\end{itemize}
\end{rem}

\paragraph{The Dirac Equation}\cite[Section~9]{bSimms68}:
Take ${\cal H}:=\z{C}^4$ and
      $\tau(A):=\left(\kern-1.5mm \begin{array}{cc}
                      A \kern-1mm &  0               \\
                      0 \kern-1mm &  \LR A^* \RR^{-1}
                 \end{array} \kern-1.5mm\right) \in \g{SL}{4,\z{C}}$
for $A\in \g{SL}{2,{\z{C}}}$. As reducible inducing \rep{} we use 
$\tau(U,a):=e^{-i\bp a}\,\tau(U)\,,\quad (U,a)\in
 \semi{\g{SU}{2}}{\z{R}^{4}}\,.$
Now, consider the orthoprojection 
\[
 \pi^{\KIn{(Dirac)}}:=\frac{\displaystyle 1}{\displaystyle 2}
           \left(\kern-1.5mm \begin{array}{cccc}
   1 \kern-1mm & 0 \kern-1mm & 1 \kern-1mm & 0             \\
   0 \kern-1mm & 1 \kern-1mm & 0 \kern-1mm & 1             \\ 
   1 \kern-1mm & 0 \kern-1mm & 1 \kern-1mm & 0             \\
   0 \kern-1mm & 1 \kern-1mm & 0 \kern-1mm & 1                                           
           \end{array} \kern-1.5mm\right) \,,
\]
which satisfies $\pi^{\KIn{(Dirac)}}\,\tau(U)
=\tau(U)\,\pi^{\KIn{(Dirac)}}$, $U\in\g{SU}{2}$, and using the isometry
$\Psi$ as well as the notation
$\varphi(p):=\LR\Psi\widetilde{\varphi}\RR\!(p)=
                           \tau(H_p)\widetilde{\varphi}(p)$ 
we can reformulate the preceding lemma in the present context as:

\begin{pro}
\label{Lem.3.2.4}
With the preceding notation we have that 
$\varphi$ satisfies the Dirac Equation\hspace{1mm} iff \hspace{1mm} 
$\widetilde{\varphi}$ satisfies 
$\pi^{\KIn{(Dirac)}}\,\widetilde{\varphi}(p)=\widetilde{\varphi}(p)$, 
$p\in\HmPlus$.
\end{pro}
\begin{beweis}
First of all we note that from Lemma~\ref{Lem.3.2.3},
$\pi^{\KIn{(Dirac)}}\,\widetilde{\varphi}(p)=\widetilde{\varphi}(p)$
iff $\pi^{\KIn{(Dirac)}}(p)\,\varphi(p)=\varphi(p)\,,$
where $\pi^{\KIn{(Dirac)}}(p):= \tau(H_p)\,\pi^{\KIn{(Dirac)}}\, 
\tau(H_p)^{-1}$. We calculate
\[
 \pi^{\KIn{(Dirac)}}(p)= \frac{\displaystyle 1}{\displaystyle 2m}
        \left(\kern-1.5mm \begin{array}{cc}
                       m\EINS    \kern-1mm &  P               \\
                      \widehat{P}  \kern-1mm &  m\EINS
        \end{array} \kern-1.5mm\right),
\]
where $P:=p_0\sigma_0+\sum\limits_{i=1}^3 p_i\sigma_i$ and
$\widehat{P}:=p_0\sigma_0-\sum\limits_{i=1}^3 p_i\sigma_i$, so that
$P \widehat{P}=p^2\EINS=m^2\EINS$. Denoting $\gamma(p):=
     \left(\kern-1.5mm \begin{array}{cc}
                          0 \kern-1.5mm &  P               \\
                \widehat{P} \kern-1.5mm &  0
     \end{array} \kern-1.5mm\right)$
we obtain $\pi^{\KIn{(Dirac)}}(p)=\frac{1}{2m}\LR m\EINS+\gamma(p)\RR$ 
and thus $\pi^{\KIn{(Dirac)}}(p)\,\varphi(p)=\varphi(p)\;\Wort{iff}\;
\gamma(p)\,\varphi(p)=m\,\varphi(p)$, which is the Dirac equation in momentum
space notation.
\end{beweis}

\paragraph{The Proca Equation}\cite{Niederer74b}:
In the present case we put ${\cal H}:=\z{C}^2\otimes\z{C}^2
\cong\z{C}^4$ and as inducing representation we use
\[
 \tau(U,a):=e^{-i\bp a}\, D^{\KIn{$(\frac12,\frac12)$}}(U)= e^{-i\bp a}\, 
 U\otimes\overline{U}\,, \quad(U,a)\in \semi{\g{SU}{2}}{\z{R}^{4}}.
\]
As an orthoprojection we take,
\[
 \pi^{\KIn{(Proca)}}:=\frac{\displaystyle 1}{\displaystyle 2}
           \left(\kern-1.8mm \begin{array}{rccc}
   1 \kern-1mm & 0 \kern-1mm & 0 \kern-1.5mm & -1            \\
   0 \kern-1mm & 2 \kern-1mm & 0 \kern-1.5mm & 0             \\ 
   0 \kern-1mm & 0 \kern-1mm & 2 \kern-1.5mm & 0             \\
  -1\kern-1mm & 0 \kern-1mm & 0 \kern-1.5mm & 1                                            \end{array} \kern-1.5mm\right),
\]
which satisfies $\pi^{\KIn{(Proca)}}\,\tau(U)
=\tau(U)\,\pi^{\KIn{(Proca)}}$, $U\in\g{SU}{2}$. Further details,
e.g.~the explicit relation between the canonical and covariant
descriptions in this particular case can be found in \cite{Boya74a}.

\subsection{Massless canonical representations}\label{Subs.2.4}
To specify massless representations with discrete helicity we choose
a character $\chi_{\bp}$, $\bp:=(1,0,0,1)\in\Lkf$ 
(the mantle of the forward light cone), i.e.
$\chi_{\bp}(a)=e^{-i\bp a}$, $a\in\R^4$ and $\bp\, a$ means the
Minkowski scalar product. A straightforward computation shows that
the isotropy subgroup is given by
${\cal G}_{{\chi}_{\bp}}=\semi{{\cal E}(2)}{\R^4}$, where
the two-fold cover of the 2-dimensional Euclidean group is
\begin{equation}
\label{Delta}
{\cal E}(2) := \left\{
            \left(\kern-1.5mm \begin{array}{cc}
            e^{\frac{i}{2}\theta} & e^{-\frac{i}{2}\theta}\ z               \\
                                0 & e^{-\frac{i}{2}\theta}
            \end{array} \kern-1.5mm\right) \in \g{SL}{2,\C}
            \mid \theta\in[0,4\pi),\; z\in\C
            \right\}.
\end{equation}
The little group ${\cal E}(2)$ is noncompact and since its commutator
subgroup is already abelian it follows that ${\cal E}(2)$ is 
solvable. Further, it has again the structure of a semi-direct product. 
(In contrast with these facts
we have that the massive little group $\g{SU}{2}$ is compact and
simple.) Since ${\cal E}(2)$ is a connected and solvable Lie group we know
from Lie's Theorem (cf.~\cite{bBarut80}) that the only
finite-dimensional irreducible \rep{s} are 1-dimensional, 
i.e.~${\cal H}:=\C$. Therefore in order to induce irreducible 
and unitary \rep{s} of the whole group that describe discrete helicity
values we define
\begin{equation}
\label{Sigma0Can}
 \tau^{\KIn{(can)}}(L,a):=e^{-i\bp a}\, \LR e^{\frac{i}{2}\theta}\RR^n,
\end{equation}
where $(L,a)\in \semi{{\cal E}(2)}{\R^4}={\cal G}_{{\chi}_{\bp}}$, 
$n\in \N$. Note that this \rep{} is not faithful. Indeed, 
the normal subgroup $\left\{
      \left(\kern-1.5mm \begin{array}{cc}
          1 & z               \\
          0 & 1
  \end{array} \kern-1.5mm\right)\mid  z\in\C\right\}$
is trivially represented (see also \cite[Section~II]{Weinberg64a}). 
Some authors associate this subgroup to certain gauge degrees of 
freedom of the system (e.g.~\cite{Han81,Kim87,Shnerb94}).
We consider next the bundles,
\[
{\cal B}_1^{\KIn{(can)}} := \LR \UPgr,\; \Wort{pr}_1,\; 
                              \Lkf \RR \quad\Wort{and}\quad
{\cal B}_2\!\LR\tau^{\KIn{(can)}}\RR
                     := \LR \UPgr\times_{\KIn{${\cal G}_{{\chi}_{\bp}}$}}{
                              \!\C},\; \Wort{pr}_2,\; \Lkf \RR,
\]
where we have used the diffeomorphism  
${\UPgr}/{{\cal G}_{{\chi}_{\bp}}}\cong\Lkf
:=\{p\in\z{R}^4\mid p^2=0\;,\;p^0>0\}$
between the factor space and the mantle of the forward 
light cone. We denote by $\mudp$ the corresponding invariant measure on 
$\Lkf$.

In contrast with the massive case the bundle ${\cal B}_1^{\KIn{(can)}}$
has no global continuous section. The reason for this lies in the
following topological obstruction: if ${\cal B}_1^{\KIn{(can)}}$ 
had a global continuous section (hence would be a trivial
bundle), then the $n$-th homotopy groups $\Pi_n(\cdot)$, $n\in\N$,
of the total space would be 
equal to the direct sum of the homotopy groups of the isotropy 
subgroup and of the base manifold. We can check in particular 
for the second homotopy group that on the one hand
\[
\Pi_2(\al G.)=\Pi_2(S^3)=0
\]
and on the other hand
\[
 \Pi_2({\cal G}_{{\chi}_{\bp}})\oplus\Pi_2(\Lkf)
 =\Pi_2(S^1)\oplus\Pi_2(S^2)=\Z
\]
(see \cite{Boya74} for further details).

Nevertheless, 
we can specify a Borel section considering a continuous one
in a chart that does not include the set $\{ p\in\Lkf\mid  p_3=-p_0\}$ 
(which is of measure zero w.r.t.~$\mudp$).
Putting $\Lk:=\Lkf\setminus \{ p\in\Lkf\mid  p_3=-p_0\}$
a (local) continuous section
is given explicitly by
\begin{equation}
\label{Section0}
 s\colon\ \Lk\longrightarrow \UPgr,\qquad s(p):=( H_p,\,0 )\in
             \semi{\g{SL}{2,{\C}}}{\R^{4}}=\UPgr,
\end{equation}
where
\begin{equation}
\label{SolHp}
    H_{p} := \frac{1}{\sqrt{2p_{0}(p_{0}+p_{3})}}
       \left(\kern-1.5mm \begin{array}{cc}
       -\sqrt{p_{0}}\ (p_{0}+p_{3})  &
       \displaystyle \phantom{-}\frac{p_{1}-ip_{2}}{\sqrt{p_{0}}}           \\
       -\sqrt{p_{0}}\ (p_{1}+ip_{2}) &
       \displaystyle -\frac{p_{0}+p_{3}}{\sqrt{p_{0}}}^{}
       \end{array} \kern-1.5mm\right).
\end{equation}
Recall that the $H_p$-matrices satisfy the equation
\begin{equation}
\label{DefHp}
H_{p}^{} \left(\kern-1.5mm \begin{array}{cc} 2\kern-1.6mm &0 \\ 
           0\kern-1.6mm &0 \end{array} \kern-1.5mm\right)
H_{p}^{*} = P, \quad \mbox{where}\quad
P = \left(\kern-1.5mm \begin{array}{cc}
    p_{0}+p_{3}  & p_{1}-ip_{2} \\
    p_{1}+ip_{2} & p_{0}-p_{3}
    \end{array} \kern-1.5mm\right)
  = p_0\sigma_0+\sum \limits_{i=1}^3 p_{i} \sigma_{i}\,.
\end{equation}
We use here, as in the massive case, the vector space isomorphism between
$\R^4$ and the set of self-adjoint elements in $\Wort{Mat}_2(\z{C})$.

\begin{rem}\label{BorelSect}
The representation spaces of the canonical representations are typically
$L^2$-spaces and therefore it is enough to consider fixed Borel sections
as above. The fact that we are allowed to choose a continuous section
of the corresponding bundle $\al B._1$ in the massive case is an other 
pleasant and characteristic feature of these models. 

From the point of view of quantum fields (to be defined 
in Section~\ref{Mqf}) what is crucial is the fact that the 
singularity of (\ref{Section0}) does not affect the continuity 
(w.r.t.~to the Schwartz and $L^2$-topologies) of the 
embedding that intertwines the covariant and the massless
canonical representations. 
Indeed, in Theorem~\ref{Wcontinuity} we give a detailed
proof of the mentioned continuity for the embedding associated to
the Weyl case and that uses explicitly the section 
defined in Eq.~(\ref{Section0}). This ensures that the 
2-point distributions defined by the corresponding 
massless quantum fields are tempered.
\end{rem}

If we consider the section in Eq.~(\ref{Section0}) fixed, then we have on 
$\mathrm{L}^2(\Lkf,\C,\mudp)$ the canonical massless representations 
(cf.~Eq.~(\ref{VfInd}))
\begin{equation}
\label{Vf0Can}
 \LR U_{\pm}(g)\varphi\RR\!(p)=e^{-ipa}\LR 
                        e^{\pm\frac{i}{2}\theta(A,p)}\RR^n\varphi (q),
\end{equation}
where $g=(A,a)\in \semi{\g{SL}{2,\C}}{\R^4}$,  
$n\in \N$, $q:=\Lambda_{A}^{-1}p$ and for 
$A=\left(\kern-1.5mm \begin{array}{cc}
          a \kern-2mm& b               \\
          c \kern-2mm& d
  \end{array} \kern-1.5mm\right)\in \g{SL}{2,\C}$ we compute
\[
  e^{-\frac{i}{2}\theta(A,p)}:=\LR H_p^{-1} A\, H_q\RR_{22}=
  \frac{-b(p_{1}+ip_{2})+d(p_{0}+p_{3})}
              {|-b(p_{1}+ip_{2})+d(p_{0}+p_{3})|}\,.
\]
$U_{\pm}$ are unitary w.r.t.~usual L$^2$-scalar product, satisfy the 
spectrum condition and the helicity of the model carrying 
one of these representations is $\pm\frac{n}{2}$.

\begin{rem}
\label{Rem.3.3.1}
In contrast with the massive case, it is now clear that in order to
relate the covariant representation with the canonical massless \rep{}
it will not be enough to consider the Fourier transformation
of suitable test functions and its
restriction to $\Lkf$ (cf.~Remark~\ref{Rem.3.2.2}). Indeed, the fibres
of ${\cal B}_2(\tau^{\KIn{(can)}})$ are 1-dimensional, while
the fibres of ${\cal B}_2(\tau^{\KIn{(cov)}})$ are at least
2-dimensional if one chooses a nontrivial inducing \rep{} 
$\tau^{\KIn{(cov)}}$. In other words, if the models describe
nontrivial helicity, then some further restriction must be performed 
on the fibres in order to 
reduce the covariant \rep{} to the unitary and irreducible canonical one
(for a more detailed analysis of this reduction see 
Subsection~\ref{FBRED}). 
There are at least three ways to perform the mentioned reduction:
\begin{itemize}
\item[(i)] One possibility of restricting the dimension of the fibres is
 to work on the space of solutions of the massless \rwe{s}
 (cf.~Remark~\ref{AllgRwe}). We will follow this alternative in 
 the following subsections and present some explicit examples.
\item[(ii)] A second possibility is to use certain semidefinite sesquilinear
 forms and the reduction is done by means of the factor spaces that can be
 naturally constructed from the degeneracy subspaces of the sesquilinear 
 form. This possibility is studied in \cite[Part~B]{Lledo95}
 and \cite{Lledo01} (cf.~also Remark~\ref{Rem.3.3.4}).
 \item[(iii)] Finally one can perform the mentioned
reduction for the bosonic models at the C*-level
by the so-called $T$-procedure of Grundling
and Hurst \cite{Grundling85} (see \cite{Lledo97,tLledo98} for details).
Imposing `quantum constraints' on the C*-algebra level will show to be
equivalent to consider the space of solutions of massless
\rwe{s} as reference space of the corresponding CCR-algebra.
\end{itemize}
\end{rem}

In the next section the embedding $\ot I.$ intertwining the covariant
and massless canonical representations (including the corresponding
fibre reduction) will be the fundamental entity. In fact,
$\ot I.$ specifies completely the corresponding net of C*-algebras
satisfying Haag-Kastler's axioms. Each possibility to carry out the 
reduction described in (i)-(iii) above, requires its own embedding.
Nontheless the 
corresponding nets of local C*-algebras turn out to be
equivalent (cf.~Remark~\ref{GenB}~(i)).

\subsection{Massless \rwe{s}}
In the present subsection we will extend Niederer and
O'Rafeartaigh's systematic analysis of massive relativistic
wave equations to the massless case 
(recall e.g.~Definition~\ref{MassRWEp}). We will point
out the fundamental differences between these two case.

We begin showing that, in contrast with the massive case 
(cf.~Subsection~\ref{Subsec2_3}), and due to the 
mathematical nature 
of the massless little group one has to give up 
the central notion of reducing projection for the inducing
representation. In fact, we will show that, in the context
of massless canonical representations, the useful objects
are the invariant (but not reducing) projections of the 
corresponding inducing representations: 
Motivated by the form
of the covariant \rep{} in Eq.~(\ref{VfCov}) (the one we want to 
reduce) we will consider as inducing \rep{s} of the isotropy subgroup
\[
 \tau(L,a):=e^{-i\bp a} D^{\KIn{$(\frac{j}{2},\frac{k}{2})$}}(L)\,,\quad
            (L,a)\in \semi{{\cal E}(2)}{\z{R}^4}\,,
\]
which act on the Hilbert space 
${\cal H}^{\KIn{$(\frac{j}{2},\frac{k}{2})$}}$ of dimension $(j+1)(k+1)$.
(In the rest of this subsection we will denote again 
$D^{\KIn{$(\frac{j}{2},\frac{k}{2})$}}(\cdot)$ simply by $D(\cdot)$ etc.) 
These \rep{s} are nonunitary and
reducible\footnote{Wigner already observes 
in \cite[p.~670]{Wigner47} using a different formal 
approach that the massless wave equations cannot
be obtained in general from the massive ones by putting $m=0$. He also
mentions that the irreducible and invariant linear manifold of states
corresponding to the massive \rep{s} turns reducible for $m=0$ and
considering the value of the spin bigger than $\frac12$.} for any
$(j,k)\neq (0,0)$. They are also not fully
decomposable \cite{Shaw65}, \cite[p.~607]{bBarut80}
in contrast to the massive case, i.e.~they can not be decomposed as 
direct sum of irreducible subrepresentations.
Therefore, no nontrivial reducing projection $\pi$ will 
exist\footnote{This is possibly a reason
why in \cite[Sections~22-25]{Niederer74b} the authors do 
not follow the elegant approach used to describe the massive 
wave equations, when they study the massless case.}
in this context (we will compute
explicitly some intertwiner spaces in the following section). 

Therefore by the general theory, the corresponding induced \rep{} 
of $\UPgr$, which is given for $g=(A,a)\in
\semi{\g{SL}{2,\z{C}}}{\z{R}^4}$, $q:=\Lambda_A^{-1}p
\in\Lk$, $\varphi\in\Lzwei{\Lkf,{\cal H},\mudp}$ by 
\begin{equation}
\label{VfCan0}
 ( \widetilde{V}(g)\varphi )(p)=e^{-ipa} D ( H_p^{-1} A\, H_q)
                       \varphi (q)\,,\quad H_p^{-1} A\,H_q\in{\cal E}(2)\,,
\end{equation}
will also be nonunitary and reducible. 

\begin{rem}
\label{Rem.3.3.3}
Before studying in detail the following examples of massless \rwe{s} we will
introduce the massless analogue to the useful space $\LR h_m,\;
\langle\cdot,\cdot\rangle_{\beta}\RR$ presented in the context of 
massive \rep{s} in Remark~\ref{Rem.3.2.1}: for $\varphi, 
\psi$ a pair of ${\cal H}$-valued functions the sesquilinear form,
\[
 \askp{\varphi}{\psi}_+:= \int\limits_{\Lkf}
 \askp{\varphi(p)}{\widetilde{\beta_+}(p)\ \psi(p)}_{\cal H} \mudp, 
\]
where $\widetilde{\beta_+}(p):=D ( H_p^{-1})^* D( H_p^{-1})$ and 
$H_p$, $p\in\Lk$, is defined in Eq.~(\ref{SolHp}). Then we put
\begin{equation}
 h_{0}:=\LG\varphi\colon\ \Lkf\longrightarrow {\cal H} \mid
  \varphi\; \Wort{is Borel and}\quad 
  \langle\varphi ,\varphi\rangle_+< \infty \RG. 
\end{equation}
Finally, the \rep{}
\begin{equation}
\label{GeneralV}
  \LR V(g)\,\varphi \RR\!(p):= e^{-ipa}\, D(A)\, \varphi(q),
\end{equation}
for $g=(A,a)\in\UPgr=\semi{\g{SL}{2,{\z{C}}}}{\z{R}^{4}}$, 
$\varphi\in h_{0}$ and $q:=\Lambda_{A}^{-1}p\in \Lk$, 
is equivalent to the \rep{} $\widetilde{V}$
defined in (\ref{VfCan0}), i.e.~there exists an isometry, $\Psi\colon\ 
\Lzwei{\Lkf,{\cal H},\mudp}\longrightarrow h_{0}$,
given by
\begin{equation}
\label{Isometry0}
 \LR\Psi\varphi\RR\!(p):=D(H_p)\,\varphi(p),
\end{equation}
such that, $\Psi\; \widetilde{V}(g)=V(g)\;\Psi$, $g\in\UPgr$. The \rep{} $V$ 
is also reducible and nonunitary w.r.t.~$\langle\cdot,\cdot\rangle_+$.
\end{rem}

\begin{rem}
\label{Rem.3.3.4}
If we require the \rep{} $V$ to leave the sesquilinear form
$\langle\cdot,\cdot\rangle_+$ invariant, then we are forced 
(cf.~\cite{Barut72})
to redefine the operator-valued function $\widetilde{\beta_+}$ 
above as
\[
  \beta_+(p):=D\!\LR \overline{H_p^{-1}}\RR^* 
         D\!\LR \left(\kern-1.5mm\begin{array}{cc}
              0 \kern-1.6mm & 0 \\ 
              0 \kern-1.6mm & 1
          \end{array} \kern-1.5mm\right) \RR
              D\!\LR \overline{H_p^{-1}}\RR
            =:D(\overline{P^{\KIn{$\dagger$}}})\,,
            \quad\Wort{with}\;\; 
                    P^{\KIn{$\dagger$}}=\frac12 \left(p_0\sigma_0
                    -\sum\limits_{i=1}^3 p_i\sigma_i\right)\,,
\]
which is only a semidefinite operator on ${\cal H}$ for each $p\in\Lk$, 
(compare with the massive case in Remark~\ref{Rem.3.2.1}). This option is
taken in \cite{Lledo01} (cf.~also \cite[Part~B]{Lledo95}), and has as a
consequence that the necessary reduction that must be performed to compare
the covariant with the canonical \rep{} (cf.~Remark~\ref{Rem.3.3.1})
is done by means of the factor spaces that can be naturally 
constructed from the degeneracy subspace of the sesquilinear form 
$\langle\cdot,\cdot\rangle_{\beta_+}$. This redefinition of the
$\beta$-functions is related with the fact of imposing massless
relativistic wave equations, which is what we will examine below.
\end{rem}

Since in the present context no nontrivial reducing projections exist
we will need to introduce the following family of invariant projections:
\begin{defi}
From the set of all projections in $\cal H$, select those
orthoprojections $\pi$ that are invariant 
w.r.t.~$D\restriction\al E.(2)$ and satisfy the 
equation\footnote{Note that this condition is trivially satisfied by
the reducing projections chosen in the massive case.}
\be\label{CondProj}
 \pi\,D(L)^*D(L)\,\pi=\pi\,,\quad L\in\al E.(2)\,.
\ee
\end{defi}

\begin{teo}\label{TInvProj}
Let $\pi$ be as in the preceding definition and put
\[
 \widetilde{h}:=\{\varphi\in\Lzwei{\Lkf,{\cal H},\mudp}\mid 
                  \pi\,\widetilde{\varphi}(p)=\widetilde{\varphi}(p)\}\,.
\]
Then $\widetilde{h}$ is a closed $\widetilde{V}$-invariant subspace 
of $\Lzwei{\Lkf,{\cal H},\mudp}$ and $\widetilde{V}(g)\restriction
\widetilde{h}$, $g\in\UPgr$, is unitary. Further if $D(\al E.(2))$
is irreducible on $\pi\al H.$, then $\widetilde{V}\restriction
\widetilde{h}$ is also irreducible.
\end{teo}
\begin{beweis}
It is obvious that $\widetilde{h}$ is a closed subspace and since
for $g=(A,a)\in\UPgr$ we have $H_p^{-1}\,A\,H_q\in\al E.(2)$, 
$q:=\Lambda_A^{-1}p$, the invariance follows for $\varphi\in
\widetilde{h}$ from 
\begin{eqnarray*}
 \pi (\widetilde{V}(g)\varphi)(p)
  &=& e^{-ipa} \pi\,D ( H_p^{-1} A\, H_q) \varphi (q)
  \kern2mm=\kern2mm e^{-ipa} \pi\,D( H_p^{-1}A\,H_q)\pi\,\varphi (q)\\
  &=& e^{-ipa} \,D ( H_p^{-1} A\, H_q)\pi\, \varphi (q) 
  \kern2mm=\kern2mm(\widetilde{V}(g)\varphi)(p)\,.
\end{eqnarray*}
Further, for $\varphi,\psi\in\widetilde{h}$ we also have
\begin{eqnarray*}
 \askp{\widetilde{V}(g)\varphi}{\widetilde{V}(g)\psi}
  &=&\int\limits_{\Lkf}\askp{
    D(H_p^{-1} A\, H_q)\,\varphi (q)}{D(H_p^{-1} A\, H_q)\,\psi (q)}_{\al H.} 
    \mudp  \\
  &=&\int\limits_{\Lkf}\askp{
     \varphi (q)}{\pi\,D(H_p^{-1}A\,H_q)^*D(H_p^{-1}A\,H_q)\,\pi\;\psi (q)}_{
     \al H.} \mudp  \\
   &=& \askp{\varphi}{\psi}\,.
\end{eqnarray*}
Here we have used Eq.~(\ref{CondProj}) and the invariance of $\mu_0$.
The irreducibility statement follows from the general theory of induced
\rep{s} stated before in this section.
\end{beweis}

The condition $\pi\,\widetilde{\varphi}(p)=\widetilde{\varphi}(p)$ used
before can be rewritten in terms of the equivalent space
$(h_{0},V,\langle\cdot,\cdot\rangle_+)$ of Remark~\ref{Rem.3.3.3}.
This will give the massless \rwe{s} written in its usual form.
\begin{lem}
\label{Lem.3.3.5}
Let $\pi$ be an invariant, 
orthoprojection w.r.t.~$\tau$, i.e.~$\pi\,\tau(g)\,\pi=\tau (g)\,\pi$ for 
all $g\in\semi{\al E.(2)}{\z{R}^{4}}$.
Then the following equations are equivalent: 
\begin{itemize}
\item[{\rm (i)}] $\pi\,\widetilde{\varphi}(p)=\widetilde{\varphi}(p),\;$ 
                 for $\;\widetilde{\varphi}\in\Lzwei{\Lkf,{\cal H},\mudp}$.
\item[{\rm (ii)}] $\pi(p)\,\varphi(p)=\varphi(p),\;$ 
                where $\pi(p):= D(H_p)\,\pi\, D(H_p)^{-1}$ and 
                $\;\varphi(p):= D(H_p)\,\widetilde{\varphi}(p)\in h_{0}$, 
                $p\in\Lk$.  
\end{itemize}
\end{lem}

\begin{defi}\label{MasslessRWEp}
Let $\pi$ be an invariant orthoprojection 
w.r.t.~$\tau^{\KIn{(can)}}\restriction (\semi{\al E.(2)}{\z{R}^{4}})$.
Then we call 
\[
 \pi(p)\,\varphi(p)=\varphi(p)\;,\quad \varphi \in h_0\,,\;p\in\Lk\,,
\]
a massless relativistic wave equation associated with $\pi$ 
(cf.~Lemma~\ref{Lem.3.3.5}~(ii)).
\end{defi}

The equation in (ii) is the generalisation to the massless case of 
Eq.~(ii) in Lemma~\ref{Lem.3.2.3} (cf.~also \cite[Eq.~(20.8)]{Niederer74b} 
or \cite[Eq.~(17) of Section~21.1]{bBarut80}). 
In the following we will consider it in
different particular cases and show in Corollary~\ref{General} that it
includes the general form of massless \rwe{s}. 

\begin{rem}\label{NonUni}
\begin{itemize}
\item[(a)] As announced in the begining of this subsection the equation
introduced in Definition~\ref{MasslessRWEp} extends neatly 
to the massless case, the work of 
Niederer and O'Rafeartaigh concerning massive relativistic wave equation
\cite{Niederer74b}. The mentioned equation contains
as special cases the conventional massless relativistic wave equations 
written in momentum space. (See the following Weyl and Maxwell 
cases below, Corollary~\ref{General} and the table in 
Subsection~\ref{Sumario}.) 

The existence of the corresponding projections $\pi$ in the massive and
massless cases is guaranteed by the fact that $\g{SU}{2}$ resp.~$\al E.(2)$
are compact resp.~solvable Lie groups. Indeed, any finite-dimensional unitary 
representation
$\tau^{\KIn{(can)}}\restriction (\semi{\g{SU}{2}}{\z{R}^{4}})$
can be decomposed as a direct sum of irreducible ones
(cf.~\cite[Theorem~27.30]{bHewittII}). Moreover, since
$\al E.(2)$ is solvable and connected, 
Lie's Theorem \cite[p.~200]{bBarut80} guarantees the existence
of one-dimensional orthoprojections invariant 
w.r.t.~$\tau^{\KIn{(can)}}\restriction (\semi{\al E.(2)}{\z{R}^{4}})$.

\item[(b)] Notice that in certain steps in this subsection we have 
make use of nonunitary representations (see e.g.~Eqs.~(\ref{VfCan0})
or (\ref{GeneralV})). Therefore these representations lie outside
of Mackey's theory of induced representations. Nevertheless in
these cases we do not use any result of this theory. The 
justification of this procedure (and the importance  of the 
massless relativistic wave equations) comes from
Theorem~\ref{TInvProj}. In fact, here we turn back to the description
of unitary and irreducible representations and these must be 
unitarily equivalent to the ones considered by Wigner.
The equivalence is given explicitly in the Weyl and Maxwell 
cases in Propositions~\ref{Lem.3.3.8} and \ref{Lem.3.3.12}.
(For the existence of projections satisfying Eq.~(\ref{CondProj})
recall the previous item.)

\item[(c)]
Note finally the fundamentally different role that massless wave equations
play (in contrast to the massive ones) when reducing the covariant \rep{}.
Indeed, massive \rwe{s} appear when we consider reducible \rep{s} of
$\g{SU}{2}$ and therefore will not be present if we choose 
e.g.~the irreducible representations given
by $D^{\KIn{$(\frac{n}{2},0)$}}\restriction\g{SU}{2}$. On the contrary 
$D^{\KIn{$(\frac{j}{2},\frac{k}{2})$}}\restriction\al E.(2)$ is 
{\em always} reducible if nontrivial helicity is admitted and therefore
the space of solutions of massless \rwe{s} is unavoidable if we 
want to work in momentum space with irreducible canonical \rep{s}.
Indeed, in Sections~\ref{TheWeyl} and \ref{TheF},
the Weyl- and Maxwell's equations will
naturally appear when considering 
$D^{\KIn{$(\frac{n}{2},0)$}}\restriction\al E.(2)$, $n=1,2$.
\end{itemize}
\end{rem}

\subsection{Conditions on the intertwining operator and fibre reduction}
\label{FBRED}

It is useful at this point to 
complete Remark~\ref{Rem.3.3.1} on the fibre reduction
and study in detail the intertwing operator $\ot I.$ between the 
covariant and massless canonical representation with nontrivial
discrete helicity. By Wigner's analysis and, in particular, due
to the dual action defined in Eq.~(\ref{DualGroupAction}),
it is clear that $\ot I.$ must contain the Fourier transformation
$~\widehat{}~$.
This is also the case in massive models (see Remark~\ref{Rem.3.2.2}; for
scalar models see also \cite[Section~IX.9]{bReedII}).
We can now decompose the intertwining operator in its constituents.
(The case with helicity $\frac12$ (hence $\al H.=\C^2$) is already 
typical.)
\[
 \al S.(\R^4,\C^2)\mathop{\longrightarrow }\limits^{\widehat{}}
 \al S.(\widehat{\R^4},\C^2)\mathop{\longrightarrow}\limits^{R}
 C^\infty((\R^3\setminus 0)\,,\,\C^2)\mathop{\longrightarrow}\limits^{M}
 C^\infty((\R^3\setminus 0)\,,\,\C)\subset L^2((\R^3\setminus 0)\,,\,\C)\,,
\]
where $R$ is the resctriction operator onto $\Lkf$,
$(R\widehat{f})(\mb p.):=
\widehat{f}(|\mb p.|,\mb p.)$, $\mb p.\in(\R^3\setminus 0)$,
 and $M$ is the operator
performing the fibre reduction mentioned above. The latter
operator is characteristic for massless models with nontrial
helicity.
(In order to keep argument transparent we work here with the 
massless representation $U$ considered in Eq.~(\ref{Vf0Can}).)
The conditions on the intertwining operator $\ot I.$, 
which is the composition of the preceding 
chain of mappings, must satisfy the following conditions:
\begin{itemize}
\item[(a)] $\ot I.$ must be linear (hence $M$ must be linear).
\item[(b)] $\ot I.$ must be continuous between the Schwartz and the Hilbert 
           space topologies (recall Remark~\ref{BorelSect}).
\item[(c)] $\ot I.$ must intertwine the covariant and massless 
           canonical representations, i.e.
\[
 \ot I.\; T(g)=U(g)\;\ot I.\,,\quad g\in\UPgr \,.
\]
\end{itemize}
Some concrete examples of intertwining operators 
satisfying the above conditions in the Weyl and Maxwell 
cases are specified in Eqs.~(\ref{IW}) and (\ref{IF}).

\subsection{The Weyl Equation}
\label{TheWeyl}
We begin with the simplest \rep{} of ${\cal E}(2)$ with dimension bigger than
one (see e.g.~\cite[Section~V.A]{Heidenreich86}). 
For $L=\left(\kern-1.5mm \begin{array}{cc}
  e^{\frac{i}{2}\theta} & e^{-\frac{i}{2}\theta}\ z          \\
                      0 & e^{-\frac{i}{2}\theta}
           \end{array} \kern-1.5mm\right)\in {\cal E}(2) $ 
we have on ${\cal H}^{\KIn{$(\frac{1}{2},0)$}}:=\z{C}^2$
the representation given by $D^{\KIn{$(\frac{1}{2},0)$}}(L):=L$. 
(Notation: In the remaining subsection we will denote the representation
$D^{\KIn{$(\frac{1}{2},0)$}}(\cdot)$ simply by $D(\cdot)$. 
We will also skip the index $\KIn{$(\frac{1}{2},0)$}$ 
from the objects associated to $D(\cdot)$, e.g.~the representation
$V$, the scalar product $\langle\cdot,\cdot\rangle$
etc., in order to keep the notation simple.)
The only nontrivial $D({\cal E}(2))$-invariant subspace is 
$\z{C} \left(\kern-1.5mm  \begin{array}{c}  
               1 \\ 0 
          \end{array} \kern-1.5mm \right)$ and we choose
the corresponding invariant orthoprojection 
          $\pi:=\left(\kern-1.5mm
          \begin{array}{cc}
              1 \kern-1.6mm & 0 \\ 0 \kern-1.6mm & 0
          \end{array} \kern-1.5mm\right)$.
In the following lemma we will establish the relation between the
equation in Lemma~\ref{Lem.3.3.5}~(ii) and the Weyl Equation.
\begin{lem}
\label{Lem.3.3.6}
Put $\varphi(p):=H_p\,\widetilde{\varphi}(p)\in h_{0}$, for all $\widetilde{\varphi}\in
\Lzwei{\Lkf,\z{C}^2,\mudp}$ $($see Eq.~$(\ref{Isometry0}))$. 
Then we have that $\widetilde{\varphi}$ satisfies the equation 
$\pi\, \widetilde{\varphi}(p)=\widetilde{\varphi}(p)$ iff 
$\;\varphi$ satisfies the Weyl Equation. 
\end{lem}
\begin{beweis}
1.~Suppose that $\widetilde{\varphi}\in\Lzwei{\Lkf,\z{C}^2,\mudp}$ satisfies
$\pi\,\widetilde{\varphi}(p)=\widetilde{\varphi}(p)$, $p\in\Lkf$. Then there
exists a scalar function $\chi\in \Lzwei{\Lkf,\z{C},\mudp}$ such that 
$\widetilde{\varphi}(p)=\left(\kern-1.5mm\begin{array}{c}
               \chi(p) \\ 0
       \end{array}\kern-1.5mm\right)$. 
Next we write the Weyl operator in momentum space as
\[ 
{\cal W}(p):=\left(p_0\sigma_0
-\sum\limits_{i=1}^3 p_i\sigma_i \right)
\]
and notice that we can rewrite ${\cal W}(p)$, $p\in\Lk$, as
\begin{equation}
\label{Wp}
  {\cal W}(p)=\left(H_p^{-1}\right)^* 
         \left(\kern-1.5mm\begin{array}{cc} 
              0 \kern-1.6mm & 0             \\
              0 \kern-1.6mm & 2
         \end{array}\kern-1.5mm\right) H_p^{-1}
            =2P^{\KIn{$\dagger$}}\,.
\end{equation}
Therefore, $\varphi(p)=H_p \left(\kern-1.5mm\begin{array}{c}
               \chi(p) \\ 0
           \end{array}\kern-1.5mm\right)$, satisfies the Weyl Equation
${\cal W}(p)\,\varphi(p)={\cal W}(p)\,
H_p \left(\kern-1.5mm\begin{array}{c}
               \chi(p) \\ 0
       \end{array}\kern-1.5mm\right)=0\,.$
Recall also that in terms of the spinorial components the Weyl equation
is usually written as
\begin{equation}\label{Components}
\sum\limits_{C=0}^{1} (P^\dagger)_{\KIn{$C'C$}}\, 
 \varphi^{\KIn{$C$}}(p)
=\sum\limits_{C=0}^{1} (P^\dagger)_{\KIn{$C'C$}}\, 
 (H_p)^{\KIn{$C$}}_{\KIn{$0$}} \,\widetilde{\varphi}^{\KIn{$0$}}(p)
=0,\quad \In{$C'$}\in\{0,1\}\,.
\end{equation}

2.~Suppose on the other hand that $\varphi(p)=H_p\,\widetilde{\varphi}(p)\in 
h_{0}$, with $\widetilde{\varphi}(p)=
    \left(\kern-1.5mm\begin{array}{c}
    \widetilde{\varphi}_1(p) \\ \widetilde{\varphi}_2(p)
       \end{array}\kern-1.5mm\right)$,
satisfies the Weyl Equation ${\cal W}(p)\,\varphi(p)=0.$ 
Then $\widetilde{\varphi}_2=0$ and 
$\widetilde{\varphi}$ satisfies the equation $\pi\, \widetilde{\varphi}(p)=
\widetilde{\varphi}(p)$.
\end{beweis}

The space of solutions of the Weyl equation is therefore given by
\[
 h_+:=\left\{ H_p \left(\kern-1.5mm\begin{array}{c}
                     \chi(p) \\ 0
 \end{array}\kern-1.5mm\right)\mid 
 \chi\in \Lzwei{\Lkf,\z{C},\mudp}\right\},
\]

\begin{lem}
\label{Lem.3.3.7}
Define on the space of solutions of the Weyl Equation $h_+$ 
the scalar product
\[
 \askp{\varphi_1}{\varphi_2}_+:=\int\limits_{\Lkf}
 \askp{\varphi_1(p)}{\LR H_p^{-1}\RR^*H_p^{-1}\,\varphi_2(p)}_{\z{C}^2}\mudp,
  \quad\Wort{for} \quad\varphi_i\in h_+,\; i=1,2.
\]
The representation given for $g=(A,a)\in \semi{\g{SL}{2,\z{C}}}{\z{R}^4}$,
$q:= \Lambda^{-1}_A p$ and $\varphi\in h_+$ by
\begin{equation}
\label{CovRep1} 
  \LR V_1(g)\,\varphi \RR\!(p)
        :=e^{-ipa}\,A\,\varphi(q),  
\end{equation}
is unitary w.r.t.~$\langle\cdot,\cdot\rangle_+$ and irreducible.
\end{lem}
\begin{beweis}
First note that for $\varphi_i(p)=
H_p \left(\kern-1.5mm\begin{array}{c}
           \chi_i(p) \\ 0
\end{array}\kern-1.5mm\right)$, $\chi_i\in \Lzwei{\Lkf,\z{C},\mudp}$, 
$i=1,2$,
\begin{equation}
\label{Isometry1}
 \askp{\varphi_1}{\varphi_2}_+=\int\limits_{\Lkf}\overline{\chi_1(p)}\,
                                          \chi_2(p)\mudp.
\end{equation}
Since $\pi\al H.$ is 1-dimensional and $\pi\,L^*L\,\pi=\pi$, 
$L\in\al E.(2)$, Theorem~\ref{TInvProj} completes the proof.
\end{beweis}

Next we establish the equivalence between the representation $\LR h_+,
\;V_1,\;\langle\cdot,\cdot\rangle_+\RR $ defined above
and the representation $\LR \Lzwei{\Lkf,\z{C},\mudp},\;U_+,\;
\langle\cdot,\cdot\rangle_{\KIn{L$^2$}}\RR$:
The canonical representation for $n=1$ (cf.~Eq.~(\ref{Vf0Can})) is
\begin{equation}
\label{CanonicalRep1} 
 \LR U_+(g)\,\chi\RR\!(p):=e^{-ipa}\,e^{\frac{i}{2} \theta(A,p)}\,
 \chi(q),\quad \chi\in\Lzwei{\Lkf,\z{C},\mudp}\,,
\end{equation}
for $q:=\Lambda^{-1}_A p$ and $g=(A,a)\in\semi{\g{SL}{2,\z{C}}}{\z{R}^4}$. 
$U_+(\cdot)$ is irreducible, strongly continuous and unitary for 
the usual ${\rm L}^2$-scalar product, 
$\langle\cdot,\cdot\rangle_{\KIn{L$^2$}}$. 
With the preceding notation we have the following equivalence of 
\rep{s}

\begin{pro}
\label{Lem.3.3.8}
The mapping $\Phi_+\colon\ h_+
\longrightarrow \Lzwei{\Lkf,\z{C},\mudp}$ defined by, 
\[
\Phi_+\left(
            H_{(\cdot)} \left(\kern-1.5mm\begin{array}{c}
                \chi(\cdot) \\  0 
          \end{array}\kern-1.5mm\right) \right)\!(p)=\chi(p)\, ,
\]
is an isometric isomorphism between $\LR h_+,\;\langle\cdot,\cdot\rangle_+
\RR$ and $\LR \Lzwei{\Lkf,\z{C},\mudp},\;\langle\cdot,\cdot\rangle_{\KIn{L$^2$}} \RR$, 
that commutes with the corresponding representations, i.e.~$\Phi_+\, V_1(g)
=U_+(g)\, \Phi_+$, $g\in\UPgr$.
\end{pro}
\begin{beweis}
That the mapping $\Phi_+$ is an isometry follows already
from Eq.~(\ref{Isometry1}) in the proof of the preceding lemma.
The intertwining property is proved by direct computation. Indeed, for 
$g=(A,a)\in \semi{\g{SL}{2,\z{C}}}{\z{R}^4}$ and putting 
$q:=\Lambda^{-1}_A p$, we have on the one hand 

\begin{eqnarray*}
 \LR U_+(A,a)\,\Phi_+(\varphi)\RR \!(p) 
    &=& e^{-ipa}\,e^{\frac{i}{2}\theta(A,p)}\,\chi(q)  \\[4mm]
 \lefteqn{\kern-5.4cm\mbox{and on the other hand}}                    \\[2mm]
 \Phi_+ \LR V_1(A,a)\,\varphi \RR \!(p)
    &=& \Phi_+\! \LR e^{-i(\cdot) a}\, AH_{\Lambda^{-1}_A (\cdot)} 
                 \left(\kern-1.5mm\begin{array}{c}
                 \chi\LR\Lambda^{-1}_A (\cdot)\RR \\  0 
                 \end{array}\kern-1.5mm\right) 
                 \RR \!(p)                                            \\    
    &=& \Phi_+\! \LR H_{(\cdot)}\,e^{-i(\cdot) a}\,H^{-1}_{(\cdot)} A 
                    H_{\Lambda^{-1}_A (\cdot)} 
                 \left(\kern-1.5mm\begin{array}{c}
                    \chi\LR\Lambda^{-1}_A (\cdot)\RR \\  0 
                 \end{array}\kern-1.5mm \right) 
                 \RR \!(p)                                             \\ 
    &=& e^{-ipa}\,e^{\frac{i}{2}\theta(A,p)}\,\chi(q),
\end{eqnarray*}
where for the last equation we have used that $\LR H^{-1}_p A
H_{q}\RR_{11}=e^{\frac{i}{2}\theta(A,p)}$.
\end{beweis}
With the preceding result we have also proved the equivalence between 
the representation $( h_+,\;V_1,\;\langle\cdot,\cdot\rangle_+ ) $ and the 
representation $( {{\got H}}'_{-},\; V'_{4},\; \langle\cdot,\cdot\rangle'_{
\beta_{-}} ) $ used in \cite[Theorem~B.2.17]{Lledo95}. 

Let us finish this subsection defining the space and the representation
corresponding to the opposite helicity. They will be denoted by the
subindex ``--" and the proofs are analogous as before.
This representation space associated to the opposite helicity
will be necessary in order to construct the reference space of the 
CAR-algebra (cf.~Subsection~\ref{WN}).

On the space 
\[
 h_-:=\left\{\overline{H_p}
 \left(\kern-1.5mm\begin{array}{c}
   \chi(p) \\ 0
 \end{array}\kern-1.5mm\right)\mid \chi\in \Lzwei{\Lkf,\z{C},\mudp}
 \right\}
\]
define for $g=(A,a)\in \semi{\g{SL}{2,\z{C}}}{\z{R}^4}$, 
$q:=\Lambda^{-1}_A p$ and $\varphi\in h_-$ the representation
\[
  \LR V_4(g)\,\varphi\RR\!(p):=e^{-ipa}\,\overline{A}\, 
                              \varphi(q). 
\]
This representation is irreducible and unitary w.r.t.~the scalar product,
\[
 \askp{\varphi_1}{\varphi_2}_-:=\int\limits_{\Lkf}\askp{\varphi_1(p)}
 {\LR\overline{H_p}^{-1}\RR^*\overline{H_p}^{-1}\,\varphi_2(p)}_{\z{C}^2}
 \mudp,\quad\Wort{for} \quad\varphi_i\in h_-,\; i=1,2. 
\]
Finally $\LR h_-,\; V_4,\;\langle\cdot,\cdot\rangle_- \RR $ is 
equivalent to $\LR \Lzwei{\Lkf,\z{C},\mudp},\;U_-,
\;\langle\cdot,\cdot\rangle_{\KIn{L$^2$}} \RR $, where the latter
representation is given by
\[
\LR U_-(g)\,\chi\RR \!(p):=e^{-ipa}
\,e^{-\frac{i}{2} \theta(A,p)}\,\chi(q)\,, 
   \quad \chi\in\Lzwei{\Lkf,\z{C},\mudp}\;,\;q:=\Lambda^{-1}_A p\,.
\]

\subsection{Maxwell Equations: ${\got F}$-Equation}
\label{TheF}

Now for any $L=\left(\kern-1.5mm \begin{array}{cc}
  e^{\frac{i}{2}\theta} & e^{-\frac{i}{2}\theta}\ z          \\
                      0 & e^{-\frac{i}{2}\theta}
           \end{array} \kern-1.5mm\right)\in {\cal E}(2) $ we have 
on ${\cal H}^{\KIn{$(1,0)$}}:=\Wort{Sym}\LR\z{C}^2\otimes\z{C}^2\RR$ 
the \rep{} 
\[
 D^{\KIn{$(1,0)$}}(L):=L\otimes L \,.
\]
(Notation: In the remaining subsection we will denote again 
the representation $D^{\KIn{$(1,0)$}}(\cdot)$ simply by $D(\cdot)$ etc.).
We select the 1-dimensional $D({\cal E}(2))$-invariant subspace 
characterised by the orthoprojection
 $ \pi:=\left(\kern-2mm\begin{array}{cc}  
                           1 \kern-3mm & 0\\ 0 \kern-3mm & 0
                     \end{array} \kern-2mm \right) 
              \!\!\otimes\!\! \left(\kern-2mm\begin{array}{cc}  
                           1 \kern-3mm & 0\\ 0 \kern-3mm & 0
                     \end{array} \kern-2mm \right)$.

\begin{lem}
\label{Lem.3.3.10}
Put $\varphi(p):=D(H_p)\, \widetilde{\varphi}(p)\in h_{0}$, 
$\widetilde{\varphi}\in
\Lzwei{\Lkf,{\cal H},\mudp}$ $($see Eq.~$(\ref{Isometry0}))$. 
Then we have that $\widetilde{\varphi}$ satisfies the equation 
$\pi\, \widetilde{\varphi}(p)=\widetilde{\varphi}(p)$ 
iff $\;\varphi$ satisfies the spinorial form of Maxwell Equation 
$(\got F$-Equation for short$)$, which in components is given
by 
\begin{equation}\label{FEquation}
\sum\limits_{C=0}^{1} (P^\dagger)_{\KIn{$C'C$}}\, 
 \varphi^{\KIn{$CB$}}(p)=0\,,
 \quad \In{$C'$}\in\{0,1\}\,,\;\In{$B$}\in\{0,1\} \,.
\end{equation}
\end{lem}
\begin{beweis}
1.~Note first that $\pi\,\widetilde{\varphi}(p)=\widetilde{\varphi}(p)$
 iff $\widetilde{\varphi}^{\KIn{$01$}}(p)=
 \widetilde{\varphi}^{\KIn{$10$}}(p)=0=\widetilde{\varphi}^{\KIn{$11$}}(p)$.
 Therefore from Eq.~(\ref{Components}) we get
\[
 \sum\limits_{C=0}^{1} (P^\dagger)_{\KIn{$C'C$}}\,\varphi^{\KIn{$CB$}}(p) 
 = \sum\limits_{C=0}^{1} (P^\dagger)_{\KIn{$C'C$}}\, 
   (H_p)^{\KIn{$C$}}_{\KIn{$0$}} (H_p)^{\KIn{$B$}}_{\KIn{$0$}}
   \,\widetilde{\varphi}^{\KIn{$00$}}(p)=0\,,
   \quad \In{$C'$}\in\{0,1\}\,,\;\In{$B$}\in\{0,1\} \,.
\] 

2.~Conversely, suppose that $\sum\limits_{C=0}^{1} (P^\dagger)_{\KIn{$C'C$}}\, 
 \varphi^{\KIn{$CB$}}(p)=0$. From the form of $P^\dagger$ 
 (see Eq.~(\ref{Wp})) and since
\[
 \widetilde{\varphi}^{\KIn{$EB$}}(p)=
    (H_p^{-1})^{\KIn{$E$}}_{\KIn{$C$}} (H_p^{-1})^{\KIn{$B$}}_{\KIn{$D$}}
     \,\varphi^{\KIn{$CD$}}(p)\,,
\]
we obtain $\widetilde{\varphi}^{\KIn{$1B$}}(p)=0$, ${\In{$B$}}\in\{0,1\}$,
hence also $\widetilde{\varphi}^{\KIn{$01$}}(p)=0$. Therefore
$\pi\,\widetilde{\varphi}(p)=\widetilde{\varphi}(p)$.
\end{beweis}

Recall that from a symmetric spinor field $\varphi^{\KIn{$CB$}}$
satisfying the ${\got F}$-Equation, one can 
construct canonically a real and antisymmetric tensor field 
$F_{\mu\nu}$ satisfying the source free Maxwell Equations
\cite[Exercise~13.3]{bWald84}, \cite[Section~5.1]{bPenrose86I}.

The space of solutions of the ${\got F}$-Equation is given by
\begin{equation}
\label{HF+}
 h_+:=\left\{ D(H_p) \left(\kern-2mm\begin{array}{c}  
                           1 \\ 0 
                     \end{array} \kern-2mm \right) 
              \!\!\otimes\!\!  \left(\kern-2mm  \begin{array}{c}  
                           1 \\ 0 
                     \end{array} \kern-2mm \right)\chi(p)
      \mid\chi\in\Lzwei{\Lkf,\z{C},\mudp}\right\},
\end{equation}

\begin{lem}
\label{Lem.3.3.11}
Define on the space of solutions of the ${\got F}$-Equation $h_+$ 
the scalar product
\[
 \askp{\varphi_1}{\varphi_2}_+:=\int\limits_{\Lkf}
 \askp{\varphi_1(p)}{D( H_p^{-1})^*D( H_p^{-1})
                     \varphi_2(p)}_{\z{C}^4}\mudp,
  \quad\Wort{for} \quad\varphi_i\in h_+,\; i=1,2.
\]
The representation given for $g=(A,a)\in \semi{\g{SL}{2,\z{C}}}{\z{R}^4}$,
$q:= \Lambda^{-1}_A p$ and $\varphi\in h_+$, by
\begin{equation}
\label{VF+} 
  \LR V_+(g)\,\varphi \RR\!(p)
        :=e^{-ipa}\,D(A)\,\varphi(q),  
\end{equation}
is unitary w.r.t.~$\langle\cdot,\cdot\rangle_+$ and irreducible.
\end{lem}
\begin{beweis}
First note that for $\varphi_i(p)=
D(H_p) \left(\kern-2mm\begin{array}{c}  
                           1 \\ 0 
                     \end{array} \kern-2mm \right) 
              \!\!\otimes\!\!  \left(\kern-2mm  \begin{array}{c}  
                           1 \\ 0 
                     \end{array} \kern-2mm \right)\chi_i(p)$,
$\chi_i\in\Lzwei{\Lkf,\z{C},\mudp}$, 
$i=1,2$,
\begin{equation}
\label{Isometry2}
 \askp{\varphi_1}{\varphi_2}_+=\int\limits_{\Lkf}\overline{\chi_1(p)}\,
                                          \chi_2(p)\mudp.
\end{equation}
Since the space $\pi\al H.$ is 1-dimensional and $\pi\,D(L)^*D(L)\,\pi=\pi$, 
$L\in\al E.(2)$, Theorem~\ref{TInvProj} completes the proof.
\end{beweis}

Next we establish the equivalence between $\LR h_+,
\;V_+,\;\langle\cdot,\cdot\rangle_+\RR $ defined above
and the representation $\LR \Lzwei{\Lkf,\z{C},\mudp} ,\;U_+,\;
\langle\cdot,\cdot\rangle_{\KIn{L$^2$}}\RR$.
Consider on the space $\Lzwei{\Lkf,\z{C},\mudp}$ the canonical 
representation for $n=2$ (cf.~Eq.~(\ref{Vf0Can})),
\begin{equation}
\label{CanonicalRep2} 
 \LR U_+(g)\,\chi\RR\!(p):=e^{-ipa}\,e^{i \theta(A,p)}\,
 \chi(q),\quad \chi\in\Lzwei{\Lkf,\z{C},\mudp}\,,
\end{equation}
for $q:=\Lambda^{-1}_A p$ and $g=(A,a)\in \semi{\g{SL}{2,\z{C}}}{\z{R}^4}$. 
$U_+(\cdot)$ is irreducible, strongly continuous and unitary for 
the usual ${\rm L}^2$-scalar product 
$\langle\cdot,\cdot\rangle_{\KIn{L$^2$}}$. 
With the preceding notation we have the following equivalence of
\rep{s}

\begin{pro}
\label{Lem.3.3.12}
The mapping $\Phi_+\colon\ h_+
\longrightarrow \Lzwei{\Lkf,\z{C},\mudp}$ defined by, 
\[
\Phi_+\left(
D\!\LR H_{(\cdot)}\RR 
\left(\kern-2mm\begin{array}{c}  
                           1 \\ 0 
                     \end{array} \kern-2mm \right) 
              \!\!\otimes\!\!  \left(\kern-2mm  \begin{array}{c}  
                           1 \\ 0 
                     \end{array} \kern-2mm \right)
                     \chi(\cdot)
\right)\!(p)=\chi(p)\, ,\quad p\in\Lk,
\]
is an isometric isomorphism between 
$\LR h_+,\;\langle\cdot,\cdot\rangle_+\RR$ and 
$\LR\Lzwei{\Lkf,\z{C},\mudp},\;\langle\cdot,\cdot\rangle_{\KIn{L$^2$}}\RR$ 
that commutes with the corresponding representations, i.e.~$\Phi_+\, V_+(g)
=U_+(g)\, \Phi_+$, $g\in\UPgr$.
\end{pro}
\begin{beweis}
That the mapping $\Phi_+$ is an isometry follows already
from Eq.~(\ref{Isometry2}) in the proof of the preceding lemma.
The intertwining property is proved by a direct computation. Indeed, for 
$g=(A,a)\in \semi{\g{SL}{2,\z{C}}}{\z{R}^4}$ and putting 
$q:=\Lambda^{-1}_A p$, we have on the one hand, 

\begin{equation}
\label{Photon+}
 \LR U_+(A,a)\,\Phi_+(\varphi)\RR \!(p)                    
    = e^{-ipa}\,e^{i\theta(A,p)}\,\chi(q)                         
\end{equation}
and on the other hand computing similarly as in the proof of 
Proposition~\ref{Lem.3.3.8}
\begin{eqnarray*}
\Phi_+ \LR V_+(A,a)\,\varphi \RR \!(p)                  
    &=& \Phi_+\! \LR D\!\LR H_{(\cdot)}\RR\,e^{-i(\cdot) a}\, 
                    D\!\LR H^{-1}_{(\cdot)} A H_{\Lambda^{-1}_A (\cdot)}\RR 
                  \left(\kern-2mm\begin{array}{c}  
                           1 \\ 0 
                     \end{array} \kern-2mm \right) 
              \!\!\otimes\!\!  \left(\kern-2mm  \begin{array}{c}  
                           1 \\ 0 
                     \end{array} \kern-2mm \right)
                 \chi\LR\Lambda^{-1}_A(\cdot)\RR\!\RR \!(p)       \\[2mm] 
    &=& e^{-ipa}\,e^{i\theta(A,p)}\,\chi(q),
\end{eqnarray*}
where for the last equation we have used that $D(H^{-1}_p A
H_{q})_{11}=e^{i\theta(A,p)}$.
\end{beweis}

Let us finish this subsection defining the space and the representation
corresponding to the opposite helicity. They will be denoted by the
subindex ``--" and the proofs are similar as before.
This representation space associated to the opposite helicity
will be necessary in order to construct the reference space of the 
CCR-algebra.

On the space
\begin{equation}
\label{HF-}
 h_-:=\left\{D\!\LR\overline{H_p}\RR
\left(\kern-2mm\begin{array}{c}  
                           1 \\ 0 
                     \end{array} \kern-2mm \right) 
              \!\!\otimes\!\!  \left(\kern-2mm  \begin{array}{c}  
                           1 \\ 0 
                     \end{array} \kern-2mm \right)
                     \chi(p)
\mid \chi\in \Lzwei{\Lkf,\z{C},\mudp}\right\}, 
\end{equation}
define for $g=(A,a)\in \semi{\g{SL}{2,\z{C}}}{\z{R}^4}$, 
$q:=\Lambda^{-1}_A p$ and $\varphi\in h_-$ the representation 
\begin{equation}
\label{VF-}
  \LR V_-(g)\,\varphi\RR\!(p):=e^{-ipa}\,D\!\LR\overline{A}\RR\, 
                               \varphi(q). 
\end{equation}
This representation is irreducible and unitary w.r.t.~the scalar product
\[
 \askp{\varphi_1}{\varphi_2}_-:=\int\limits_{\Lkf}\askp{\varphi_1(p)}{
 D\!\LR\overline{H_p}^{-1}\RR^*D\!\LR\overline{H_p}^{-1}\RR\,
 \varphi_2(p)}_{\z{C}^4} \mudp,\quad\Wort{for} \quad\varphi_i\in h_-,\; 
 i=1,2. 
\]
Finally $\LR h_-,\; V_-,\;\langle\cdot,\cdot\rangle_- \RR $ is
equivalent to $\LR \Lzwei{\Lkf,\z{C},\mudp},\;U_-,
\;\langle\cdot,\cdot\rangle_{\KIn{L$^2$}} \RR $, 
where the latter representation is given by 
\begin{equation}
\label{Photon-}
 \LR U_-(g)\,\chi\RR \!(p):=e^{-ipa}
 \,e^{-i \theta(A,p)}\,\chi(q)\,,\quad \chi\in\Lzwei{\Lkf,\z{C},\mudp}
  \,,q:=\Lambda^{-1}_A p\,.
\end{equation}

\paragraph{Generalisation to arbitrary helicity:}
We collect in this paragraph the obvious generalisation of the 
previous analysis of the Weyl resp.~Maxwell equations (which 
correspond to helicities $\pm\frac12$ resp.~$\pm 1$) to the 
systems carrying arbitrary discrete helicity.

We begin characterising general massless \rwe{s} 
corresponding to helicity $\frac{n}{2}$
(cf.~\cite[p.~375]{bPenrose86I}). The following result
contains as special cases Lemmas~\ref{Lem.3.3.6} 
and \ref{Lem.3.3.10}.
\begin{cor}\label{General}
Consider the $D^{\KIn{$(\frac{n}{2},0)$}}({\cal E}(2))$-invariant subspace 
characterised by the one-dimensional orthoprojection
$\pi_n:=\mathop{\otimes}\limits^n
               \left(\kern-2mm\begin{array}{cc}  
               1 \kern-3mm & 0\\ 0 \kern-3mm & 0
               \end{array} \kern-2mm \right)$, $n\in\N$,
and put 
\[
 \varphi(p):=D^{\KIn{$(\frac{n}{2},0)$}}(H_p)\, 
 \widetilde{\varphi}(p)\in h_{0}\,, \quad\widetilde{\varphi}\in
 \Lzwei{\Lkf,{\cal H},\mudp}\quad (cf.~Eq.~(\ref{Isometry0})) .
\] 
Then $\widetilde{\varphi}$ satisfies the equation 
$\pi\, \widetilde{\varphi}(p)=\widetilde{\varphi}(p)$ 
iff $\;\varphi$ satisfies the massless \rwe{} corresponding to
helicity $\frac{n}{2}$. The latter equation is written in momentum
space for the spinorial components as 
\[
 \sum\limits_{C=0}^{1} (P^\dagger)_{\KIn{$C'C$}}\, 
 \psi^{\KIn{$CC_1\dots C_{n-1}$}}(p)=0,
 \quad \In{$C_1,\dots,C_{n-1}$}\in\{0,1\},\;\In{$C'$}\in\{0,1\}\,.
\]
\end{cor}

\begin{rem}\label{GenRWE}
\begin{itemize}
\item[(i)] The way of presenting relativistic wave equations written in 
momonetum space is justified by the group theoretical approach which 
is one of the basic ingredients of the present paper. 
However, to give a more complete picture of these equations we need to
comment on them also as PDEs in position space, since
they usually appear in the literature in this form. 
A general massless relativistic wave equations on position space is 
given by 
\begin{equation}\label{PDEm0}
 \sum\limits_{C=0}^{1} \partial_{\KIn{$C'C$}}\, 
 \psi^{\KIn{$CC_1\dots C_{n-1}$}}(x)=0,
 \quad \In{$C_1,\dots,C_{n-1}$}\in\{0,1\},\;\In{$C'$}\in\{0,1\}\,,
\end{equation}
where $\partial_{\KIn{$C'C$}}$ is the first order differential 
operator on spinor fields corresponding to the usual gradient
$\partial_\mu$, $\mu=0,\dots ,3$ \cite[Eq.~13.1.64]{bWald84}. 
It can be shown that 
Eq.~(\ref{PDEm0}) is equivalent to the usual wave equation
$\Box\psi^{\KIn{$CC_1\dots C_{n-1}$}}(x)=0$, which is an 
hyperbolic equation, together with Eq.~(\ref{PDEm0}) holding only as
an initial value constraint on a Cauchy surface (e.g.~$x^0=0$)
(for details see \cite[pp.~376-377]{bWald84}). 
This fact confirms the point of view
already stated in the introduction
(see also Remark~\ref{Rem.3.3.1}~(i)) that massless relativistic
wave equations can be seen as constraint equations restricting 
the fibre degrees of freedom. 

The results cited above show that Eq.~(\ref{PDEm0}) has a 
well-posed initial value formulation is relevant if one wants
to construct quantum fields on more general 
(globally hyperbolic) space-times, where
the group theoretical approach is not possible due 
to the lack of symmetry.

\item[(ii)]
We want now complete the generalisation to include the 
corrsponding spaces of solutions of the 
relativistic wave equations, 
the representations and the associated isometric isomorphisms.
For this one needs only to replace the labels
$(1,0)$ resp.~$(0,1)$ by $(\frac{n}{2},0)$ 
resp.~$(0,\frac{n}{2})$, $n\geq 3$, 
in the present subsection. In particular we 
obtain in this way a characterisation of the 
Wigner massless Hilbert spaces with discrete helicity in terms
of the space of solutions of the corresponding massless \rwe{}
(cf.~Corollary~\ref{General}).
\end{itemize}
\end{rem}

\subsection{Maxwell Equations: \mbox{${\got A}$-Equation}}
\label{TheA}
For completeness we will include in our group theoretical context the 
discussion of Maxwell Equations in terms of the vector potential field
(${\got A}$-Equation). We will see that some techniques used
in the previous subsection for treating Maxwell Equations in terms of 
the field strength (${\got F}$-Equation) will not be applicable anymore
(cf.~Remark~\ref{Rem.3.3.16}). Instead we will use a Gupta-Bleuler
like procedure will allow to establish an isometric isomorphism to 
the previous representation space. For a detailed treatment of 
quantum electromagnetism in terms of the vector potential 
(including constraints) we refer to \cite{Lledo00}.

For $L=\left(\kern-1.5mm \begin{array}{cc}
  e^{\frac{i}{2}\theta} & e^{-\frac{i}{2}\theta}\ z          \\
                      0 & e^{-\frac{i}{2}\theta}
           \end{array} \kern-1.5mm\right)\in {\cal E}(2) $ 
we have on ${\cal H}^{\KIn{$(\frac12,\frac12)$}}
            :=\z{C}^2 \otimes\z{C}^2\cong\z{C}^4$ 
the representation 
 \[
  D^{\KIn{$(\frac12,\frac12)$}}(L):=L\otimes\overline{L}\cong
        \left(\kern-1.5mm \begin{array}{cccc}
   1 & e^{i\theta}\ \overline{z} & e^{-i\theta}\ z & |z|^2        \\
   0 & e^{i \theta}              & 0               & z            \\ 
   0 & 0                         & e^{-i \theta}   & \overline{z} \\
   0 & 0                         & 0               & 1   
       \end{array} \kern-1.5mm\right)\,.
\]
(Notation: In the remaining subsection we will denote 
when no confusion arises again the 
representation $D^{\KIn{$(\frac12,\frac12)$}}(\cdot)$ simply by $D(\cdot)$). 
In order to include the nontrivial phases of the diagonal of $D(L)$ 
(recall that now we want to describe both helicity values $\pm 1$) one
is forced to consider in this context the 3-dimensional 
space characterised by the $D({\cal E}(2))$-invariant  
orthoprojection 
$ \pi:=\left(\kern-1.5mm \begin{array}{cccc}
   1 \kern-1mm & 0 \kern-1mm & 0 \kern-1mm & 0             \\
   0 \kern-1mm & 1 \kern-1mm & 0 \kern-1mm & 0             \\ 
   0 \kern-1mm & 0 \kern-1mm & 1 \kern-1mm & 0             \\
   0 \kern-1mm & 0 \kern-1mm & 0 \kern-1mm & 0                                
           \end{array} \kern-1.5mm\right)$.

In the following lemma we will write the divergence equation
$p_0\psi_0(p)-\sum\limits_{i=1}^3p_i\psi_i(p)=0$ in an equivalent and 
for us more convenient form. 

\begin{lem}
\label{Lem.3.3.14}
The vector $\varphi(p):=\left(\kern-1.5mm  \begin{array}{c}  
    \varphi_0(p) \\ \varphi_1(p) \\ \varphi_2(p) \\ \varphi_3(p)
                        \end{array} \kern-1.5mm \right)$, $p\in\Lk$, 
satisfies the equation
\begin{equation}
\label{AEq}
-(p_0-p_3)\varphi_0(p)+(p_1+ip_2)\varphi_1(p)+
(p_1-ip_2)\varphi_2(p)-(p_0+p_3)\varphi_3(p)=0
\end{equation}
iff the vector $\psi(p):=W_s\, \varphi(p)$, 
where $W_s:=\frac{\displaystyle 1}{\sqrt{\displaystyle 2}}
       \left(\kern-1.5mm \begin{array}{ccrr}
   1 \kern-1mm & 0 \kern-1.6mm & 0  \kern-1.6mm & 1             \\
   0 \kern-1mm & 1 \kern-1.6mm & 1  \kern-1.6mm & 0             \\ 
   0 \kern-1mm & i \kern-1.6mm & -i \kern-1.6mm & 0             \\
   1 \kern-1mm & 0 \kern-1.6mm & 0  \kern-1.6mm & -1           
       \end{array} \kern-1.5mm\right)$
is a unitary matrix acting on $\z{C}^4$, satisfies the equation
$p_0\psi_0(p)-\sum\limits_{i=1}^3p_i\psi_i(p)=0$.
\end{lem}
\begin{beweis}
The proof is straightforward since it uses essentially a unitary
transformation acting on $\cal H$. We write it explicitly down in 
order to introduce some useful notation for later on. Put
\be\label{SpinEta}
\eta:= W_s^{-1}\;\eta_{\KIn{Mink}}\;W_s=
      \left(\kern-1.5mm \begin{array}{rccc}
   0 \kern-1mm & 0 \kern-1mm & 0 \kern-1.6mm & -1            \\
   0 \kern-1mm & 1 \kern-1mm & 0 \kern-1.6mm & 0             \\ 
   0 \kern-1mm & 0 \kern-1mm & 1 \kern-1.6mm & 0             \\
  -1 \kern-1mm & 0 \kern-1mm & 0 \kern-1.6mm & 0               
           \end{array} \kern-1.5mm\right)\,,\quad\mathrm{with}\quad
\eta_{\KIn{Mink}}:=
      \left(\kern-1.5mm \begin{array}{rccc}
   1 \kern-1mm & 0 \kern-1mm & 0 \kern-1.6mm & 0             \\
   0 \kern-1mm &-1 \kern-1mm & 0 \kern-1.6mm & 0             \\ 
   0 \kern-1mm & 0 \kern-1mm &-1 \kern-1.6mm & 0             \\
   0 \kern-1mm & 0 \kern-1mm & 0 \kern-1.6mm & -1              
           \end{array} \kern-1.5mm\right)\,.
\ee
Now for $p=(p_0,p_1,p_2,p_3)\in\Lkf$ and recalling that 
$\langle\cdot,\cdot\rangle_{\KIn{$\z{C}^4$}}$ is
antilinear in the first argument we have that the equations
\[
 \askp{p}{\eta_{\KIn{Mink}}\;\psi(p)}_{\KIn{$\z{C}^4$}}
   = \askp{p}{W_s\,\eta\,W_s^{-1}\;\psi(p)}_{\KIn{$\z{C}^4$}}
   = \askp{W_s^{-1}\,p}{\eta\;\varphi(p)}_{\KIn{$\z{C}^4$}}
\]
conclude the proof, since the last term is precisely 
$\In{$\frac{1}{\sqrt{2}}$}$ times the l.h.s.~of Eq.~(\ref{AEq}).
\end{beweis}

In the following lemma we will establish the relation between the equation
in Lemma~\ref{Lem.3.3.5}~(ii) and the ${\got A}$-Equation (\ref{AEq}).

\begin{lem}
\label{Lem.3.3.15}
Put $\varphi(p):=D(H_p)\, 
\widetilde{\varphi}(p)\in h_{0}$, for all $\widetilde{\varphi}\in
\Lzwei{\Lkf,{\cal H},\mudp}$ $($see Eq.~$(\ref{Isometry0}))$. 
Then we have that $\widetilde{\varphi}$ satisfies the equation 
$\pi\, \widetilde{\varphi}(p)=\widetilde{\varphi}(p)$ 
iff $\;\varphi$ satisfies the ${\got A}$-Equation $(\ref{AEq})$.
\end{lem}
\begin{beweis}
1.~Suppose that $\widetilde{\varphi}$ satisfies the equation
$\pi\,\widetilde{\varphi}(p)=\widetilde{\varphi}(p)$, $p\in\Lk$. Then 
there exist three scalar functions
$\chi_i\in\Lzwei{\Lkf,\z{C},\mudp} $, $i=0,1,2$, such that 
$\widetilde{\varphi}(p)=
\left(\kern-1.5mm\begin{array}{c}  
    \chi_0(p) \\ \chi_1(p) \\ \chi_2(p) \\ 0
\end{array} \kern-1.5mm \right)$. But using Eq.~(\ref{SolHp}) one can
explicitly check that $\varphi(p)=D(H_p)\,\widetilde{\varphi}(p)$ satisfies
Eq.~(\ref{AEq}).

2.~Suppose now that $\varphi$ satisfies Eq.~(\ref{AEq}). Then we can
compute:
\begin{eqnarray*}
\widetilde{\varphi}_3(p) &=& \LR D(H_p)^{-1}\varphi(p)\RR_3         \\[1mm]
                     &=& (p_0-p_3)\varphi_0(p)-(p_1+ip_2)\varphi_1(p)-
                         (p_1-ip_2)\varphi_2(p)+(p_0+p_3)\varphi_3(p) \\[2mm]
                     &=& 0 \, ,
\end{eqnarray*}
and therefore $\widetilde{\varphi}$ satisfies the equation 
$\pi\,\widetilde{\varphi}(p)=\widetilde{\varphi}(p)$, $p\in\Lk$.
\end{beweis}

Denote by
\be
\label{SolutionPot}
 h_{\KIn{$+-$}}:=\left\{ D(H_p) 
            \left(\kern-1.5mm\begin{array}{c}
 \chi_0(p) \\ \chi_1(p) \\  \chi_2(p) \\ 0
            \end{array}\kern-1.5mm\right)\!\mid  
 \chi_i\in\Lzwei{\Lkf,\z{C},\mudp},\; i=0,1,2\right\}
\ee
the space of solutions of the ${\got A}$-Equation
and recall that from the general definition of $V$ given in 
Eq.~(\ref{GeneralV}) we have here,
\[
 \LR V_{\KIn{$+-$}}(g)\,\varphi \RR\!(p):=e^{-ipa}\,
                           D^{\KIn{$(\frac12,\frac12)$}}(A)\, \varphi(q),
\] 
where $g=(A,a)\in \semi{\g{SL}{2,\z{C}}}{\z{R}^4}$,
$q:=\Lambda^{-1}_A p$ and $\varphi\in h_{\KIn{$+-$}}$.

\begin{rem}
\label{Rem.3.3.16}
In analogy with the Weyl case or with the ${\got F}$-Equation case, 
we can try to define on $h_{\KIn{$+-$}}$ the following sesquilinear form:
\[
 \askp{\varphi}{\varphi^{\bullet}}
      := \int\limits_{\Lkf}\askp{\varphi(p)}
          {D(H_p^{-1})^*D(H_p^{-1})\,\varphi^{\bullet}(p)}_{\z{C}^4}\mudp 
      = \int\limits_{\Lkf}\sum\limits_{i=0}^{2}\overline{\chi_i(p)} 
             \chi_i^{\bullet}(p) \mudp,
\]
where $\varphi(p):=D(H_p)\left(\kern-1.5mm\begin{array}{c}
     \chi_0(p) \\ \chi_1(p) \\ \chi_2(p) \\ 0
                  \end{array}\kern-1.5mm\right)$,
$\varphi^{\bullet}(p):=D(H_p)\left(\kern-1.5mm\begin{array}{c}
   \chi_0^{\bullet}(p) \\ \chi_1^{\bullet}(p) \\ \chi_2^{\bullet}(p) \\ 0
                    \end{array}\kern-1.5mm\right)\in h_{\KIn{$+-$}}$.
This sesquilinear form is positive definite, but it will not be
$V_{\KIn{$+-$}}$-invariant, since in the present case we have in general 
that $\pi\,D(H^{-1}_p A H_{q})^*D(H^{-1}_p A H_{q})\,\pi\neq\pi$,
cf.~Eq.~(\ref{CondProj}). Roughly speaking, we have allowed 
too many degrees of freedom on the fibre in order to apply the 
arguments used in the Weyl case or in the ${\got F}$-Equation case
which are based on Theorem~\ref{TInvProj}.
\end{rem}

Motivated nevertheless by the Lorentz-invariance of the 
Minkowski scalar product we introduce the following sesquilinear form: 
for $\varphi$, $\varphi^{\bullet}$ as in the preceding lemma,
\begin{eqnarray}
\label{Askp4}\askp{\varphi}{\varphi^{\bullet}}_{\KIn{$+-$}}
      & := & \int\limits_{\Lkf}\askp{\varphi(p)}
             {\eta\,\varphi^{\bullet}(p)}_{\z{C}^4}\mudp \\
\label{IsometryA}
      &\phantom{:}=& \int\limits_{\Lkf}\sum\limits_{i=1}^{2}
                     \overline{\chi_i(p)}\, \chi_i^{\bullet}(p) \mudp,
\end{eqnarray}
where $\eta$ is given in the proof of
Lemma~\ref{Lem.3.3.14} and can be seen as the spinorial form of the 
Minkowski metric.
Note that since
\be
\label{SpinorialMetricRelation}
D(A)^*\;\eta\;D(A)=\eta
\ee
the above sesquilinear form is
$V_{\KIn{$+-$}}$-invariant, but now 
$\langle\cdot,\cdot\rangle_{\KIn{$+-$}}$ 
is only semidefinite and the corresponding 
(degenerate) space of zero vectors is easily seen to be,
\be
\label{HA0}
  h_{d}:=\left\{ D(H_p) 
            \left(\kern-1.5mm\begin{array}{c}
         \chi(p) \\ 0 \\ 0 \\ 0
            \end{array}\kern-1.5mm\right)
           \mid \chi\in \Lzwei{\Lkf,\z{C},\mudp} \right\}.
\ee
Since $h_{d}$ is a $V_{\KIn{$+-$}}$-invariant space
denote by $V_{\KIn{${\got A}$}}$ the natural definition of $V_{\KIn{$+-$}}$ 
on the factor space $h_{\KIn{${\got A}$}}:=
{h_{\KIn{$+-$}}}/{h_{d}}$, which is Hilbert space 
w.r.t.~the scalar product, $\langle\cdot,\cdot\rangle_{\KIn{${\got A}$}}$,
defined as the lift of
$\langle\cdot,\cdot\rangle_{\KIn{$+-$}}$.\footnote{Notice 
that we can not restrict $V_{\KIn{$+-$}}$ to the space 
$\left\{ D(H_p) 
            \left(\kern-1.5mm\begin{array}{c}
          0 \\ \chi_1(p) \\  \chi_2(p) \\ 0
            \end{array}\kern-1.5mm\right)
           \mid \chi_i\in\Lzwei{\Lkf,\z{C},\mudp},\; i=1,2 \right\}$,
because it is not $V_{\KIn{$+-$}}$-invariant. Indeed, this follows from the 
fact that the space 
$\left\{\left(\kern-1.5mm\begin{array}{c}
                   0 \\ a \\ b \\ 0
  \end{array}\kern-1.5mm\right) \mid a,b\in \z{C} \right\}$
is not $D(L)$-invariant, $L\in{\cal E}(2)$ (cf.~\cite[Section~5.B.1]{Asorey85}).}
The elements of the factor space are written as $\left[ \varphi\right]_0$,
where $\varphi\in h_{\KIn{$+-$}}$ and the bracket, $[\cdot]_0$, specify the
corresponding equivalence class. The preceding situation with the
appearance of factor spaces is typical when dealing with not
fully decomposable \rep{s} of a Lie group. This situation is studied in
general terms by Araki and our construction above using the
space of solutions of the ${\got A}$-Equation is a special 
case of Theorem~1 in \cite{Araki85}. 

The following statement justifies
the use of the index $\In{$+-$}$ in $h_{\KIn{$+-$}}$, 
since this space carries a \rep{} that contains the 
irreducible \rep{s} describing helicity $+1$ and $-1$.
In the next result we will show the equivalence of the \rep{s}
$(V_{\KIn{${\got A}$}},h_{\KIn{${\got A}$}})$ and the direct sum of 
the canonical \rep{s} $U_+$ and $U_-$ given in Eqs.~(\ref{Photon+}) 
and (\ref{Photon-}) respectively.

\begin{pro}
\label{Lem.3.3.17}
The mapping, $\Phi_{\KIn{${\got A}$}}\colon\ 
h_{\KIn{${\got A}$}}\longrightarrow \Lzwei{\Lkf,\z{C},\mudp}
\oplus\Lzwei{\Lkf,\z{C},\mudp}$, defined by
\[
  \LR\Phi_{\KIn{${\got A}$}}\LR \LE D\!\LR H_{(\cdot)}\RR 
      \left(\kern-1.5mm\begin{array}{c}
    \chi_0(\cdot) \\ \chi_1(\cdot) \\  \chi_2(\cdot) \\ 0
      \end{array}\kern-1.5mm\right)  \RE_0 \RR\RR\!(p):=
      \chi_1(p)\oplus\chi_2(p)\, ,
\]
is an isometric isomorphism that commutes with the \rep{s} 
$V_{\KIn{${\got A}$}}$ and
$U_+\oplus U_-$, i.e.~the equation, $\Phi_{\KIn{${\got A}$}}\, 
V_{\KIn{${\got A}$}}(g)=
\LR U_+(g)\oplus U_-(g)\RR\Phi_{\KIn{${\got A}$}}$ holds for all $g\in\UPgr$. 
The \rep{} $V_{\KIn{${\got A}$}}$ on $\LR h_{\KIn{${\got A}$}}, 
\langle\cdot,\cdot\rangle_{\KIn{${\got A}$}}\RR$ is unitary, 
strongly continuous and reducible.
\end{pro}
\begin{beweis}
The unitarity of $V_{\KIn{${\got A}$}}$ follows from the 
$V_{\KIn{$+-$}}$-invariance of the sesquilinear form 
$\langle\cdot,\cdot\rangle_{\KIn{$+-$}}$ and from the
construction of the factor space 
${h_{\KIn{$+-$}}}/{h_{d}}$.

The isometry property of $\Phi_{\KIn{${\got A}$}}$ follows already from 
Eq.~(\ref{IsometryA}) and the intertwining property can be checked 
by direct computation as in Proposition~\ref{Lem.3.3.12}. Use, for instance, 
the relation
\begin{eqnarray*}
 \lefteqn{\LR\Phi_{\KIn{${\got A}$}}\LR\LE D\!\LR H_{(\cdot)}\RR e^{-i(\cdot)a}\; 
  D\!\LR H^{-1}_{(\cdot)} A H_{\Lambda^{-1}_A (\cdot) }\RR
          \left(\kern-1.5mm\begin{array}{c}
    \chi_0(\Lambda^{-1}_A\cdot) \\ \chi_1(\Lambda^{-1}_A\cdot) \\  
    \chi_2(\Lambda^{-1}_A\cdot) \\              0
          \end{array}\kern-1.5mm\right)  \RE_0 \RR\RR\!\!(p)}\kern8cm \\[2mm]
 \kern6cm &=& \LR U_+(g)\chi_1\RR\!(p)\oplus \LR U_-(g)\chi_2\RR\!(p)\, . 
 \end{eqnarray*}

Recall further that, for $g=(A,a)\in 
\semi{\g{SL}{2,\z{C}}}{\z{R}^4}$, $q:=\Lambda^{-1}_A p$ and 
$\chi\in \Lzwei{\Lkf,\z{C},\mudp}$, the canonical representations 
(cf.~Eq.~(\ref{Vf0Can}) in the case $n=2$),
\[
 \LR U_{\pm}(A,a)\,\chi\RR\!(p):= e^{-ipa}\,e^{\pm i\theta(A,p)}\,\chi(q) \,,
\]
are unitary, strongly continuous and irreducible. They correspond to
systems with opposite helicity. 
\end{beweis}

Finally, we will prove a theorem that relates the pair 
$( h_{\KIn{${\got A}$}},\;V_{\KIn{${\got A}$}})$ 
defined above with some combinations of the spaces and \rep{s} 
used in Subsection~\ref{TheF}. Concretely, using the 
definitions\footnote{Recall that in Subsection~\ref{TheF}, 
$D(\cdot)$ means $D^{\KIn{$(1,0)$}}(\cdot)$. We have also written 
$D^{\KIn{$(0,1)$}}(A)$ instead of using the notation
$D^{\KIn{$(1,0)$}}(\overline{A})$, $A\in\g{SL}{2,\z{C}}$.} 
(\ref{HF+}), (\ref{VF+}), (\ref{HF-}) and (\ref{VF-}) we consider
the following Hilbert space, scalar product and unitary 
\rep{} w.r.t.~it
\begin{Klammer}
\label{ReferenzraumF}
  h_{\KIn{${\got F}$}}                   &:=& h_+ \oplus h_-          \\[2mm]
  \langle\cdot,\cdot\rangle_{\KIn{${\got F}$}} 
                                         &:=& \langle\cdot,\cdot\rangle_+
                                  \oplus \langle\cdot,\cdot\rangle_- \\[3mm]  
  V_{\KIn{${\got F}$}}                   &:=& V_+ \oplus V_- \, .
\end{Klammer}
Then we have the following equivalence
between $(V_{\KIn{${\got F}$}},h_{\KIn{${\got F}$}})$ and 
$(V_{\KIn{${\got A}$}},h_{\KIn{${\got A}$}})$:

\begin{teo}
\label{Teo.3.3.18}
The mapping $\Phi_{\KIn{$\got A\got F$}}\colon\ h_{\KIn{${\got F}$}} 
\longrightarrow h_{\KIn{${\got A}$}}$, defined by
\[
 \Phi_{\KIn{$\got A\got F$}}\!\LR\!
 D^{\KIn{$(1,0)$}} (H_{(\cdot)})\!
         \left(\kern-2mm  \begin{array}{c}  
                           1 \\ 0 
                     \end{array} \kern-2mm \right) 
              \!\!\otimes\!\!  \left(\kern-2mm  \begin{array}{c}  
                           1 \\ 0 
                     \end{array} \kern-2mm \right)
                     \!\chi_+(p) 
      \!\oplus\! D^{\KIn{$(0,1)$}}(H_{(\cdot)})\!
        \left(\kern-2mm  \begin{array}{c}  
                           1 \\ 0 
                     \end{array} \kern-2mm \right) 
              \!\!\otimes\!\!  \left(\kern-2mm  \begin{array}{c}  
                           1 \\ 0 
                     \end{array} \kern-2mm \right)
                     \!\chi_-(p)\! \RR\!\!(p)
:= \!\LE \!D^{\KIn{$(\frac12,\frac12)$}} (H_p)\!\left(\kern-2mm\begin{array}{c}
           0 \\ \chi_+(p) \\ \chi_-(p) \\ 0
            \end{array}\kern-2mm\right) \!\RE_0 \!\! ,
\]
with $\varphi_+(p):= D^{\KIn{$(1,0)$}} (H_p)\!
       \left(\kern-2mm  \begin{array}{c}  
                           1 \\ 0 
                     \end{array} \kern-2mm \right) 
              \!\otimes\!  \left(\kern-2mm  \begin{array}{c}  
                           1 \\ 0 
                     \end{array} \kern-2mm \right)
                     \!\chi_+(p) \in h_+$ and
$\varphi_-(p):= D^{\KIn{$(0,1)$}}({H_p})\!
\left(\kern-2mm  \begin{array}{c}  
                           1 \\ 0 
                     \end{array} \kern-2mm \right) 
              \!\otimes\!  \left(\kern-2mm  \begin{array}{c}  
                           1 \\ 0 
                     \end{array} \kern-2mm \right)
                     \!\chi_-(p) \in h_-$, 
is an isometric isomorphism that commutes with the corresponding \rep{s}, 
i.e.~$\Phi_{\KIn{$\got A\got F$}}\,
V_{\KIn{${\got F}$}}(g)=V_{\KIn{${\got A}$}}(g)\, 
\Phi_{\KIn{$\got A\got F$}}$, $g\in\UPgr$.
\end{teo}
\begin{beweis}
The isometry property follows from the equations,
\begin{eqnarray*}
 \askp{\Phi_{\KIn{$\got A\got F$}}(\varphi_+\oplus\varphi_-)}{
         \Phi_{\KIn{$\got A\got F$}}(\varphi_+\oplus
         \varphi_-)}_{\KIn{${\got A}$}} 
 &=& \int\limits_{\Lkf}\!\!\LR |\chi_+(p)|^2+|\chi_-(p)|^2\RR\!\!\mudp \\[1mm]
 &=& \askp{\varphi_+\oplus\varphi_-}{\varphi_+\oplus\varphi_-}_{
      \KIn{${\got F}$}}\, .
\end{eqnarray*}
The intertwining property is a consequence of Proposition~\ref{Lem.3.3.12} 
(and the corresponding result for the opposite helicity) and of 
Proposition~\ref{Lem.3.3.17}. Indeed, note that $\Phi_{\KIn{$\got A\got F$}}=
\Phi_{\KIn{${\got A}$}}^{-1} \Phi_{\KIn{${\got F}$}}$ and, therefore,
\[
 \Phi_{\KIn{$\got A\got F$}}\,V_{\KIn{${\got F}$}}(g)=
 \Phi_{\KIn{${\got A}$}}^{-1} \Phi_{\KIn{${\got F}$}}\;
     V_{\KIn{${\got F}$}}(g)=
 \Phi_{\KIn{${\got A}$}}^{-1}\LR U_+(g)\oplus U_-(g)\RR 
     \Phi_{\KIn{${\got F}$}}=
 V_{\KIn{${\got A}$}}(g)\; \Phi_{\KIn{${\got A}$}}^{-1} 
     \Phi_{\KIn{${\got F}$}}=
 V_{\KIn{${\got A}$}}(g)\, \Phi_{\KIn{$\got A\got F$}}\,,\;\;g\in\UPgr\,,
\]
and the proof is concluded.
\end{beweis}

\begin{rem}\label{Landsman1}
Landsman and Wiedemann \cite[Theorem~1]{Landsman95b} 
interpret the space $( h_{\KIn{${\got A}$}},\; \langle\cdot,
\cdot\rangle_{\KIn{${\got A}$}}, \;V_{\KIn{${\got A}$}})$ 
(written in tensorial language) in the context of Marsden-Weinstein 
reduction theory. 
They also mention its equivalence to the triplet $(h_{\KIn{${\got F}$}},\; 
\langle\cdot,\cdot\rangle_{\KIn{${\got F}$}},\;V_{\KIn{${\got F}$}})$.
For the relation between the two preceding spaces in terms of tensors
see also \cite{Bertrand71} or \cite[Section~V.B]{Heidenreich86}.
\end{rem}

\newpage
\subsection{Summary}\label{Sumario}
In the present section we have described covariant and massive/massless
canonical representations of the Poincar\'e group in the general frame
of induced representation theory. Relativistic wave equations appear
in the context of canonical representations if one considers reducible
representations of the corresponding little groups. Due to the different
nature of the massive resp.~massless little groups, the corresponding\
\rwe{s} play also a different role and are characterised by 
reducing resp.~invariant projections. 
We will summerise in the following table some of the results concerning
massive and massless \rwe{s}.\vspace{.8cm}

\label{Table1}
\begin{tabular}{|c|c|c|}
\hline
\multicolumn{3}{|c|}{\rule[-3mm]{0mm}{10mm}\Large{\bf massive}, $m>0$}\\
\hline\hline
   \mbox{\rule[-3mm]{0mm}{8mm}{\sc rel.~wave equation}} & 
   \mbox{\rule[-3mm]{0mm}{8mm}{\sc Inducing rep.~of SU(2):}} & 
   \mbox{\rule[-3mm]{0mm}{8mm}{\sc reducing projection}} \\[-3.5mm]
   \mbox{\rule[-3mm]{0mm}{8mm}{}} & 
   \mbox{\rule[-3mm]{0mm}{8mm}{Unitary and fully decomposable}} & 
   \mbox{\rule[-3mm]{0mm}{8mm}{}} \\
\hline\hline\rule[-1.2cm]{0mm}{2.7cm}
  Dirac & $\tau(U):=U\oplus U$ &
   $\pi^{\KIn{(Dirac)}}:=\frac12
           \left(\kern-1.5mm \begin{array}{cccc}
   1 \kern-1mm & 0 \kern-1mm & 1 \kern-1mm & 0             \\
   0 \kern-1mm & 1 \kern-1mm & 0 \kern-1mm & 1             \\ 
   1 \kern-1mm & 0 \kern-1mm & 1 \kern-1mm & 0             \\
   0 \kern-1mm & 1 \kern-1mm & 0 \kern-1mm & 1                        
     \end{array} \kern-1.5mm\right)$ \\
\hline \rule[-1.2cm]{0mm}{2.7cm}
Proca & $D^{\KIn{$(\frac12,\frac12)$}}(U)=U\otimes \overline{U}$ &
   $\pi^{\KIn{(Proca)}}:=\frac12
           \left(\kern-1.5mm \begin{array}{cccc}
   1 \kern-1mm & 0 \kern-1mm & 0 \kern-1mm & -1            \\
   0 \kern-1mm & 2 \kern-1mm & 0 \kern-1mm & 0             \\ 
   0 \kern-1mm & 0 \kern-1mm & 2 \kern-1mm & 0             \\
  -1 \kern-1mm & 0 \kern-1mm & 0 \kern-1mm & 1                        
      \end{array} \kern-1.5mm\right) $\\
\hline\hline
\multicolumn{3}{|c|}{\rule[-3mm]{0mm}{10mm}\Large{\bf massless}, $m=0$}\\
\hline\hline
   \mbox{\rule[-3mm]{0mm}{8mm}{\sc rel.~wave equation}} & 
   \mbox{\rule[-3mm]{0mm}{8mm}{\sc Inducing rep.~of ${\cal E}(2):$}} & 
   \mbox{\rule[-3mm]{0mm}{8mm}{\sc invariant projection}} \\[-3.5mm]
   \mbox{\rule[-3mm]{0mm}{8mm}{}} & 
   \mbox{\rule[-3mm]{0mm}{8mm}{{\em Non}unitary and {\em non}}} & 
   \mbox{\rule[-3mm]{0mm}{8mm}{}} \\[-3.5mm]
   \mbox{\rule[-3mm]{0mm}{8mm}{}} & 
   \mbox{\rule[-3mm]{0mm}{8mm}{fully decomposable}} & 
   \mbox{\rule[-3mm]{0mm}{8mm}{}} \\
\hline\hline  \rule[-7mm]{0mm}{1.6cm}
  Weyl & $D^{\KIn{$(\frac{1}{2},0)$}}(L):=L$ &
   $\pi_{\KIn{${\got W}$}}:=
    \left(\kern-1.5mm\begin{array}{cc}
         1 \kern-1.6mm & 0 \\
         0 \kern-1.6mm & 0
    \end{array}\kern-1.5mm\right)$ \\        
\hline \rule[-7mm]{0mm}{1.6cm}
  Maxwell: $\got F$-Eq.~& $D^{\KIn{$(1,0)$}}(L):=L\otimes L$ &
   $\pi_{\KIn{${\got F}$}}:=
    \left(\kern-1.5mm\begin{array}{cc}
         1 \kern-1.6mm & 0 \\
         0 \kern-1.6mm & 0
    \end{array}\kern-1.5mm\right)\otimes
     \left(\kern-1.5mm\begin{array}{cc}
         1 \kern-1.6mm & 0 \\
         0 \kern-1.6mm & 0
    \end{array}\kern-1.5mm\right)$ \\
\hline \rule[-1.2cm]{0mm}{2.7cm}
Maxwell: $\got A$-Eq. & 
$D^{\KIn{$(\frac12,\frac12)$}}(L)=L\otimes \overline{L}$ &
   $\pi_{\KIn{${\got A}$}}:=
           \left(\kern-1.5mm \begin{array}{cccc}
   1 \kern-1mm & 0 \kern-1mm & 0 \kern-1mm & 0             \\
   0 \kern-1mm & 1 \kern-1mm & 0 \kern-1mm & 0             \\ 
   0 \kern-1mm & 0 \kern-1mm & 1 \kern-1mm & 0             \\
   0 \kern-1mm & 0 \kern-1mm & 0 \kern-1mm & 0                        
      \end{array} \kern-1.5mm\right) $\\
\hline \rule[-10mm]{0mm}{2.0cm}
    $\begin{array}{c} \mathrm{General~massless} \\[2mm] 
    (\mathrm{helicity}: \frac{n}{2},n\geq 1)\end{array}$ &  
$D^{\KIn{$(\frac{n}{2},0)$}}(L)=\mathop{\otimes}\limits^{n}L$ &
   $\pi_{\KIn{$n$}}:=\mathop{\otimes}\limits^n
    \left(\kern-1.5mm\begin{array}{cc}
         1 \kern-1.6mm & 0 \\
         0 \kern-1.6mm & 0
    \end{array}\kern-1.5mm\right)$ \\
\hline
\end{tabular}
\newpage

\section{Massless free nets and \rwe{s}}\label{AxiomsFreeNets}

We begin this section stating some core axioms of local quantum physics
\cite{bHaag92,Haag64,bBaumgaertel95,bBaumgaertel92,Schroer99,FredenhagenIn91}.
Here the point of view is that the correspondence $\al O.\mapsto
\al A.(\al O.)$ between Minkowski space regions $\al O.$ and local
algebras of observables $\al A.(\al O.)$ characterises intrinsically the
theory. Following Haag's suggestive idea, quantum 
fields (which are the central objects in other
formulations of QFT) can be seen in the present setting just as
`coordinates' of the preceding net, in the sense that one may use
different quantum fields to describe the same abstract net. 
In order to avoid any concrete representation of the C*-algebra we 
will construct the net directly following the strategy presented 
in \cite{Lledo95} (see also \cite{bSegal63}), 
i.e.~basing the construction on group-theoretical 
arguments and standard CAR or CCR-theory (see for the 
latter algebras \cite[Chapter~8]{bBaumgaertel92} and 
references cited therein; in the Fermi case we will use
Araki's self-dual approach to the CAR-algebra). 
We will call the result of this 
construction a {\em free net} and the fundamental object that
characterises it is the so-called embedding that reduces the
covariant representation in terms of the corresponding canonical
ones (cf.~Remark~\ref{Rem.3.3.1}). As reference spaces of the 
CAR- and CCR-algebras we will use in this paper 
the space of solutions of the 
Weyl- and the $\got F$-Equation introduced in 
Subsections~\ref{TheWeyl} and \ref{TheF}, respectively,
and will therefore call the corresponding already typical nets of 
local C*-algebras Weyl- resp.~$\got F$-net. Some relations
of the latter with the vector potential will also be mentioned.
We will also relate the present construction to the 
nets specified in \cite{Lledo01}, since 
in this reference the nets were given without mentioning 
explicitly the corresponding \rwe{s}. Concretely, 
we will show that the nets associated to the 
Weyl- and the $\got F$-Equation are isomorphic to the nets
constructed in \cite[cases $n=1$ and $n=2$]{Lledo01},
respectively. This isomorphy may then be easily generalised to
arbitrary $n$. Thus producing the same net of local C*-algebras
the generalisation of the Weyl- and the $\got F$-net constructed
in the following will present a new methodological aspect
w.r.t.~\cite{Lledo95,Lledo01}, namely we will 
show explicitly the relation to the corresponding \rwe{s}. 
In particular the embeddings used here are different 
from those in \cite{Lledo95,Lledo01} and to construct them we
will make essential use of distinguished elements of certain
intertwiner spaces associated to representations of the little group
$\al E.(2)$. By means of these elements the embedding will map any 
(vector-valued) test function into the space of solutions of the 
corresponding massless \rwe{}. This procedure illuminates another
aspect of the reduction of the (fibre) degrees of freedom that is
necessary when considering massless representations of nontrivial 
helicity (cf.~Remark~\ref{Rem.3.3.1}).

Denote by $\BR$ the family of open and bounded regions in Minkowski space
partially ordered by inclusion `$\subset$'. $(\BR,\,\subset)$ is 
then a directed index set which is stable under the action of the
Poincar\'e group \cite[Sections~5.1 and 7.1]{bBaumgaertel92}. 

\begin{defi}
\label{HKnet}
A correspondence $\BR\ni\al O.\mapsto \al A.(\al O.)$ where the {\em 
local algebras} $\al A.(\al O.)$ are 
$($abstract$)$ unital {\rm C*}-algebras with common unit $\EINS$, is 
called a Haag-Kastler net $($HK-net for short$)$ if the following conditions
are satisfied:
\begin{itemize}
\item[{\rm (i)}] $($Isotony$)$
 If $\B_1\subseteq\B_2$, then 
 $\al A.(\B_1)\subseteq\al A.(\B_2)$, $\al O._1,\al O._2\in\BR$.
 We denote by $\al A.:=\mathop{{\rm lim}}\limits_{\longrightarrow}\al A.(\B)$
 the corresponding inductive limit which is called the {\em quasi local
 algebra}.
\item[{\rm (ii)}] $($Additivity$)$
 Let $\{\al O._\lambda\}_{\lambda\in\Lambda}\subset\BR$ with
 $\cup_\lambda\al O._\lambda\in\BR$. The net $\al O.\mapsto \al A.(\al O.)$
 satisfies additivity if for any such 
 $\{\al O._\lambda\}_\lambda$ the following equation holds in $\al A.$:
 $\al A.(\cup_\lambda\al O._\lambda)=
      \Wort{C}^*\!\LR\cup_\lambda \al A.(\al O._\lambda)\RR\!$.
\item[{\rm (iii)}] $($Causality$)$
 For any $\B_1\in\BR$ space-like separated 
 w.r.t.~$\B_2\in\BR$ $($we denote this by $\B_1\perp\B_2)$, then
 $\al A.(\B_1)$ commutes elementwise with $\al A.(\B_2)$ in $\al A.$.
\end{itemize}
Suppose further that there exists a representation $\alpha_{(\cdot)}$ of the 
$\UPgr$ in terms of *-automorphisms of $\al A.$,
i.e.~$\UPgr\ni g\mapsto \alpha_g\in{\rm Aut}\,\al A.$.
\begin{itemize}
\item[{\rm (iv)}] $($Covariance$)$
 The net $\B\mapsto \al A.(\B)$
 transforms covariantly w.r.t.~$\alpha$, if for every $\B\in\BR$ we have
 $\alpha_g (\al A.(\B))=\al A.(g\B)$, $g\in\UPgr$,
 where $g\B:=\{gx\mid x\in\B \}$.
\end{itemize}
\end{defi}

Next we will introduce the notion of isomorphic HK-nets 
(cf.~\cite[Section~3]{Buchholz95}).

\begin{defi}
\label{IsoHK}
Two HK-nets $(\al A.^{\KIn{(i)}}(\B),
\alpha_{\KIn{$(\cdot)$}}^{\KIn{(i)}})_{\KIn{$\B\in\BR$}}$
with quasi local algebras $\al A.^{(i)}$, $i=1,2$, are called 
{\em isomorphic} if there exists a *-isomorphism
$\Lambda\colon\ \al A.^{(1)}\to\al A.^{(2)}$ which preserves localisation,
i.e.~$\Lambda(\al A.^{(1)}(\B))=\al A.^{(2)}(\B)$, $\B\in\BR$, and 
intertwines between the corresponding actions of the Poincar\'e group,
i.e.~$\Lambda\,\alpha_g^{\KIn{(1)}}=\alpha_g^{\KIn{(2)}}\,\Lambda$, 
$g\in\UPgr$.
\end{defi}

Following the strategy suggested in \cite[Section~8.3]{bBaumgaertel92}
we will now study a subclass of HK-nets, namely those where the local 
C*-algebras are certain C*-subalgebras of the CAR- resp.~CCR-algebras.
Due to the nice functorial properties of these
algebras it is possible to encode the axioms of isotony, 
additivity, causality and covariance
of the HK-net at the level of the respective reference spaces. 
We will call the result of this construction a free net.
(The indices F/B below denote the Fermi/Bose cases.)

\begin{defi}
\label{Tuples}
We consider the following tuples associated to $(\R^4,\perp,\UPgr)$,
where $\perp$ is the causal disjointness relation given by the 
Minkowski metric on $\R^4$.
\begin{itemize}
\item[{\rm (i)}]In the {\em Fermi or CAR case} we have    
  $( h_{\KIn{F}},\, \langle\cdot,\cdot\rangle,\, \Gamma,\, V_{\KIn{F}},\,
  {\cal T}_{\KIn{F}},\, T_{\KIn{F}},\, {\got I}_{\KIn{F}})$, where
  $( h_{\KIn{F}},\, \langle\cdot,\cdot\rangle)$ 
  is a complex Hilbert space and $\Gamma$ an anti-unitary involution 
  on it. $V_{\KIn{F}}$ denotes a unitary representation of ${\UPgr }$ on 
  $( h_{\KIn{F}},\,\langle\cdot,\cdot\rangle )$.
  Further, ${\cal T}_{\KIn{F}}$
  is the set of test functions on $\R^4$ with compact
  support and $T_{\KIn{F}}$ is a representation of ${\UPgr }$
  on ${\cal T}_{\KIn{F}}$ satisfying the following support property: 
  if $f\in{\cal T}_{\KIn{F}}$ with $\Wort{supp}f\subset {\cal O}\in
  {\cal B}({\R^4})$, then $T_{\KIn{F}}(g) f\in {\cal T}_{\KIn{F}}$
  with $\Wort{supp}\,T_{\KIn{F}}(g) f\subset g{\cal O}$, $g\in{\UPgr }$. 
  Finally, we require for the linear embedding
  ${\got I}_{\KIn{F}}\colon\ {\cal T}_{\KIn{F}}\longrightarrow h_{\KIn{F}}$ 
  the following properties:
 \begin{itemize}
   \item[$(${\bf F1}$)$]$(\Gamma$-invariance.$)$ For an arbitrary 
    $f\in{\cal T}_{\KIn{F}}$ with $\Wort{supp}f\subset{\cal O}\in
    {\cal B}({\R^4 })$, there exists a $k\in{\cal T}_{\KIn{F}}$ 
    such that $\Gamma\, {\got I}_{\KIn{F}}f= 
    {\got I}_{\KIn{F}}k$ and $\Wort{supp}k\subset{\cal O}$.
   \item[$(${\bf F2}$)$] $($Causality.$)$ For all 
    $f,k\in {\cal T}_{\KIn{F}}$ such that 
    $\Wort{supp}f \bot\,\Wort{supp}k$, we have 
    $\langle{\got I}_{\KIn{F}}f\,,\,{\got I}_{\KIn{F}}k\rangle=0$.
   \item[$(${\bf F3}$)$] $($Covariance.$)$ 
    $\Gamma\, V_{\KIn{F}}(g)= V_{\KIn{F}}(g)\,
    \Gamma$ and ${\got I}_{\KIn{F}}\, T_{\KIn{F}}(g)=V_{\KIn{F}}(g)\, 
    {\got I}_{\KIn{F}}$, for all $g\in{\UPgr }$.
 \end{itemize}

\item[{\rm (ii)}] In the {\em Bose or CCR case} we have     
  $( h_{\KIn{B}},\, \sigma,\, V_{\KIn{B}},\, 
  {\cal T}_{\KIn{B}},\, T_{\KIn{B}},\, {\got I}_{\KIn{B}})$, 
  where $V_{\KIn{B}}$ is a symplectic representation
  of ${\UPgr }$ on the real symplectic space $( h_{\KIn{B}},\, \sigma )$,
  i.e.~$V_{\KIn{B}}(g)$, $g\in \UPgr $, is a bijection of 
  $h_{\KIn{B}}$ that leaves $\sigma$ invariant. ${\cal T}_{\KIn{B}}$ 
  is again the set of test functions on ${\R^4}$ with compact
  support and $T_{\KIn{B}}$ is a representation of ${\UPgr }$ on 
  ${\cal T}_{\KIn{B}}$ satisfying the same support property as in the 
  fermionic case. We require for the linear embedding 
  ${\got I}_{\KIn{B}}\colon\ {\cal T}_{\KIn{B}} 
  \longrightarrow h_{\KIn{B}}$ the following properties:
 \begin{itemize}
\item[$(${\bf B1}$)$] $($Causality.$)$ For all $f,k\in {\cal T}_{\KIn{B}}$
  such that $\Wort{supp}f \bot\Wort{supp}k$, we have 
  $\sigma({\got I}_{\KIn{B}}f,\,{\got I}_{\KIn{B}}k)=0$.
\item[$(${\bf B2}$)$] $($Covariance.$)$ 
     ${\got I}_{\KIn{B}}\, T_{\KIn{B}}(g)=
     V_{\KIn{B}}(g)\, {\got I}_{\KIn{B}}$, for all $g\in{\UPgr }$.
 \end{itemize}
\end{itemize}
\end{defi}

Next we will show that the tuples that were specified in the preceding
definition characterise in a canonical way a HK-net.

\begin{teo}
\label{TeoFreeNet}
Assume the notation given in Definition~$\ref{Tuples}$ and consider the
following nets of local linear submanifolds of the corresponding reference
spaces:
\bea
 \BR\ni {\cal O}\longmapsto h_{\KIn{F}}({\cal O})
     &:=& \LG{\got I}_{\KIn{F}}f\mid  f\in {\cal T}_{\KIn{F}}
          \quad \Wort{and}\quad 
          \Wort{supp}f \subset {\cal O}\RG\subset h_{\KIn{F}}\,,        \\
 \BR\ni {\cal O} \longmapsto h_{\KIn{B}}({\cal O})
     &:=& \LG{\got I}_{\KIn{B}}f\mid  f\in {\cal T}_{\KIn{B}}
          \quad \Wort{and}\quad 
          \Wort{supp}f \subset {\cal O}\RG\subset h_{\KIn{B}}.
\eea
Then we have 
\begin{itemize}
\item[{\rm (i)}] The net of local {\rm C*}-algebras given by
\[
 \BR\ni {\cal O}\longmapsto {\cal A}_{\KIn{F}}({\cal O})
 :=\Wort{C}^*\LR \{\ot a.(\varphi)\mid  \varphi\in h_{\KIn{F}}({\cal O})\}
 \RR^{\KIn{$\z{Z}_2$}} \subset \car{h_{\KIn{F}},\Gamma}
\]
is a HK-net. Here $\ot a.(\cdot)$ denote the generators of 
$\car{h_{\KIn{F}},\Gamma}$ and
${\cal A}^{\;\KIn{$\z{Z}_2$}}$ means the fixed point subalgebra of the 
C*-algebra $\cal A$ w.r.t.~Bogoljubov automorphism associated to the 
unitarity $-\EINS$. The covariance of this net of local C*-algebras is 
realised
by the Bogoljubov automorphisms $\alpha_g$ associated to the Bogoljubov 
unitaries $V_{\KIn{F}}(g)$, $g\in{\UPgr }$. 
\item[{\rm (ii)}] The net of local {\rm C*}-algebras given 
 by
\[
 \BR\ni {\cal O}\longmapsto {\cal A}_{\KIn{B}}({\cal O}) 
 :=\Wort{C}^*\LR \{W(\varphi)\mid  \varphi\in h_{\KIn{B}}({\cal O})\}\RR
 \subset \ccr{h_{\KIn{B}},\sigma}\,
\]
is a HK-net. Here $W(\cdot)$ denote the Weyl elements 
(generators) of $\ccr{h_{\KIn{B}},\sigma}$. The 
covariance of the net is given by the Bogoljubov 
automorphisms $\alpha_g$ associated to  $V_{\KIn{B}}(g)$, 
$g\in{\UPgr }$.
\end{itemize}
We call the nets of C*-algebras given in {\rm (i)} and {\rm (ii)} 
above {\em free nets}.
\end{teo}
\begin{beweis}
First note that from the $\Gamma$-invariance property (F1) we have that
$\Gamma\,h_{\KIn{F}}({\cal O})=h_{\KIn{F}}({\cal O})$, $\B\in\BR$, which 
implies ${\cal A}_{\KIn{F}}({\cal O})^*={\cal A}_{\KIn{F}}({\cal O})$
(as a set). The isotony of the local C*-algebras in the CAR and the CCR 
case follows immediately from the isotony property of the corresponding 
nets of linear submanifolds $\B\mapsto h_{\KIn{F/B}}({\cal O})$.

To prove additivity we will show first 
that for $\{\al O._\lambda\}_{\lambda\in\Lambda}\subset\BR$
as in Definition~\ref{HKnet}~(ii) we have for the nets of 
local linear submanifolds
\[
 h_{\KIn{F/B}}(\cup_\lambda\al O._\lambda)=
   \Wort{span}\,\{ h_{\KIn{F/B}}(\al O._\lambda)\mid \lambda\in\Lambda\} .
\]
Indeed, the inclusion `$\supseteq$' follows from the linearity of 
$\got I$ and the fact that if $\Wort{supp}\,f_{\lambda_l}\subset
\al O._{\lambda_l}$, $l=1,\ldots,L$, then 
$\Wort{supp}\LR \sum_l \mu_{\lambda_l}  f_{\lambda_l}\RR\subset
\cup_{\lambda_l}\al O._{\lambda_l}$ for $f_{\lambda_l}\in\al T._{\KIn{F/B}}$
and $\mu_{\lambda_l}\in\z{C}$. To show the converse inclusion
take $f\in\al T._{\KIn{F/B}}$ with $\Wort{supp}\,f\subset
\cup_\lambda\al O._\lambda$. By compactness there exists a finite
subcovering such that $\Wort{supp}\,f\subset\cup_{l=1}^L\al O._{\lambda_l}$
and using a subordinate smooth partition of unity 
(which exists since $\z{R}^4$ is paracompact) we can
write $f=f_{\lambda_1}+\ldots +f_{\lambda_L}$, 
where $f_{\lambda_l}\in\al T._{\KIn{F/B}}$
and $\Wort{supp}\,f_{\lambda_l}\subset\al O._{\lambda_l}$, $l=1,\ldots,L$.
Therefore 
\[
 {\got I}f={\got I}f_{\lambda_1}+\ldots +{\got I}f_{\lambda_L}\in
     \Wort{span}\,\{ h_{\KIn{F/B}}(\al O._\lambda)\mid \lambda\in\Lambda\} .
\]
Now additivity follows from the properties of the generators
of the CAR- and CCR-algebras, cf.~\cite[Section~8.3]{bBaumgaertel92}.

For the causality property take $\B_1$, $\B_2\in\BR$ with $\B_1\perp
\B_2$. Now in the CAR case define the sets ${\got P}_i$ of polynomials in
the generators $\ot a.(\varphi_i)$, $\varphi_i\in h_{\KIn{F}}({\cal O}_i)$,
$i=1,2$, such that the degree of the corresponding monomials is even.
From property (F2) and the CAR's we have that 
$\ot a.(\varphi_1)\,\ot a.(\varphi_2)+
 \ot a.(\varphi_2)\,\ot a.(\varphi_1)=0$ 
for all $\varphi_i\in h_{\KIn{F}}(\B_i)$, 
$i=1,2$, and therefore $[{\got P}_1,{\got P}_2]=0$. Now, since
${\got P}_i$ is dense in ${\cal A}_{\KIn{F}}({\cal O}_i)$, $i=1,2$,
we obtain that 
$[{\cal A}_{\KIn{F}}(\B_1),{\cal A}_{\KIn{F}}(\B_2)]=0$ in $\cal A$.
In the Bose case note that from (B2) and the Weyl relation we have
$W(\varphi_1)\,W(\varphi_2)=W(\varphi_2)\,
W(\varphi_1)$ for all 
$\varphi_i\in h_{\KIn{B}}({\cal O}_i)$ and since $\;\Wort{span}\,
\{W(\varphi_i)\mid \varphi_i\in h_{\KIn{F}}(\B_i)\}\;$ is dense in 
${\cal A}_{\KIn{B}}(\B_i)$, $i=1,2$, we also obtain in this case that 
$[{\cal A}_{\KIn{B}}(\B_1),{\cal A}_{\KIn{B}}(\B_2)]=0$ in $\cal A$.

Finally, to prove the covariance property denote by 
$\alpha_g^{\KIn{(F)}}$ resp.~$\alpha_g^{\KIn{(B)}}$ the Bogoljubov
automorphisms associated to $V_{\KIn{F}}(g)$ resp.~$V_{\KIn{B}}(g)$,
$g\in\UPgr$. Now by the support properties of 
$T_{\KIn{F}}(g)$ and $T_{\KIn{B}}(g)$ as well as by (F3) and (B2)
we have 
\begin{equation}\label{CovRS}
 V_{\KIn{F}}(g)\,h_{\KIn{F}}({\cal O})=h_{\KIn{F}}(g{\cal O})\quad
 \Wort{and}\quad 
 V_{\KIn{B}}(g)\,h_{\KIn{B}}({\cal O})=h_{\KIn{B}}(g{\cal O})\,.
\end{equation}
Now from the way the Bogoljubov automorphisms act on the corresponding 
generators of the CAR/CCR-algebras
it follows from the preceding equations that
\[
 \alpha^{\KIn{(F)}}_g (\al A._{\KIn{F}}(\B))=\al A._{\KIn{F}}(g\B)
  \quad\Wort{and}\quad
 \alpha^{\KIn{(B)}}_g (\al A._{\KIn{B}}(\B))=\al A._{\KIn{B}}(g\B)
  \,,\quad g\in\UPgr \,,
\]
which concludes the proof. (See further Theorem~3.6 in \cite{Lledo01} 
for a detailed proof of the covariance property.)
\end{beweis}

We will need later on the notion of isomorphic free nets explicitly.
The isomorphy can be transcribed in terms of the corresponding
reference spaces:

\begin{pro}
\label{IsoFreeNet}
\begin{itemize}
\item[{\rm (i)}]
Consider two tuples $(h_i,\,\Gamma_i,
 \,\langle\cdot,\cdot\rangle_i,\,V_i,\,{\cal T},\, T,\,{\got I}_i)$,
$i=1,2$, over the same test function space and
satisfying the properties of Definition~$\ref{Tuples}$~{\rm (i)}. 
Suppose that there exists a unitary linear mapping 
$\lambda_{\KIn{F}}\colon\ h_1\to h_2$ 
(i.e.~$\langle\lambda_{\KIn{F}}(\varphi),\lambda_{\KIn{F}}(\psi)\rangle_2
   =\langle\varphi, \psi\rangle_1$, $\varphi,\psi\in h_1$) satisfying
\[
  \lambda_{\KIn{F}}\, V_1(g) = V_2(g)\, \lambda_{\KIn{F}}\,,\;
  \;\; \lambda_{\KIn{F}}\, \Gamma_1(g) = \Gamma_2(g)\, \lambda_{\KIn{F}}
  \;\,,\,g\in\UPgr\,,
  \quad\Wort{and}\quad \lambda_{\KIn{F}}\ot I._1=\ot I._2\,.
\]
Then the corresponding fermionic free nets are isomorphic.
\item[{\rm (ii)}]
Consider two tuples $(h_i,\,\sigma_i,\,V_i,\,{\cal T},\, T,\,{\got I}_i)$,
$i=1,2$, over the same test function space and
satisfying the properties of Definition~$\ref{Tuples}$~{\rm (ii)}. 
Suppose that there exists a $($real$)$ linear symplectic bijection 
$\lambda_{\KIn{B}}\colon\ h_1\to h_2$ 
(i.e.~$\sigma_2\!\LR \lambda_{\KIn{B}}(\varphi),\lambda_{\KIn{B}}(\psi)\RR
   =\sigma_1(\varphi, \psi)$, $\varphi,\psi\in h_1$) satisfying
\be
\label{IsoMf}
  \lambda_{\KIn{B}}\, V_1(g) = V_2(g)\, \lambda_{\KIn{B}}\,,\;g\in\UPgr\,,
  \quad\Wort{and}\quad \lambda_{\KIn{B}}\ot I._1=\ot I._2\,.
\ee
Then the corresponding bosonic free nets are isomorphic.
\end{itemize}
\end{pro}
\begin{beweis}
The proof of (ii) is typical:
denote by $W(\varphi_i)$, $\varphi_i\in h_i$, 
the Weyl elements of the corresponding C*-algebras
$\ccr{h_i,\sigma_i}$, $i=1,2$. Then the mapping 
$\Lambda(W(\varphi_1)):=W(\lambda(\varphi_1))$, 
$\varphi_1\in h_1$, extends uniquely to an
isomorphism (also denoted by $\Lambda$) of the corresponding
CCR-algebras. Further, the equation $\lambda_{\KIn{B}}\ot I._1=\ot I._2$
implies $\lambda_{\KIn{B}}h_1(\al O.)=h_2(\al O.)$,
so that for the local C*-subalgebras we have $\Lambda(\al A._1(\B))
=\al A._2(\B)$, ${\cal O}\in \BR$. 
Finally, the intertwining property of $\lambda$ 
in Eq.~(\ref{IsoMf}) implies that 
$\Lambda\, \alpha_g^{\KIn{(1)}} = \alpha_g^{\KIn{(2)}}\,\Lambda$,
$g\in\UPgr$.
\end{beweis}

\begin{rem}
Part~(ii) of Definition~\ref{Tuples} contains some aspects of Segal's
notion of quantisation for bosonic systems
(cf.~\cite[p.~750]{SegalIn81},\cite[p.~106]{bBaez92}). With 
Definition~\ref{Tuples} and concretely through the requirements on the
embedding $\got I$ we incorporate to this program the axioms of 
local quantum physics.
Note nevertheless that since Haag-Kastler axioms are stated in terms of
abstract C*-algebras we do not require initially (in contrast with 
Segal's approach) that the abstract CCR-algebra is represented in 
any Hilbert space nor the specification of any regular state. 
(For further reasons on this last point see also 
 \cite{Grundling88c,Lledo00}.)
\end{rem}

We consider next also the spectrality condition, which
in the context of free nets can be stated in terms of the tuples
considered in Definition~\ref{Tuples}.

\begin{defi}
\label{Spectrality}
With the notation of Definition~$\ref{Tuples}$ we require respectively:
\begin{itemize}
\item[$(${\bf F4}$)$] There exists a basis projection 
$P$ on $h_{\KIn{F}}$ (i.e.~an orthoprojection satisfying 
$P+\Gamma P\Gamma=\EINS$) reducing the
representation $V_{\KIn{F}}$, i.e.~$P\,V_{\KIn{F}}(g)=V_{\KIn{F}}(g)\,P$,
$g\in\UPgr$, and such that the corresponding representation on $Ph_{\KIn{F}}$
is strongly continuous and satisfies the spectrality 
condition $($i.e.~the spectrum of the 
corresponding generators of the space-time translations is contained
in the forward light cone
$\overline{\cal V}_+)$.

\item[$(${\bf B3}$)$] There exists a  real scalar product $s$ on 
$h_{\KIn{B}}$ and an internal complexification $J$
satisfying $J^2=-\EINS$, $\sigma(\varphi,J\psi)=-\sigma(J\varphi,\psi)$,
$\sigma(\varphi,J\varphi)=s(\varphi,\varphi)$
and $|\sigma(\varphi,\psi)|^2\leq s(\varphi,\varphi)s(\psi,\psi)$,
$\varphi,\psi\in h_{\KIn{B}}$. W.r.t.~this complexification
$V_{\KIn{F}}$ is a strongly continuous unitary representation on
the one particle Hilbert space $(h_{\KIn{B}},k_{\KIn{$J$}})$ satisfying the 
spectrality condition. Here $k_{\KIn{$J$}}=s+i\sigma$ denotes the 
corresponding complex scalar product.
\end{itemize}
\end{defi}

\begin{rem}\label{SpecRem}
Recall first that the basis projection $P$ resp.~the complexification $J$
characterise Fock states of the CAR- resp.~CCR-algebras.
We will show in this remark that the preceding definition implies the 
existence of a covariant representation of the C*-dynamical systems
$(\car{h_{\KIn{F}},\Gamma}, \R^4,\alpha_{\KIn{$(\cdot)$}})$ and
$(\ccr{h_{\KIn{B}},\sigma}, \R^4,\alpha_{\KIn{$(\cdot)$}})$ 
satisfying the spectrality condition (cf.~\cite{BorchersIn87} and 
\cite[Teorems~A.4.2 and A.4.5]{Lledo95}). Compare also with the notion 
of covariant representations introduced in 
\cite{Polley81,Kraus81}.

\begin{itemize}
\item[(i)]
From (F4) and from standard results of the CAR theory \cite{ArakiIn87}
it can be shown that the Bogoljubov automorphisms $\alpha_g$, 
corresponding to $V_{\KIn{F}}(g)$, $g\in\UPgr$, are uniquely implemented by
unitary operators $Q_g$ on $\al F._a(Ph_{\KIn{F}})$ (the antisymmetric
Fock space over $Ph_{\KIn{F}}$) that leave the Fock vacuum 
$\Omega$ invariant. Now it is straightforward calculation to show
that on the set of finite particle vectors (which is dense in 
$\al F._a(Ph_{\KIn{F}})$) the following equations hold
for all $\varphi\in h_{\KIn{F}}$ and $g\in\UPgr$:
\[
 \pi_{P} (\alpha_{g}(\ot a.(\varphi)))=Q(PV_{\KIn{F}}(g))\,
 \pi_{P}(\ot a.(\varphi))\,Q(PV_{\KIn{F}}(g))^{-1}\quad\Wort{and}\quad
 Q(PV_{\KIn{F}}(g))\,\Omega=\Omega\,,
\]
where $\pi_{P}$ is the Fock representation characterised by $P$ and
$Q(PV_{\KIn{F}}(g))$ denotes the second quantisation of the corresponding 
subrepresentation on $\al F._a(Ph_{\KIn{F}})$. This implies that 
$Q_g=Q(PV_{\KIn{F}}(g))$ and since $PV_{\KIn{F}}(g)$ satisfies the 
spectrality condition on $Ph_{\KIn{F}}$, $Q(PV_{\KIn{F}}(g))$ will also 
satisfy it on $\al F._a(Ph_{\KIn{F}})$ \cite{ArakiIn87}.

\item[(ii)]
In the CCR-case we obtain from (B3) and from the definition of the 
generating functional 
$h_{\KIn{F}}\ni\varphi\to e^{-\frac14 s(\varphi,\varphi)}$ that the 
Bogoljubov automorphisms $\alpha_g$, corresponding to $V_{\KIn{B}}(g)$, 
$g\in\UPgr$, are uniquely implemented by
unitary operators $Q_g$ on $\al F._s(h_{\KIn{B}})$ (the symmetric
Fock space over the one particle Hilbert space $h_{\KIn{B}}$) 
that leave the Fock vacuum $\Omega$ invariant
(see \cite[Section~8.2]{bBaumgaertel92}).
But again a straightforward calculation shows
that on the set of coherent vectors (which is total in 
$\al F._s(h_{\KIn{B}})$ cf.~\cite[Chapter~2]{bGuichardet72}) 
the following equations hold for all $\varphi\in h_{\KIn{B}}$ and $g\in\UPgr$:
\[
 \pi_{J} (\alpha_{g}(W(\varphi)))=Q(V_{\KIn{B}}(g))\,
 \pi_{J}(W(\varphi))\,Q(V_{\KIn{B}}(g))^{-1}\quad\Wort{and}\quad
 Q(V_{\KIn{B}}(g))\,\Omega=\Omega\,,
\]
where $\pi_{J}$ is the Fock representation characterised by $J$ and
$Q(V_{\KIn{B}}(g))$ denotes the second quantisation of $V_{\KIn{B}}$ 
on $\al F._s(h_{\KIn{B}})$. This shows that $Q_g=Q(V_{\KIn{B}}(g))$.
Now by the property of Fock states 
(cf.~\cite[Section~8.2.7]{bBaumgaertel92}) that any positive operator 
on $(h,k_{\KIn{$J$}})$ has a positive second quantisation on 
$\al F._s(h_{\KIn{B}})$, we get finally that the spectrality condition 
of $V_{\KIn{B}}(g)$ on the one particle Hilbert space implies the 
spectrality condition for $Q(V_{\KIn{B}}(g))$ on $\al F._s(h_{\KIn{B}})$.
\end{itemize}
\end{rem}

\begin{rem}
\label{Rem.4.2.1}
The existence of the structures given in Definition~\ref{Tuples}~(i) or 
(ii) satisfying (F1)~-~(F4) or (B1)~-~(B3) (and therefore
the existence of free nets) is shown in the context of 
Minkowski space in \cite{Lledo01,Lledo95}. 
In this paper we construct free nets of 
local C*-algebras associated to massive 
(massless) systems with arbitrary spin (helicity) \cite{Wigner39}. 
The embedding, which is the central object of the free net construction, is
given for example in the massive case of \cite{Lledo95} as a direct sum of
those mappings that reduce the covariant \rep{} into the irreducible 
massive canonical \rep{} (cf.~also Remark~\ref{Rem.3.2.2}). 
In other words the embedding
selects from the algebraically reducible covariant \rep{} two
irreducible components.\footnote{These types of embeddings play 
also an important role in the (rigorous) context of quantised fields, 
defined mathematically as operator-valued distributions 
(see e.g.~\cite[Theorem~X.42]{bReedII} or \cite[Appendix~B]{bBaez92} 
in the example of the Klein-Gordon field).}
Summing up, we have transcribed Haag-Kastler's axioms in terms of the
embeddings $\got I$ and given a neat group theoretical interpretation 
of it in the context of the Poincar\'e group. 
Note finally that the free net construction avoids (in the spirit of
local quantum physics) any explicit use of the notion of quantum field.
\end{rem}

We finish this section adapting Lemma~A.1.4 in \cite{Lledo95} to the
present massless case. This result will be essential 
for proving the causality property of the following models of free nets.
\begin{lem}
\label{Lem.4.2.2}
Let $x\in\z{R}^4$ be a spacelike vector and $\beta_n(\cdot)$ a 
matrix-valued function on $\Lkf$ such that at each point $p\in\Lkf$ the
matrix elements of $\beta_n(p)$ are homogeneous polynomials of degree n 
in $p_\mu$, $\mu=0,1,2,3$. Then we have
\begin{eqnarray}
    \label{BoseGleichung}
    \int \limits_{\Lkf} \left( e^{ipx}-e^{-ipx} \right) \beta_{n}(p)\;
    \mudp &=&0, \qquad n \mbox{ even,}                                     \\
    \label{FermiGleichung}
    \int \limits_{\Lkf} \left( e^{ipx}+e^{-ipx} \right) \beta_{n}(p)\;
    \mudp &=&0, \qquad n \mbox{ odd.}
    \end{eqnarray}
\end{lem}
\begin{beweis}
It is well-known that for $x^2<0$ 
the Pauli-Jordan function
  \begin{equation}
    \Delta(x) = \int\limits_{\Lkf} e^{-ipx} \mudp
  \end{equation}
    is an even C$^\infty$ function, i.e.~$\Delta(x) = \Delta(-x)$ (see
    \cite[pgs.~71~and~107]{bReedII}). Let $\alpha$ be a multi-index and
    $|\alpha|:=\alpha_0+\alpha_1+\alpha_2+\alpha_3$. Then
\[
  \frac{\partial^{|\alpha|}\Delta}{\partial x^\alpha}\bigg|_x
    = (-i)^{|\alpha|} \int\limits_{\Lkf}
      p_\alpha\kern.2em e^{-ipx} \mudp 
      \kern1em\mbox{is an}\kern.5em
      \left\{ \begin{array}{l} 
        \mbox{even function, if $|\alpha|$ is even.}  \\[3mm] 
        \mbox{odd function, if $|\alpha|$ is odd.}
      \end{array}\right.
\]
But from hypothesis the matrix elements
of $\beta_n(p)$ are homogeneous polynomials in $p_\mu$ of degree
$n$ and therefore the last expression implies
Eqs.~(\ref{BoseGleichung}) and (\ref{FermiGleichung}). 
\end{beweis}

We will denote the objects of the constructions in the following
Sections with the subindex $\got W$ or ${\got F}$ 
depending if the net is associated to the Weyl Equation or to the
${\got F}$-Equation, respectively.
All the mentioned models will have a particular 
function $\beta(\cdot)$ (recall the preceding lemma) that characterises 
the corresponding scalar products and symplectic forms.

\subsection{Weyl net}\label{WN}
The following construction will illustrate the fermionic 
axioms (F1) - (F4) making explicit use of the Weyl
equation. Further it will provide the simplest nontrivial
example where certain intertwiner spaces are explicitly introduced in 
order to define the corresponding embeddings that satisfy
the conditions already stated in Subsection~\ref{FBRED}. 
Indeed, making
use of the notation and results of Subsection~\ref{TheWeyl} 
and of the particular structure of the intertwiner space 
associated to the finite-dimensional representations 
$D^{\KIn{$(\frac{1}{2},0)$}}$ and $D^{\KIn{$(0,\frac{1}{2})$}}$ 
restricted to the massless little group ${\cal E}(2)$,
we will construct the free net associated to the Weyl equation. 
The free net resulting from this construction is isomorphic
to the one given in \cite{Lledo01} for $n=1$ 
(cf.~Remark~\ref{GenF}~(i)). 

We will see later on in the section that this construction procedure can
be easily adapted to the bosonic case. Recall that given 
two representations $V,V'$ of a group $\cal G$ on finite dimensional
Hilbert spaces $\al H.,\al H.'$ the corresponding intertwiner space
is defined as
\[
 (V(\al G.),V'(\al G.)):= \{\Psi\colon\ \al H.\to\al H.'\mid
 \psi~\Wort{is~linear~and~}\Psi\,V(g)=V'(g)\,\Psi\,,\; g\in\al G. \}\,.
\]

\begin{lem}
\label{Lem.4.3.0}
With the notion above we compute the following intertwiner spaces:
\begin{eqnarray*}
\LR D^{\KIn{$(\frac{1}{2},0)$}}({\cal E}(2)),\; 
    D^{\KIn{$(0,\frac{1}{2})$}}({\cal E}(2)) \RR 
  &=& \LG  \left(\kern-1.5mm\begin{array}{cc}
                           0 \kern-2mm & s \\
                           0 \kern-2mm & 0
                  \end{array}\kern-1.5mm\right)\mid  s\in\z{R}  \RG ,  \\
\LR D^{\KIn{$(\frac{1}{2},0)$}}({\cal E}(2)),\; 
     D^{\KIn{$(\frac{1}{2},0)$}}({\cal E}(2)) \RR
  &=& \z{C}\EINS \,.
\end{eqnarray*}
\end{lem}
\begin{beweis}
The first intertwiner space consists of 
all $M\in\Wort{Mat}_2(\z{R})$ such that $M\,L=\overline{L}\,M$ for 
all $L\in {\cal E}(2)$. It is now immediate to check that 
$M=\left(\kern-1.5mm\begin{array}{cc}
                           0 \kern-2mm & s \\
                           0 \kern-2mm & 0
   \end{array}\kern-1.5mm\right)$, $s\in\z{R}$. 
The triviality of the second intertwiner space is a 
straightforward computation.
\end{beweis}

\subsubsection{CAR-algebra}
First recall the definitions associated to the Weyl Equation given in
Subsection~\ref{TheWeyl} and the form of the $H_p$-matrices, $p\in\Lk$,
given in Eq.~(\ref{SolHp}).  
We consider the complex Hilbert space
$h_{\KIn{$\got W$}}:= h_+\oplus h_-\oplus h_+\oplus h_-$ with the 
scalar product given by $\langle\cdot,\cdot\rangle_{\KIn{$\got W$}}:=
\langle\cdot,\cdot\rangle_+ \oplus \langle\cdot,\cdot\rangle_-
\oplus\langle\cdot,\cdot\rangle_+ \oplus \langle\cdot,\cdot\rangle_-$. 
To define the anti-linear involution on 
$h_{\KIn{$\got W$}}$ consider first the mapping
$\Gamma_1\colon\  h_+ \longrightarrow  h_-$ given by
\[
 \LR\Gamma_1\,\varphi_+ \RR\!(p):=\overline{H_p}\;\Gamma_0\;H^{-1}_p\,
        \varphi_+(p)=\overline{H_p}\;\Gamma_0
          \left(\kern-1.5mm\begin{array}{c}
                 \chi_+(p) \\ 0 
             \end{array}\kern-1.5mm\right),
\]
where $\varphi_+\in h_+$ and $\Gamma_0\colon\al H.^{\KIn{$(\frac{1}{2},0)$}}
\to \al H.^{\KIn{$(0,\frac{1}{2})$}}$ is an
anti-unitary involution (conjugation). It can be easily shown that 
$\Gamma_1$ is anti-linear and that it satisfies the equation 
$\askp{\Gamma_1\,\varphi_+}{\Gamma_1\,\psi_+}_-=
\askp{\psi_+}{\varphi_+}_+$ for all $\varphi_+$, $\psi_+
\in  h_+$.

Finally, define in terms of $\Gamma_1$
the anti-unitary involution on $h_{\KIn{$\got W$}}$ as
\[
 \Gamma_{\KIn{$\got W$}}\LR\varphi_+\oplus\varphi_-
         \oplus\psi_+\oplus\psi_-\RR:=
 \Gamma_1^{-1}\,\psi_-\oplus \Gamma_1\,\psi_+\oplus
 \Gamma_1^{-1}\,\varphi_-\oplus\Gamma_1\,\varphi_+. 
\]
$\Gamma_{\KIn{$\got W$}}$ is anti-linear and it can easily be 
checked that, 
\[
 \Gamma_{\KIn{$\got W$}}^2=\EINS \qquad\Wort{and}\qquad
 \askp{\Gamma_{\KIn{$\got W$}}\,\varphi^{(1)}}{\Gamma_{\KIn{$\got W$}}\,
 \varphi^{(2)}}_{\KIn{$\got W$}}=\askp{\varphi^{(2)}}{\varphi^{(1)}}_{ 
 \KIn{$\got W$}}\, , 
\]
$\varphi^{(i)}\in  h_{\KIn{$\got W$}}$, $i=1,2$. 
The C*-algebra $\car{h_{\KIn{$\got W$}},
\Gamma_{\KIn{$\got W$}}}$ is therefore uniquely given.

\subsubsection{Existence theorem for the local algebras}

We consider here on the test function space
spaces $\al T._{\KIn{$\got W$}}:=
\Test{\z{R}^{4},{\cal H}^{\KIn{$(0,\frac{1}{2})$}}}$, 
$h_+$ and $h_-$ the following covariant and canonical representations of 
$\UPgr=\semi{\g{SL}{2,{\z{C}}}}{\z{R}^{4}}\ni g=(A,a)$: for 
$f\in\al T._{\KIn{$\got W$}}$, $\varphi_\pm\in h_\pm$
\[
 \LR T_{\KIn{$\got W$}}(g)\kern.2emf \RR\! (x)
   := \overline{A}\, f\!\left( \Lambda_{A}^{-1}(x-a) \right),\\[-.35cm]
\]
\begin{eqnarray*}
( V_1(g)\kern.2em\varphi_+)(p)
&:=& e^{-ipa}\ A\, \varphi_+( \Lambda_{A}^{-1}p ),\qquad
( V_3(g)\kern.2em\varphi_+ )(p)
 \kern3mm:=\kern3mm e^{ipa}\ A\,\varphi_+(\Lambda_{A}^{-1}p),\\
( V_2(g)\kern.2em\varphi_-)(p)
&:=& e^{-ipa}\ \overline{A}\,\varphi_-( \Lambda_{A}^{-1}p ),\qquad
 (V_4(g)\kern.2em\varphi_-)(p)
 \kern3mm:=\kern3mm e^{ipa}\ \overline{A}\,\varphi_-( 
                                              \Lambda_{A}^{-1}p ).
\end{eqnarray*}
Note that the covariant \rep{} $T_{\KIn{$\got W$}}$, satisfies the support
property mentioned in part (i) of Definition~\ref{Tuples}.
We consider next the following reducible representation of $\UPgr$ over
$h_{\KIn{$\got W$}}$:
\[
V_{\KIn{$\got W$}}:=V_1\oplus V_2\oplus V_3\oplus V_4
\] 

\begin{lem}
\label{Lem.4.3.1}
The equation $\Gamma_{\KIn{$\got W$}}\, V_{\KIn{$\got W$}}(g)=
V_{\KIn{$\got W$}}(g)\,\Gamma_{\KIn{$\got W$}}$ holds for all $g\in\UPgr$.
\end{lem}
\begin{beweis}
The equation is based on the following intertwining properties of
$\Gamma_1$: 
\begin{Klammer}
  \Gamma_1\, V_1(g) & = & V_4(g)\,\Gamma_1   \\[1mm]
  \Gamma_1\, V_3(g) & = & V_2(g)\,\Gamma_1 \, ,
\end{Klammer}
which are a direct consequence of the definitions.
\end{beweis}

Further we consider the embeddings ${\got I}_{1,3}\colon\ 
\al T._{\KIn{$\got W$}}\longrightarrow  h_+$ and 
${\got I}_{2,4}\colon\ \al T._{\KIn{$\got W$}}\longrightarrow  h_-$ 
defined for all $f \in \al T._{\KIn{$\got W$}}$ by,
\begin{eqnarray*}
   \LR{\got I}_1 f\RR\!(p)   :=  H_p \left(\kern-1.5mm\begin{array}{cc}
                 0 \kern-3mm & 1 \\
                 0 \kern-3mm & 0
  \end{array}\kern-1.5mm\right) \overline{H_p}^{-1}\,\widehat{f}(p)  \;, 
                              & &
   \LR{\got I}_3 f\RR\!(p)   :=  H_p \left(\kern-1.5mm\begin{array}{cc}
                 0 \kern-3mm & 1 \\
                 0 \kern-3mm & 0
  \end{array}\kern-1.5mm\right) \overline{H_p}^{-1}\,\widehat{f}(-p) \\[1mm] 
  \LR{\got I}_2 f\RR\!(p)   :=  \overline{H_p} 
            \left(\kern-1.5mm\begin{array}{cc}
                 0 \kern-3mm & 1 \\
                 0 \kern-3mm & 0
  \end{array}\kern-1.5mm\right) H_p^{-1}\,\widehat{\Gamma_0f}(p) \;,
                              & &
   \LR{\got I}_4 f\RR\!(p)   :=  \overline{H_p} 
            \left(\kern-1.5mm\begin{array}{cc}
                 0 \kern-3mm & 1 \\
                 0 \kern-3mm & 0
  \end{array}\kern-1.5mm\right) H_p^{-1}\,\widehat{\Gamma_0 f}(-p),
\end{eqnarray*}
where the `hat' $\widehat{f}$ means the 
Fourier transformation and $p$ is
restricted to $\Lk$. Note that since
$\LR D^{\KIn{$(0,\frac{1}{2})$}}({\cal E}(2)),\; 
D^{\KIn{$(\frac{1}{2},0)$}}({\cal E}(2)) \RR\ni
\left(\kern-1.5mm\begin{array}{cc}
                 0 \kern-3mm & 1 \\
                 0 \kern-3mm & 0
         \end{array}\kern-1.5mm\right)\in 
\LR D^{\KIn{$(\frac{1}{2},0)$}}({\cal E}(2)),\; 
D^{\KIn{$(0,\frac{1}{2})$}}({\cal E}(2)) \RR$ the above definitions
are consistent. 

Finally, the embedding that specifies the net structure is given by
\begin{equation}\label{IW}
 {\got I}_{\KIn{$\got W$}}\colon\ \al T._{\KIn{$\got W$}}
 \longrightarrow h_{\KIn{$\got W$}},\quad
 {\got I}_{\KIn{$\got W$}} f:= 
 {\got I}_1f\oplus {\got I}_2  f\oplus {\got I}_3f 
 \oplus {\got I}_4 f\,.
\end{equation}
\begin{rem}
Note for instance that
\[
  H_p \left(\kern-1.5mm\begin{array}{cc}
                 0 \kern-3mm & 1 \\
                 0 \kern-3mm & 0
  \end{array}\kern-1.5mm\right) \overline{H_p}^{-1}
= \frac12 \left(\kern-1.5mm \begin{array}{cc}
    -(p_{1}-ip_{2})  &  p_{0}+p_{3}\\
   -(p_{0}-p_{3})  & p_{1}+ip_{2}
    \end{array} \kern-1.5mm\right)\,.
\]
Therefore, the matrix elements of the previous expression correspond
on position space to differential operators and $\ot I._{\KIn{$\got W$}}$  
will not change the localisation properties of $f$. Further, the
embeddings (say $\ot I._1$) that specify 
${\got I}_{\KIn{$\got W$}}$ can be written in components as 

\[
 (\ot I._1f)^{\KIn{$C$}}(p)=(H_p)^{\KIn{$C$}}_{\KIn{$B$}}\;
        Q^{\KIn{$BC'$}}\;
    \varepsilon_{\KIn{$C'B'$}}\;
    (\overline{H_p}^ {-1})^{\KIn{$B'$}}_{\KIn{$E'$}}
    \,\widehat{f^{\KIn{$E'$}}}(p)
    \quad (\mr sum~over~repeated~indices.) \,,
\]
where $(\varepsilon_{\KIn{$C'B'$}}):=
    \left(\kern-1.5mm \begin{array}{rc}
        0  &\kern-2mm  1\\
        -1 &\kern-2mm  0
    \end{array} \kern-1.5mm\right)$
and
$(Q^{\KIn{$BC'$}}):=\left(\kern-1.5mm \begin{array}{cc}
        1  & \kern-2mm 0\\
        0  & \kern-2mm  0
    \end{array} \kern-1.5mm\right)$    
is the matrix corresponding to the point $\frac12 (1,0,0,1)$ in the 
positive light cone (recall Subsection~\ref{Subs.2.4}). Moreover,
$Q$ can be seen as the 1-dimensional projection characterising the 
Weyl equation (see Subsection~\ref{TheWeyl}).
\end{rem}

The covariance property of the net characterised by the preceding
embedding is guaranteed by the following result

\begin{lem}
\label{Lem.4.3.2}
The equation, ${\got I}_{\KIn{$\got W$}}\, T_{\KIn{$\got W$}}(g)=
V_{\KIn{$\got W$}}(g)\,{\got I}_{\KIn{$\got W$}}$, holds for all $g\in\UPgr$.
\end{lem}
\begin{beweis}
First recall that for $g=(A,a)\in \semi{\g{SL}{2,{\z{C}}}}{\z{R}^{4}}$,
$p\in\Lk$ and $q:=\Lambda^{-1}_A p$ the matrix
$H^{-1}_p A H_q\in\al E.(2)$. Thus by Lemma~\ref{Lem.4.3.0} we have
\[
 H^{-1}_p A H_q\left(\kern-1.5mm\begin{array}{cc}
                                   0 \kern-3mm & 1 \\
                                   0 \kern-3mm & 0
                                \end{array}\kern-1.5mm\right) 
    =  \left(\kern-1.5mm\begin{array}{cc}
             0 \kern-3mm & 1 \\
             0 \kern-3mm & 0
         \end{array}\kern-1.5mm\right)\overline{H^{-1}_p A H_q}
         \kern3mm = \kern3mm e^{\frac{i}{2}\theta(A,a)}
                    \left(\kern-1.5mm\begin{array}{cc}
                   0 \kern-3mm & 1          \\
                   0 \kern-3mm & 0
              \end{array}\kern-1.5mm\right)\,.
\]
From this, the relations 
\begin{Klammer}
\label{IntertwiningEq}
  {\got I}_{1,3}\; T_{\KIn{$\got W$}}(g) 
             & = & V_{1,3}(g)\; {\got I}_{1,3}   \\[1mm]
  {\got I}_{2,4}\; \Gamma_0 T_{\KIn{$\got W$}}(g) 
             & = & V_{2,4}(g)\; {\got I}_{2,4}\Gamma_0 \; ,
\end{Klammer}
can be easily shown and the intertwining equation of the statement is
proved.
\end{beweis}

The next result will ensure causality 
for the net characterised by the embedding 
${\got I}_{\KIn{$\got W$}}$.
\begin{lem}
\label{Lem.4.3.3}
If $\Wort{supp}\,f \perp\Wort{supp}\, k$ for 
$f,k \in\al T._{\KIn{$\got W$}}$, then the equation 
$\askp{{\got I}_{\KIn{$\got W$}}f}{{\got I}_{
\KIn{$\got W$}}k}_{\KIn{$\got W$}}=0$ holds.
\end{lem}
\begin{beweis}
First put $\widetilde{\beta_{+}}(p):=(H_p^{-1})^* H_p^{-1}$ and 
$\widetilde{\beta_{-}}(p)= \Gamma_0\widetilde{\beta_{+}}(p)
\Gamma_0$, $p\in\Lk$ (see also Remark~\ref{Rem.3.3.3}).
Then we compute
\begin{eqnarray*}
\lefteqn{\askp{{\got I}_{\KIn{$\got W$}}f}{{\got I}_{\KIn{$\got W$}} 
 k}_{\KIn{$\got W$}}}                                               \\[3mm] 
 &=& \int\limits_{\Lkf}\! \askp{({\got I}_1 f)(p)}
        {\widetilde{\beta_{+}}(p)\; ({\got I}_1 k)(p)}_{\z{C}^2}\!\mudp \,+\,
     \int\limits_{\Lkf}\! \askp{({\got I}_2 \Gamma_0 f)(p)}
   {\widetilde{\beta_{-}}(p)\; ({\got I}_2 \Gamma_0k)(p)}_{\z{C}^2}\!\mudp \\
 & &  \kern-1em+
     \int\limits_{\Lkf}\! \askp{({\got I}_3 f)(p)}
    {\widetilde{\beta_{+}}(p)\; ({\got I}_3 k)(p)}_{\z{C}^2}\!\mudp\,+\,
     \int\limits_{\Lkf}\! \askp{({\got I}_4 \Gamma_0 f)(p)}
{\widetilde{\beta_{-}}(p)\;({\got I}_4\Gamma_0 k)(p)}_{\z{C}^2}\!\mudp\\[2mm]
&=& \int\limits_{\Lkf} \askp{{\widehat f}(p)}
     {\beta_{+}(p)\; {\widehat k}(p)}_{\z{C}^2} \mudp  \;+\;
        \int\limits_{\Lkf} \askp{\widehat{\Gamma_0 f}(p)}
     {\beta_{-}(p)\; \widehat{\Gamma_0k}(p)}_{\z{C}^2} \mudp  \\
 & &  \kern-1em+
        \int\limits_{\Lkf} \askp{{\widehat f}(-p)}
    {\beta_{+}(p)\; {\widehat k}(-p)}_{\z{C}^2} \mudp    \;+\;
        \int\limits_{\Lkf} \askp{\widehat{\Gamma_0 f}(-p)}
    {\beta_{-}(p)\; \widehat{\Gamma_0k}(-p)}_{\z{C}^2} \mudp   \\[3mm]    
&=& \int\limits_{\Lkf} \askp{{\widehat f}(p)}
     {\beta_{+}(p)\; {\widehat k}(p)}_{\z{C}^2} \mudp  \;+\;
        \int\limits_{\Lkf} \askp{{\widehat k}(-p)}
     {\beta_{+}(p)\; {\widehat f}(-p)}_{\z{C}^2} \mudp  \\
 & &  \kern-1em+
        \int\limits_{\Lkf} \askp{{\widehat f}(-p)}
    {\beta_{+}(p)\; {\widehat k}(-p)}_{\z{C}^2} \mudp    \;+\;
        \int\limits_{\Lkf} \askp{{\widehat k}(p)}
    {\beta_{+}(p)\; {\widehat f}(p)}_{\z{C}^2} \mudp   \\[3mm]   
&=& 0\; ,
\end{eqnarray*}
where $\beta_+(p):=\LR \overline{H_p}^{-1}\RR^*\! 
          \left(\kern-1.5mm\begin{array}{cc}
              0 \kern-1.6mm & 0 \\ 
              0 \kern-1.6mm & 1
          \end{array} \kern-1.5mm\right)
               \overline{H_p}^{-1}=\overline{P^{\KIn{$\dagger$}}}$ 
and $\beta_-(p):=\Gamma_0\,\beta_+(p)\,\Gamma_0$
(recall Remark~\ref{Rem.3.3.4}).
The last equation follows from Lemma~\ref{Lem.4.2.2} and
the fact that the matrix elements of 
$P^{\KIn{$\dagger$}}$ are homogeneous polynomials of degree 1.
\end{beweis}

We can now prove the existence of a free net associated to the Weyl
Equation, which we call {\em Weyl net} for short.
\begin{teo}
\label{Teo.4.3.4}
Consider the net of local linear submanifolds of $h_{\KIn{$\got W$}}$ 
given for $\al O.\in\BR$ by
\[
 \z{R}^4\supset{\cal O}\longmapsto h_{\KIn{$\got W$}}({\cal O})
  :=\LG {\got I}_{\KIn{$\got W$}}f\mid  
    f\in \TestO{\z{R}^{4}, \al H.^{\KIn{$(0,\frac{1}{2})$}}},\;
    \Wort{supp}f\subset {\cal O} \RG .
\]
Then the net of local C*-algebras defined by
\[
 \z{R}^4\supset{\cal O}\longmapsto {\cal A}_{\KIn{$\got W$}}({\cal O}):=
 \Wort{C}^*\LR\LG \ot a.(\varphi)\mid  \varphi\in 
 h_{\KIn{$\got W$}}({\cal O})\RG\RR^{\z{Z}_2},
\]
where the $\ot a.(\cdot)$'s denote the generators of the C*-algebra
$\car{h_{\KIn{$\got W$}},\Gamma_{\KIn{$\got W$}}}$,
is a HK-net.
\end{teo}
\begin{beweis}
First note that the local linear submanifolds satisfy the 
$\Gamma_{\KIn{$\got W$}}$-invariance property (F1) in
Definition~\ref{Tuples}. Indeed, from the relations
$\Gamma_1({\got I}_1f)={\got I}_4(\Gamma_0f)$ and 
$\Gamma_1({\got I}_3f)={\got I}_2(\Gamma_0f)$, 
$f\in \al. T_{\KIn{$\got W$}}$, it follows that $\Gamma_{\KIn{$\got W$}}\,
{\got I}_{\KIn{$\got W$}}f={\got I}_{\KIn{$\got W$}}f$ (which for 
the generators implies $\ot a.({\got I}_{\KIn{$\got W$}}f)^*
=\ot a.({\got I}_{\KIn{$\got W$}}f)$).

Now from Lemmas~\ref{Lem.4.3.1}, \ref{Lem.4.3.2} and 
\ref{Lem.4.3.3} we have that 
$\LR h_{\KIn{$\got W$}},\, \langle\cdot,\cdot\rangle_{\KIn{$\got W$}},\, 
 \Gamma_{\KIn{$\got W$}},\, V_{\KIn{$\got W$}},\,{\cal T}_{\KIn{$\got W$}},
 \, T_{\KIn{$\got W$}},\, {\got I}_{\KIn{$\got W$}}\RR $
satisfies all conditions stated in Definition~\ref{Tuples}~(i) and by 
Theorem~\ref{TeoFreeNet}~(i) we get that the net of local C*-algebras
above is a HK-net.
\end{beweis}

\begin{rem}
\label{GenF}
\begin{itemize}
\item[(i)]
We will show next that the net constructed 
above is isomorphic to the fermionic net defined in 
\cite[Section~3, case $n=1$]{Lledo01}. Using the notation
and results of the latter reference we specify the unitary
$\lambda_{\KIn{${\got W}$}}
 \colon\ h_{\KIn{${\got W}$}}\rightarrow h_1$
(recall Proposition~\ref{IsoFreeNet}~(i)): for
$\chi_\pm$,$\omega_\pm\in\Lzwei{\Lkf,\z{C},\mudp}$ put
\beaO
  \lefteqn{\lambda_{\KIn{F}}\!\LR 
      H_{(\cdot)} \left(\kern-2mm\begin{array}{c}  
                           \chi_+(\cdot) \\ 0 
                  \end{array} \kern-2mm \right)
                \oplus 
      \overline{H_{(\cdot)}} \left(\kern-2mm\begin{array}{c}  
                         \chi_-(\cdot)   \\ 0 
                  \end{array} \kern-2mm \right)
                 \oplus
      H_{(\cdot)} \left(\kern-2mm\begin{array}{c}  
                         \omega_+(\cdot)   \\ 0 
                  \end{array} \kern-2mm \right)
                \oplus 
      \overline{H_{(\cdot)}} \left(\kern-2mm\begin{array}{c}  
                          \omega_-(\cdot)  \\ 0 
                  \end{array} \kern-2mm \right)
           \RR\!(p)} \hspace{4cm}\\
  &:=& \LE \overline{H_p}
       \left(\kern-2mm\begin{array}{c}  
                           0 \\ \chi_+(p)  
       \end{array} \kern-2mm \right)  \RE_+ 
               \oplus 
       \LE H_p
       \left(\kern-2mm\begin{array}{c}  
                           0 \\ \chi_-(p) 
       \end{array} \kern-2mm \right)  \RE_- \\
  & &     \!\!\!\!  \oplus 
       \LE \overline{H_p}
       \left(\kern-2mm\begin{array}{c}  
                           0 \\ \omega_+(p)
       \end{array} \kern-2mm \right) \RE_+ 
               \oplus 
       \LE H_p
       \left(\kern-2mm\begin{array}{c}  
                           0 \\ \omega_-(p) 
       \end{array} \kern-2mm \right)  \RE_- 
\eeaO
where $[\cdot]_\pm$ denote the classes of the factor spaces ${\got H}'_\pm$
defined in \cite[Section~3]{Lledo01}. Using the statements in the proof
of \cite[Lemma~3.2]{Lledo01} it is straightforward to show that 
$\lambda_{\KIn{F}}$ satisfies the properties required in 
Proposition~\ref{IsoFreeNet}~(i).

\item[(ii)] From the construction given in Remark~\ref{SpecRem}
it can be easily shown that 
$P:=
           \left(\kern-1.5mm \begin{array}{cccc}
 \EINS \kern-1mm & 0 \kern-1mm & 0 \kern-1mm & 0             \\
   0 \kern-1mm & \EINS \kern-1mm & 0 \kern-1mm & 0             \\ 
   0 \kern-1mm & 0 \kern-1mm & 0 \kern-1mm & 0             \\
   0 \kern-1mm & 0 \kern-1mm & 0 \kern-1mm & 0                        
      \end{array} \kern-1.5mm\right) $
is a basis projection on $h_{\KIn{$\got W$}}$ 
(i.e.~$P+\Gamma_{\KIn{$\got W$}}\,P\,\Gamma_{\KIn{$\got W$}}=\EINS$) 
that characterises a Fock
state on $\car{h_{\KIn{$\got W$}},\Gamma_{\KIn{$\got W$}}}$
satisfying the spectrality condition (recall 
Definition~\ref{Spectrality}~(F4)). Note that $V_1(a)\oplus V_2(a)$,
$a\in\z{R}^4$, satisfies the spectrality condition on the one particle
Hilbert space $P\,h_{\KIn{$\got W$}}= h_+\oplus h_-$.

\item[(iii)] It is also straightforward to generalise the present 
 construction to higher (half-integer) helicity values, just
 replacing in the preceding construction 
 the indeces $\In{$(\frac12,0)$}$ by $\In{$(\frac{n}{2},0)$}$
 and $\In{$(0,\frac12)$}$ by $\In{$(0,\frac{n}{2})$}$ with
 $n\geq 3$ and odd. Adapting part (i) above we get the isomorphy 
 to the corresponding nets in \cite{Lledo01}. 
\end{itemize}
\end{rem}

From the isomorphy given in (i) of the previous remark we can assume the
structural results of \cite[Section~5]{Lledo01} (see also \cite{Brunetti93}). 
For example we have:
\begin{cor}
The net of von Neumann algebras $\al O.\mapsto\al M._{\KIn{$\got W$}}(\al O.)$
obtained from the Weyl net using the canonical
Fock space given in Remark~\ref{GenF}~(ii)
transforms in addition covariantly w.r.t.~the (fourthfold covering)
of the conformal group. Moreover it satisfies essential duality 
as well as timelike duality for the forward/backward cones.
\end{cor}

\begin{rem}
The formulas for the graph of the modular operator and the modular 
conjugation associated to double cones given for fermionic models in 
\cite[Theorem~5.10]{Lledo02a} can be also applied to the present 
construction.
\end{rem}

\subsection{$\ot F.$-net}
\label{FNet}
The construction bellow will illustrate the bosonic axioms
(B1) - (B3) of Section~\ref{AxiomsFreeNets}, making now
use of the ${\got F}$-Equation (\ref{FEquation})
(recall also the definitions and results in Subsection~\ref{TheF}).
As in the Weyl case the following computation of intertwiner spaces
will be essential for the construction of the corresponding embedding.
The proof of the following result is simmilar as in 
Lemma~\ref{Lem.4.3.0}.

\begin{lem}
\label{Lem.4.4.0}
Recalling the notion of intertwiner space in Subsection~\ref{WN} we have: 
\begin{eqnarray*}
\LR D^{\KIn{$(1,0)$}}({\cal E}(2)),\; D^{\KIn{$(0,1)$}}({\cal E}(2)) \RR
   &=& \LG s\, 
\left(\kern-2mm\begin{array}{cc}  
                           0 \kern-2mm & 1\\ 0 \kern-2mm & 0
                     \end{array} \kern-2mm \right) 
              \!\!\otimes\!\! \left(\kern-2mm\begin{array}{cc}  
                           0 \kern-2mm & 1\\ 0 \kern-2mm & 0
                     \end{array} \kern-2mm \right)\mid  s\in\z{R}\RG , \\
\LR D^{\KIn{$(1,0)$}}({\cal E}(2)),\; D^{\KIn{$(1,0)$}}({\cal E}(2)) \RR
  &=& \z{C}\EINS \;.
\end{eqnarray*}
\end{lem}

Next, consider the space 
\[
 h_{\KIn{${\got F}$}}:=h_+\oplus h_-,
\]
as a real space with nondegenerate symplectic form given by
\[
 \sigma_{\KIn{${\got F}$}}(\varphi,\psi):=
   \Wort{Im}\askp{\varphi}{\psi}_{\KIn{${\got F}$}}
 =\frac{1}{2i}\LR \askp{\varphi}{\psi}_{\KIn{${\got F}$}}-
  \askp{\psi}{\varphi}_{\KIn{${\got F}$}}\RR,
\]
where $\langle\cdot,\cdot\rangle_{\KIn{${\got F}$}}:=
\langle\cdot,\cdot\rangle_+\oplus\,
\langle\cdot,\cdot\rangle_-$ and $\varphi,\psi\in h_{\KIn{${\got F}$}}$. 
The C*-algebra
$\ccr{h_{\KIn{${\got F}$}},\sigma_{\KIn{${\got F}$}}}$ 
is simple and uniquely given by \cite{Manuceau73}.

The reducible \rep{}
\[
 V_{\KIn{${\got F}$}}:=V_+\oplus V_-,
\]
where for $g=(A,a)\in\semi{\g{SL}{2,{\z{C}}}}{\z{R}^{4}}$,
$\varphi\in h_+$ and $\psi\in h_-$ we define
\[
 \LR V_+(g)\kern.2em\varphi\RR\!(p)
  := e^{-ipa}\ D^{\KIn{$(1,0)$}}(A)\,\varphi
      (\Lambda_{A}^{-1}p)\; {\rm and}\;
 \LR V_-(g)\kern.2em\psi \RR\!(p)
  := e^{-ipa}\ D^{\KIn{$(0,1)$}}(A)\, \psi
      ( \Lambda_{A}^{-1}p ),
\]
leaves the real-bilinear form $\langle\cdot,\cdot\rangle_{\KIn{${\got F}$}}$
invariant and, therefore, the symplectic form $\sigma_{\KIn{${\got F}$}}$ 
is also $V_{\KIn{${\got F}$}}$-invariant.
In the rest of the section we will also write the finite-dimensional \rep{}
$D^{\KIn{$(1,0)$}}(A)$ simply as $D(A)$ and $D^{\KIn{$(0,1)$}}(A)$ as 
$D(\overline{A})$, $A\in\g{SL}{2,{\z{C}}}$.

Define also the covariant \rep{} for the present model (which satisfies
the support properties required in Definition~\ref{Tuples}~(i)):
\[
   ( T_{\KIn{${\got F}$}}(g)\kern.2emf ) (x)
   := D( \overline{A}) f( \Lambda_{A}^{-1}(x-a))\,,
   \quad g=(A,a)\in\semi{\g{SL}{2,{\z{C}}}}{\z{R}^{4}}\,,
   f\in\Test{\z{R}^{4},{\cal H}^{\KIn{$(0,1)$}}}=:
   \al T._{\KIn{${\got F}$}}.
\]

In analogy to the Weyl case we introduce the following embeddings
${\got I}_1\colon\ \Test{\z{R}^{4},{\cal H}}\longrightarrow  h_+$ and 
${\got I}_2\colon\ \Test{\z{R}^{4},{\cal H}}\longrightarrow  h_-$ 
defined for all $f\in\Test{\z{R}^{4},{\cal H}}$ by
\begin{eqnarray*}
   \LR{\got I}_1 f\RR\!(p)  & := & D(H_p)\; 
    D\!\LR \left(\kern-1.5mm\begin{array}{cc}
                 0 \kern-3mm & 1 \\
                 0 \kern-3mm & 0
    \end{array}\kern-1.5mm\right)\RR 
           D\!\LR\overline{H_p}\RR^{-1}\,\widehat{f}(p)\,,\quad p\in\Lk\,, \\
   \LR{\got I}_2 f\RR\!(p)  & := & D\!\LR\overline{H_p}\RR 
    D\!\LR \left(\kern-1.5mm\begin{array}{cc}
                 0 \kern-3mm & 1 \\
                 0 \kern-3mm & 0
    \end{array}\kern-1.5mm\right)\RR 
          D(H_p)^{-1}\,\widehat{\Gamma_0f}(p)\,,\quad p\in\Lk\,,
\end{eqnarray*}
where $D\!\LR\left(\kern-1.5mm\begin{array}{cc}
                 0 \kern-3mm & 1 \\
                 0 \kern-3mm & 0
         \end{array}\kern-1.5mm\right)\RR
\in\LR D^{\KIn{$(1,0)$}}({\cal E}(2)),\; D^{\KIn{$(0,1)$}}({\cal E}(2)) \RR$ 
and where the `hat' $\,\widehat{}\,$ means the 
Fourier transformation which is restricted to $\Lk$ 
as in the Weyl case. Further
$\Gamma_0\colon\ {\cal H}^{\KIn{$(0,1)$}}\to{\cal H}^{\KIn{$(1,0)$}}$
is again an anti-unitary involution (conjugation).

Finally, the embedding that specifies the net structure and
satisfies the conditions stated in Subsection~\ref{FBRED}
is given by
\begin{equation}\label{IF}
 {\got I}_{\KIn{${\got F}$}} \colon\ 
     {\cal T}_{\KIn{${\got F}$}}\longrightarrow h_{\KIn{${\got F}$}},\;  
{\rm with}\quad {\got I}_{\KIn{${\got F}$}}f :=
     {\got I}_1f\oplus {\got I}_2 f\,.
\end{equation}

The covariance property of the net characterised by the preceding
embedding is guaranteed by the following result:
\begin{lem}
\label{Lem.4.4.1}
Using the notation introduced above the equation 
  ${\got I}_{\KIn{${\got F}$}}\, T_{\KIn{${\got F}$}}(g)=
  V_{\KIn{${\got F}$}}(g)\, {\got I}_{\KIn{${\got F}$}}$
holds for all $g\in\UPgr$.
\end{lem}
\begin{beweis}
The proof is done similarly as in Lemma~\ref{Lem.4.3.2}.
The intertwining equation is now based on the relations, 
\begin{Klammer}
\label{IntertwiningFEq}
  {\got I}_1\; T_{\KIn{${\got F}$}}(g) 
      & = & V_+(g)\; {\got I}_1   \\[1mm]
  {\got I}_2\; \Gamma_0 T_{\KIn{${\got F}$}}(g) 
      & = & V_-(g)\; {\got I}_2 \Gamma_0\; ,
\end{Klammer}
for any $g\in\UPgr$.
\end{beweis}

The next result will ensure the causality property of the net associated to
embedding ${\got I}_{\KIn{${\got F}$}}$.
\begin{lem}
\label{Lem.4.4.2}
Suppose that $\Wort{supp}\,f \perp\Wort{supp}\, k$ for 
$f$,$k\in\al T._{\KIn{${\got F}$}}$. Then 
$ \sigma_{\KIn{${\got F}$}} \!\LR
 {\got I}_{\KIn{${\got F}$}}f,\,
 {\got I}_{\KIn{${\got F}$}}k\RR=0 $
holds.
\end{lem}
\begin{beweis}
First note that ${\got I}_2\!\LR\Gamma_0 f\RR\!(p)=\Gamma_0
\LR{\got I}_1 f\RR\!(-p)$ 
for all $f\in\Test{\z{R}^{4},{\cal H}}$. Then, computing similarly
as in Lemma~\ref{Lem.4.3.3}
(putting now
$\widetilde{\beta_+}(p):=D(H_p^{-1})^* D(H_p^{-1})$
and $\widetilde{\beta_-}(p):=\Gamma_0\,{\beta_+}(p)\,\Gamma_0$), we get
\begin{eqnarray*}
\lefteqn{\sigma_{\KIn{${\got F}$}}\!
     \LR {\got I}_{\KIn{${\got F}$}}f,\,
     {\got I}_{\KIn{${\got F}$}}k\RR \kern2mm=\kern2mm
        \sigma_{\KIn{${\got F}$}}\!\LR 
     {\got I}_1f\oplus{\got I}_2(\Gamma_0 f)\, ,\,
     {\got I}_1k\oplus{\got I}_2(\Gamma_0 k)\RR }     \\ 
 &=& \frac{1}{2i}\LR\;\int\limits_{\Lkf} 
     \askp{{\widehat f}(p)}{\beta_{+}(p)\;
                         {\widehat k}(p)}_{\z{C}^4} \mudp \;+\;
     \int\limits_{\Lkf} \askp{{\widehat k}(-p)}
    {\beta_{+}(p)\; {\widehat f}(-p)}_{\z{C}^4}\mudp \Rdummy \\[2mm]
 & &\kern.5cm  \Ldummy -\int\limits_{\Lkf} \askp{{\widehat k}(p)}
    {\beta_{+}(p)\; {\widehat f}(p)}_{\z{C}^4} \mudp    \;-\;
        \int\limits_{\Lkf} \askp{{\widehat f}(-p)}
   {\beta_{+}(p)\; {\widehat k}(-p)}_{\z{C}^4} \mudp\RR \\[3mm]   
 &=& 0\; ,
\end{eqnarray*}
where the last equation follows from the fact that the matrix elements of 
the operator-valued function (recall Remark~\ref{Rem.3.3.4})
\begin{eqnarray*}
 \beta_+(p) &:=&D\!\LR \LR \overline{H_p^{-1}}\RR^* 
          \left(\kern-1.5mm\begin{array}{cc}
              0 \kern-1.6mm & 0 \\ 
              0 \kern-1.6mm & 1
          \end{array} \kern-1.5mm\right)\;\overline{H_p}^{-1}\RR \\[4mm]
            &\cong& \frac14
       \In{$ \left(\kern-1.5mm \begin{array}{cccc}
  (p_0-p_3)^2 \kern-1mm          &-(p_0-p_3)(p_1+ip_2) \kern-1mm & 
 -(p_0-p_3)(p_1+ip_2)\kern-1mm   & (p_1+ip_2)^2                       \\[2mm]
 -(p_0-p_3)(p_1+ip_2)\kern-1mm   &\phantom{-} (p_0+p_3)(p_0-p_3)\kern-1mm & 
 \phantom{-}(p_0+p_3)(p_0-p_3)\kern-1mm &  -(p_0+p_3)(p_1+ip_2)       \\[2mm] 
 -(p_0-p_3)(p_1+ip_2)\kern-1mm   &\phantom{-}(p_0+p_3)(p_0-p_3) \kern-1mm  & 
 \phantom{-}(p_0+p_3)(p_0-p_3) \kern-1mm   &  -(p_0+p_3)(p_1+ip_2)    \\[2mm]
   (p_1-ip_2)^2  \kern-1mm       & -(p_0+p_3)(p_1-ip_2) \kern-1mm & 
  -(p_0+p_3)(p_1-ip_2) \kern-1mm & (p_0+p_3)^2        
           \end{array} \kern-1.5mm\right)$}
\end{eqnarray*}
are homogeneous polynomials of degree 2 
(see Lemma~\ref{Lem.4.2.2}).
\end{beweis}

We will show the existence of a free net associated to the $\got
F$-Equation, which we call {\em ${\got F}$-net} for short.

\begin{teo}
\label{Teo.4.4.3}
Consider the net of local linear submanifolds of $h_{\KIn{${\got F}$}}$ 
given for $\al O.\in\BR$ by
\[
 \z{R}^4\supset{\cal O}\longmapsto h_{\KIn{${\got F}$}}({\cal O}):=
 \LG  {\got I}_{\KIn{${\got F}$}}f\mid  f\in 
 \TestO{\z{R}^{4},{\cal H}},\;\Wort{supp}f\subset {\cal O} \RG .
\]
Then the net of local C*-algebra defined by
\[
 \z{R}^4\supset{\cal O}\longmapsto {\cal A}_{\KIn{${\got F}$}}({\cal O}):=
 \Wort{C}^*\LR\LG W(\varphi)\mid  \varphi\in h_{\KIn{${\got F}$}}({\cal O})
 \RG\RR\, ,
\]
where the $W(\cdot)$'s denote the generators of the C*-algebra 
$\ccr{h_{\KIn{${\got F}$}},\sigma_{\KIn{${\got F}$}}}$, is a HK-net.
\end{teo}
\begin{beweis}
By Lemmas~\ref{Lem.4.4.1} and \ref{Lem.4.4.2} we have that the tuple
$ \LR h_{\KIn{$\got F$}},\, \sigma_{\KIn{$\got F$}},\, 
 V_{\KIn{$\got F$}},\,{\cal T}_{\KIn{$\got F$}},
 \, T_{\KIn{$\got F$}},\, {\got I}_{\KIn{$\got F$}}\RR $
satisfies all conditions stated in Definition~\ref{Tuples}~(ii) and again by 
Theorem~\ref{TeoFreeNet}~(ii) we get that the net of local C*-algebras
above is a HK-net.
\end{beweis}

\begin{rem}
\label{GenB}
\begin{itemize}
\item[(i)]
We will show next that the net constructed 
above is isomorphic to the bosonic net defined in 
\cite[Section~3, case~$n=2$]{Lledo01}. Using the notation
and results of the latter reference we specify the 
symplectic bijection $\lambda_\mathrm{B}
 \colon\ h_{\KIn{${\got F}$}}\rightarrow h_2$
(recall Proposition~\ref{IsoFreeNet}~(ii)): for
$\chi_\pm\in\Lzwei{\Lkf,\z{C},\mudp}$ 
put
\beaO
  \lefteqn{\lambda_\mathrm{B}\!\LR 
  D\!\LR H_{(\cdot)}\RR
          \left(\kern-2mm\begin{array}{c}  
                           1 \\ 0 
            \end{array} \kern-2mm \right) 
          \!\!\otimes\!\!  \left(\kern-2mm  \begin{array}{c}  
                           1 \\ 0 
          \end{array} \kern-2mm \right)\chi_+(\cdot)
                \oplus 
  D\!\LR \overline{H_{(\cdot)}}\RR
        \left(\kern-2mm\begin{array}{c}  
                           1 \\ 0 
              \end{array} \kern-2mm \right) 
              \!\!\otimes\!\!  \left(\kern-2mm  \begin{array}{c}  
                           1 \\ 0 
        \end{array} \kern-2mm \right)\chi_-(\cdot)\RR\!(p)} \hspace{4cm}\\
  &:=& \LE D(H_p)  
        \left(\kern-2mm\begin{array}{c}  
                           0 \\ 1 
             \end{array} \kern-2mm \right) 
              \!\!\otimes\!\!  \left(\kern-2mm  \begin{array}{c}  
                           0 \\ 1 
        \end{array} \kern-2mm \right)\chi_+(p)
       \RE_+ \oplus \LE D(H_p)\,  
        \left(\kern-2mm\begin{array}{c}  
                           0 \\ 1 
            \end{array} \kern-2mm \right) 
          \!\!\otimes\!\!  \left(\kern-2mm  \begin{array}{c}  
                           0\\ 1 
          \end{array} \kern-2mm \right)\chi_-(p)\RE_-\,,
\eeaO
where $[\cdot]_\pm$ denote the clases of the factor spaces ${\got H}'_\pm$
defined in \cite[Section~3]{Lledo01}. 
Using again the statements in the proof
of \cite[Lemma~3.2]{Lledo01} it is straightforward to prove that 
$\lambda_{\KIn{F}}$ satisfies the properties required in 
Proposition~\ref{IsoFreeNet}~(ii). (An isometry to the free net
constructed in \cite[Part~B]{Lledo95} is given in 
\cite[Remark~3.3.5]{tLledo98}.)

\item[(ii)] 
The natural complexification of $h_{\KIn{${\got F}$}}$ given by 
$J(\varphi_+\oplus\varphi_-):=i\,\varphi_+\oplus i\,\varphi_-$, $\varphi_+
\oplus\varphi_-\in h_{\KIn{${\got F}$}}$, already defines a 
Fock state satisfying the spectrum condition 
(cf.~Definition~\ref{Spectrality}~(B3) and 
\cite[Subsection~8.2.3]{bBaumgaertel92})
and where the one-particle Hilbert space
carries the \rep{} usually considered in 
the literature for describing photons 
with both helicities \cite[Section~2]{Hislop88}. 

\item[(iii)] It is now obvious that as in the 
 Fermi case we may generalise the preceding construction to arbitrary 
 values of the integer helicity parameter
 $n\in\z{N}$. Replace the index $\In{$(1,0)$}$ by $\In{$(n,0)$}$
 and $\In{$(0,1)$}$ by $\In{$(0,n)$}$ etc.
 Thus we have produced (considering Remark~\ref{GenF}~(i))
 isomorphic nets to the ones
 given in \cite{Lledo01}. Note also that the use of the direct sum of 4 
 reference spaces in the Fermi case was forced by the self-dual
 approach to the CAR-algebra. Nevertheless in the Bose {\em and} Fermi 
 cases the corresponding one-particle Hilbert
 spaces given by the canonical 
 Fock states (cf.~Remark~\ref{GenF}~(ii))
 are of the form $h_+\oplus h_-$.
\end{itemize}
\end{rem}

\begin{rem} From Theorem~\ref{Teo.3.3.18} we can show the equivalence of the
C*-dynamical systems 
$(\ccr{h_{\KIn{${\got F}$}},\sigma_{\KIn{${\got F}$}}},\alpha_g^\ot F.,
\UPgr)$ and 
$(\ccr{h_{\KIn{${\got A}$}},\sigma_{\KIn{${\got A}$}}},\alpha_g^\ot A.,
\UPgr)$,
where $\sigma_{\KIn{${\got A}$}}:=\mathrm{Im}
\langle\cdot,\cdot\rangle_{\KIn{${\got A}$}}$ and $\alpha_g^\ot A.$
is the Bogoljubov automorphism associated to $V_{\KIn{${\got A}$}}$
(recall Subsection~\ref{TheA}). But due to the specific form of the
factor space $h_{\KIn{${\got A}$}}$ and the corresponding covariant 
representation $T_{\KIn{${\got A}$}}$ there does {\em not} exist
a nontrivial embedding $\ot I._{\KIn{${\got A}$}}$ satisfying the 
corresponding intertwining property with $V_{\KIn{${\got A}$}}$
(cf.~axiom~(B2)). The impossibility of constructing the free 
net associated to the vector potential is the analogue in our context 
of the well-known Strocchi no-go theorems, that are formulated in the
quantum field theoretical context (cf.~\cite{StrocchiIn73}).
For a detailed treatment of the nets associated to the electromagnetic
vector potential (including a general analysis of the localised constraints)
see \cite{Lledo00}.
\end{rem}

\section{Massless quantum fields}\label{Mqf}

In the previous section we have seen that the embeddings
that characterise the massless free nets
naturally reduce the degrees
of freedom in the fibre (cf.~Remark~\ref{Rem.3.3.1} and Eqs.~(\ref {IW}),
(\ref{IF})) by using elements
$\KIn{$\left(\kern-1.5mm\begin{array}{cc}
                 0 \kern-3mm & 1 \\
                 0 \kern-3mm & 0
         \end{array}\kern-1.5mm\right)$}$
of intertwiner space between the little group $\al E.(2)$ and its conjugate
$\overline{\al E.(2)}$. This choice shows explicitly that the  
embeddings map the test functions into the space of solutions of 
massless \rwe{s}. Now using the canonical Fock states associated
to the CAR- and CCR-algebras
(recall Remarks~\ref{GenF} and \ref{GenB}~(ii)) we will obtain 
in a natural way quantum fields that satisfy in the distributional sense
the Weyl and Maxwell equations. Since these two cases are 
typical (see Remarks~\ref{GenRWE}~(ii), \ref{GenF} and \ref{GenB}) 
the following procedure establishes a neat way to define
massless fields for any helicity value. 
This construction is considerably simpler 
than what is done usually in QFT, where so-called $2j+1$ quantum fields 
are introduced (a clear reminiscence of the massive case) 
and then constrained by imposing suitable equations on them 
\cite{WeinbergIn65},\cite[Section~2]{Hislop88}.

If one considers the canonical Fock states mentioned before, then one 
can also interpret the embeddings $\ot I.$, 
that where used to completely characterise the 
free nets in Theorems~\ref{Teo.4.3.4} and \ref{Teo.4.4.3},
as a one particle Hilbert structure. Indeed, $\ot I.$ can
be seen as a real linear map from 
$\al T.:=\Test{\z{R}^{4},{\cal H}^{\KIn{$(0,\frac{n}{2})$}}}$ into
the (complex) one-particle Hilbert space 
\[
 \ot H._1\subset h_+\oplus h_-
\]
(with scalar product $\langle\cdot,\cdot\rangle:=
\langle\cdot,\cdot\rangle_+\oplus\langle\cdot,\cdot\rangle_-$).
By Propositions~\ref{Lem.3.3.8}, \ref{Lem.3.3.12} and Remark~\ref{GenRWE}~(ii)
the one particle Hilbert space $\ot H._1$ carries 
representations of the Poincar\'e group equivalent to the 
massless, positive energy, Wigner representations
with helicities $\pm\frac{n}{2}$. Thus we can 
use $\ot I.$ to construct canonically massless free quantum fields.
We will treat the Weyl (fermionic) and the Maxwell (bosonic) case
separately. The fermionic/bosonic fields are defined on the 
antisymmetric/symmetric Fock space $\al F._a(\ot H._1)/\al F._s(\ot H._1)$ 
over the corresponding one-particle Hilbert spaces $\ot H._1$. 
We denote the corresponding vacua simply by $\Omega$ and the 
scalar products by $\langle\cdot,\cdot\rangle_{a/s}$.

\paragraph{Free Weyl quantum field:}

Consider the C*-algebra $\car{h_{\KIn{$\got W$}},\Gamma_{\KIn{$\got W$}}}$
defined in Section~\ref{WN} and the basis projection 
$P:=
           \left(\kern-1.5mm \begin{array}{cccc}
 \EINS \kern-1mm & 0 \kern-1mm & 0 \kern-1mm & 0           \\
   0 \kern-1mm & \EINS \kern-1mm & 0 \kern-1mm & 0         \\ 
   0 \kern-1mm & 0 \kern-1mm & 0 \kern-1mm & 0             \\
   0 \kern-1mm & 0 \kern-1mm & 0 \kern-1mm & 0                        
      \end{array} \kern-1.5mm\right) $
specified in Remark~\ref{GenF}~(ii) (see also 
Remark~\ref{SpecRem}~(i)). 
Recall that in this context the creation
and annihilation operators are given as follows:\\ 
for $\psi,\psi_1,\ldots,\psi_n\in \ot H._1:=P
(\overline{\ot I._{\KIn{$\got W$}}(\al T._{\KIn{$\got W$}})})
\subset h_+\oplus h_-$ we put
\begin{eqnarray*}
c(\psi)\Omega &:=& 0\\  
c(\psi)(\psi_1\wedge\dots\wedge\psi_n)
   &:=&\sqrt{n}\; \sum_{l=1}^n (-1)^{l-1}\,\langle\psi,\psi_l\rangle\,
       \psi_1\wedge\dots
       \hat{\psi_l}\dots\wedge\psi_n\\
c(\psi)^*\Omega &=& \psi \\
c(\psi)^*  (\psi_1\wedge\dots\wedge\psi_n)
   &=& \frac{1}{\sqrt{n+1}}\,\psi\wedge\psi_1\wedge\dots\wedge\psi_n\,,
\end{eqnarray*}
where the wedges mean the antisymmetrised tensor product
\[
 \psi_1\wedge\dots\wedge\psi_n:=
    \sum_{\sigma\in\P_n}\mr sgn.(\sigma)\,
   \psi_{\sigma(a_1)}\otimes\ldots\otimes \psi_{\sigma(a_n)}\,.
\]
The previous creation and annihilation operators
are mutually adjoint w.r.t.~$\langle \cdot,\cdot\rangle_a$ 
and satisfy the usual anticommutation relations: 
for $\psi,\psi'\in\ot H._1$ one has
\[
 [c(\psi) ,c(\psi')^*]_+=\langle\psi,\psi'\rangle_a\, \1\,,
\]
where $[\cdot,\cdot]_+$ denotes the anticommutator.

In this context we may define the free Weyl quantum field as follows
(recall that $\Gamma_{\KIn{$\got W$}}\ot I._{\KIn{$\got W$}}=
\ot I._{\KIn{$\got W$}}$):

\begin{defi}\label{FQF}
Let $\omega_P$ be the Fock state corresponding to the basis projection
$P$ and denote by $(\al F._a(\ot H._1),\Pi_P,\Omega)$ the 
corresponding GNS-data.
We define the free Weyl quantum field acting on $\al F._a(\ot H._1)$ by 
\[
 \phi_{\KIn{$\got W$}}(f)
        :=\frac{1}{\sqrt{2}}\left(
          \Pi_P\Big(\ot a.(\ot I._{\KIn{$\got W$}}f)\Big)\right)
         = \frac{1}{\sqrt{2}}\left(
          c\Big(P(\ot I._{\KIn{$\got W$}}f)\Big)^*
          +c\Big(P(\ot I._{\KIn{$\got W$}}f)\Big)\right)\;,
\]
where $f\in \al T._{\KIn{$\got W$}}
:=\Test{\z{R}^{4},{\cal H}^{\KIn{$(0,\frac{1}{2})$}}}$
and $\ot a.(\cdot)$ denote the (abstract) generators of
$\car{h_{\KIn{$\got W$}},\Gamma_{\KIn{$\got W$}}}$.
\end{defi}

\begin{teo}\label{Wcontinuity}
The embedding ${\got I}_{\KIn{$\got W$}}$ (cf.~Eq.~(\ref{IW}))
is continuous w.r.t.~the corresponding 
Schwartz and Hilbert space topologies.
\end{teo}
\begin{beweis}
It is enough to show the continuity of $\ot I._1$. Recall that
for $p\in\Lkf$ the scalar product is characterised by the 
positive matrix-valued function
$\beta_{+}(p)=\frac12 \left(\kern-1.5mm \begin{array}{cc}
                             p_{0}-p_{3}  &  -p_{1}+ip_{2}\\
                             -p_{1}-ip_{2}  & p_{0}+p_{3}
              \end{array} \kern-1.5mm\right)$.
Then we have the estimates
\begin{eqnarray*}
\|\ot I._1 f\|_+^2 
  & =  & \int\limits_{\Lkf} \askp{{\widehat f}(p)}
         {\beta_{+}(p)\; {\widehat f}(p)}_{\z{C}^2} \mudp\\
  &\leq& \sum_{\KIn{$C',C=0$}}^1\;\int\limits_{\Lkf}
         |\beta_{+,\,\KIn{$C'C$}}(p)|\cdot|\widehat{f^{\KIn{$C'$}}}(p)|
         \cdot|\widehat{f^{\KIn{$C$}}}(p)|\mudp\\
  &\leq& \sum_{\KIn{$C',C=0$}}^1\;
         \int\limits_{\R^3\setminus\{0\}} |\mb p.|
         \cdot|\widehat{f^{\KIn{$C'$}}}(|\mb p.|,\mb p.)|
         \cdot|\widehat{f^{\KIn{$C$}}}(|\mb p.|,\mb p.)|
         \;\,\frac{\mr d.^3\mb p.}{|\mb p.|}\\
  & =  & \sum_{\KIn{$C',C=0$}}^1\;
         \int\limits_{\R^3\setminus\{0\}} 
         |\widehat{f^{\KIn{$C'$}}}(|\mb p.|,\mb p.)|
         \cdot|\widehat{f^{\KIn{$C$}}}(|\mb p.|,\mb p.)|
         \cdot\frac{(1+|\mb p.|^2)^4}{(1+|\mb p.|^2)^4}
         \;\mr d.^3\mb p.\\
  &\leq& M\; \Big(\sum_{\KIn{$C',C=0$}}^1 
            \|\widehat{f^{\KIn{$C'$}}}\|_{\KIn{4,0}} \cdot
            \|\widehat{f^{\KIn{$C$}}}\|_{\KIn{4,0}}\Big)\,,
\end{eqnarray*}
where $M=\int\limits_{\R^3}\frac{1}{(1+|\mb p.|^2)^4}
         \;\mr d.^3\mb p.$ and 
$\|\widehat{f^{\KIn{$C$}}}\|_{\KIn{4,0}}:=\mr sup._{p\in\R^4}\,
\{(1+|p|^2)^2\;|\widehat{f^{\KIn{$C$}}}(p)| \}$ is a particular seminorm
corresponding to the Schwartz space topology.
Suppose now that $f_n\to 0$ in the topology of 
$\al T._{\KIn{$\got W$}}:=
\Test{\z{R}^{4},{\cal H}^{\KIn{$(0,\frac{1}{2})$}}}$. Then by the continuity
of Fourier transformation and the previous estimates we conclude that
$\ot I._1(f_n)\to 0$ in $h_+$ and the proof is concluded.
\end{beweis}

In the following theorem we will show that the quantum field defined 
previously satisfies the Wightman axioms as well as the Weyl equation
in the distributional sense.

\begin{teo}\label{WigWeyl}
The Weyl quantum field $\phi_{\KIn{$\got W$}}(f)$, 
$f\in \al T._{\KIn{$\got W$}}$, defined on $\al F._a(\ot H._1)$ is a bounded,
self-adjoint operator. Moreover we have
\begin{itemize}
\item[(i)] {\bf (Weyl equation)} $\phi_{\KIn{$\got W$}}(\cdot)$ satisfies
 the Weyl equation in the distributional sense:
\[
 \phi_{\KIn{$\got W$}}\Big(\partial^{\KIn{$\;CC'$}} f_{\KIn{$C$}}\Big)=0
  \;,\quad f\in \Test{\z{R}^{4},{\cal H}^{\KIn{$(\frac{1}{2},0)$}}}\,.
\]
\item[(ii)] {\bf (Poincar\'e invariance and spectral condition)} 
  $\phi_{\KIn{$\got W$}}(\cdot)$ transforms covariantly under the Poincar\'e
  group: Let $T_{\KIn{$\got W$}}$ and $Q$ be the covariant and the second
  quantisation of the canonical representation 
  $PV_{\KIn{$\got W$}}=V_1\oplus V_2$ on 
  $\al F._a(\ot H._1)$, respectively (recall Section~\ref{WN}). Then 
\begin{equation}\label{CovQF} 
 Q(g)\,\phi_{\KIn{$\got W$}}(f)\,Q(g)^{-1}
          =\phi_{\KIn{$\got W$}}(T_{\KIn{$\got W$}}f)\;,\quad
           f\in \al T._{\KIn{$\got W$}}\;,\;\;g\in \UPgr\,.
\end{equation}
Further, the representation
$V_1\oplus V_2$ satisfies the spectral condition on $\ot H._1$ and
$Q(g)\Omega=\Omega$, $g\in \UPgr$.
\item[(iii)] {\bf (Anticommutation relations)} For
 $f,k\in\al T._{\KIn{$\got W$}}$ such that 
 $\Wort{supp}\,f$ and $\Wort{supp}\, k$ are space-like separated, the 
 anticommutator of the corresponding smeard fields vanishes:
\[
 [\phi_{\KIn{$\got W$}}(f),\phi_{\KIn{$\got W$}}(k)]_+=0\,.
\]
\item[(iv)] {\bf (Regularity)} The map
\[
 \al T._{\KIn{$\got W$}}\ni f\mapsto 
      \langle\psi_1, \phi_{\KIn{$\got W$}}(f) \psi_2\rangle_a
      \;,\quad \psi_1,\psi_2\in\al F._a(\ot H._1)\,,
\]
is a tempered distribution.
\end{itemize}

\end{teo}
\begin{beweis}
The boundedness and self-adjointness of the field follows from the 
same properties of the generator $\ot a.(\ot I._{\KIn{$\got W$}}f)$ 
of the CAR-algebra.

To show (i) recall that e.g.~the embedding $\ot I._1$ used to specify
$\ot I._{\KIn{$\got W$}}$ maps into the space  of solutions of 
Weyl equation (cf.~(\ref{IW})). Indeed, we will show that
$\ot I._1(\partial^{\KIn{$\;CC'$}} f_{\KIn{$C$}})=0$,
$f\in\Test{\z{R}^{4},{\cal H}^{\KIn{$(\frac{1}{2},0)$}}}$:
for $\In{$B$}\in\{0,1\}$ and summing over repeated indices we have
\begin{eqnarray*}
 \ot I._1^{\KIn{$B$}}(\partial^{\KIn{$\;CC'$}} f_{\KIn{$C$}})(p)
     &=& \left(H_p \left(\kern-1.5mm\begin{array}{cc}
                 0 \kern-3mm & 1 \\
                 0 \kern-3mm & 0
            \end{array}\kern-1.5mm\right) \overline{H_p}^{-1}
            \right)^{\KIn{$B$}}_{\KIn{$C'$}}\;
            \widehat{\partial^{\KIn{$\;CC'$}} f_{\KIn{$C$}}}(p)\\
     &=& -i\;\left(H_p \left(\kern-1.5mm\begin{array}{cc}
                 0 \kern-3mm & 1 \\
                 0 \kern-3mm & 0
            \end{array}\kern-1.5mm\right) \overline{H_p}^{-1}
            \right)^{\KIn{$B$}}_{\KIn{$C'$}}\; 
            \widehat{P}^{\KIn{$\;CC'$}}\;\widehat{f_{\KIn{$C$}}}(p)\\
     &=& 0\;, 
\end{eqnarray*}
where the last equation follows from the fact that 
\[
 (\widehat{P}^{\KIn{$\;CC'$}})=\left(\kern-1.5mm \begin{array}{cc}
                                      p_{0}+p_{3}  &  p_{1}+ip_{2}\\
                                      p_{1}-ip_{2}  & p_{0}-p_{3}
                               \end{array} \kern-1.5mm\right)
     =\overline{H_p} \left(\kern-1.5mm\begin{array}{cc}
                 2 \kern-3mm & 0 \\
                 0 \kern-3mm & 0
            \end{array}\kern-1.5mm\right) \overline{H_p}^*\,.
\] 
Similarly we obtain $\ot I._2(\partial^{\KIn{$\;CC'$}} f_{\KIn{$C$}})=0$,
hence $P(\ot I._{\KIn{$\got W$}}(\partial^{\KIn{$\;CC'$}} f_{\KIn{$C$}}))
=0$ and the field satisfies the Weyl equation as required:
\[
  \phi_{\KIn{$\got W$}}\Big(\partial^{\KIn{$\;CC'$}} f_{\KIn{$C$}}\Big)
=c\Big(P(\ot I._{\KIn{$\got W$}}(\partial^{\KIn{$\;CC'$}}f_{\KIn{$C$}}))\Big)^*
+c\Big(P(\ot I._{\KIn{$\got W$}}(\partial^{\KIn{$\;CC'$}} f_{\KIn{$C$}}))\Big)
=0 \,.
\] 

The property (ii) follows from Remark~\ref{SpecRem}~(i). The anticommutation
of the field in (iii) is a consequence of the anticommutation 
of the generators $\ot a.(\ot I._{\KIn{$\got W$}}f)$ and 
$\ot a.(\ot I._{\KIn{$\got W$}}k)$ of the CAR-algebra 
(cf. Lemma~\ref{Lem.4.3.3} 
and Theorem~\ref{Teo.4.3.4}).

The regularity (iv) of the field follows from the continuity of
the embedding $\ot I._{\KIn{$\got W$}}$ (see Theorem~\ref{Wcontinuity}).
Indeed, let $f_n\to 0$ in the Schwartz topology of 
$\al T._{\KIn{$\got W$}}$. Then by Theorem~\ref{Wcontinuity} we have
$\ot I._{\KIn{$\got W$}}f_n\to 0$ in the Hilbert space topology. Now
for any $\psi\in\al F._a(\ot H._1)$
\[
 \|\phi_{\KIn{$\got W$}}(f_n)\psi \|
    \leq \|\ot a.(\ot I._{\KIn{$\got W$}}f_n)\|_{\KIn{C$^*$}}\;\|\psi\|
    \leq \|\ot I._{\KIn{$\got W$}}f_n \| \;\|\psi\|
\]
and therefore $\mr s-.\!\lim_{n\to\infty}\phi_{\KIn{$\got W$}}(f_n)=0$. The
strong continuity of the field implies finally the temperedness of the 
distribution.
\end{beweis}

We show in the next theorem some additional properties satisfied by the 
Weyl quantum field.

\begin{teo}\label{WAddition}
The field $\phi_{\KIn{$\got W$}}$ transforms in addition 
covariantly w.r.t.~the (fourfold cover of) conformal group in 
Minkowski space $\g{SU}{2,2}$.
\end{teo}
\begin{beweis}
The extension of the covariance property to the conformal group follows
from the results in \cite[Section~5]{Lledo01} (recall also the isomorphy 
between the massless free nets of Section~\ref{WN} and those constructed
in \cite{Lledo01} given Remark~\ref{GenF}~(i)).
\end{beweis}

\begin{rem}
As mentioned above, the Weyl case is typical for 
fermionic models with nontrivial (half-integer) helicity. Hence replacing
for example $\In{$(0,\frac12)$}$ by $\In{$(0,\frac{n}{2})$}$ 
with $n\geq 3$ and odd, 
one can similarly define the massless fermionic free quantum field with
helicity $\frac{n}{2}$ by
\[
\phi_n(f)
        :=
          \frac{1}{\sqrt{2}} \left(\Pi_P\left(\ot a.(\ot I._nf)\right)\right)
         = \frac{1}{\sqrt{2}} \left(
           c\left(P(\ot I._{\KIn{$\got W$}}f)\right)^*
          +c\left(P(\ot I._{\KIn{$\got W$}}f)\right)\right)\;,\quad
  f\in \al T._n
  :=\Test{\z{R}^{4},{\cal H}^{\KIn{$(0,\frac{n}{0})$}}}\,.
\]
These fields also satisfy Wightman axioms and the adapted version of
Theorem~\ref{WAddition}. In particular, it satisfies the corresponding
massless relativistic wave equation in the weak sense:
\[
\phi_n\Big(\partial^{\KIn{$\;CC'$}} f_\KIn{$C$}^\KIn{$~~C_1'\dots C_{n-1}'$}
      \Big)=0
\]
(cf.~Corollary~\ref{General}).
\end{rem}

\paragraph{Free Maxwell quantum field:}

Consider the simple C*-algebra
$\ccr{h_{\KIn{${\got F}$}},\sigma_{\KIn{${\got F}$}}}$
given in Section~\ref{FNet} as well as the 
canonical Fock state $\omega_J$ on 
$\ccr{h_{\KIn{${\got F}$}},\sigma_{\KIn{${\got F}$}}}$
specified by the internal complexification 
$J(\varphi_+\oplus\varphi_-):=i\,\varphi_+\oplus i\,\varphi_-$, $\varphi_+
\oplus\varphi_-\in h_{\KIn{${\got F}$}}$ 
(recall Remarks~\ref{SpecRem} and \ref{GenB}~(ii)). 
Putting $\ot H._1:=\overline{\ot I._{\KIn{${\got F}$}}
(\al T._{\KIn{${\got F}$}})}\subset h_+\oplus h_-$
the generating functional is given by 
\[
 \ot H._1\ni\psi \mapsto e^{-\frac14 \|\psi\|^{\,2}} \;.
\]

\begin{defi}\label{BQF}
Let $\omega_J$ be the Fock state associated to the internal complexification
$J$ given above and denote by $(\al F._s(\ot H._1),\Pi_J,\Omega)$ the 
corresponding GNS-data. 
$\overline{\Phi_J(\ot I._{\KIn{$\got F$}}f)}$,
$f\in \al T._{\KIn{$\got F$}}:=\Test{\z{R}^{4},{\cal H}^{\KIn{$(0,1)$}}}$,
is the infinitesimal generator of the strongly continuous unitary group
\[
 \z{R}\ni t\mapsto \Pi_J\big( 
 W(t\,{\got I}_{\KIn{${\got F}$}}f)\big)
 =e^{\KIn{$-it\,\overline{\Phi_J(\ot I._{\KIn{$\got F$}}f)}$}}\,,
\] 
where $W(\cdot)$ are the (abstract) generators of
$\ccr{h_{\KIn{$\got F$}},\sigma_{\KIn{$\got F$}}}$.
Then we define the free Maxwell quantum field acting on 
$\al F._s(\ot H._1)$ by 
\[
 \phi_{\KIn{$\got F$}}(f)
        := \overline{\Phi_J(\ot I._{\KIn{$\got F$}}f)}\,,\quad
           f\in \al T._{\KIn{$\got F$}}\,.
\]
\end{defi}

Notice that the free Maxwell quantum field is, as a consequence of 
the uniqueness of the GNS representation and Nelson's analytic vector 
theorem, the closure of the essentially self-adjoint operator
\[
\Phi_J(\ot I._{\KIn{$\got F$}}f)
        =\frac{1}{\sqrt{2}}\left(
         a((\ot I._{\KIn{$\got F$}}f))^*
        +a((\ot I._{\KIn{$\got F$}}f))\right)\,,
\]
on the set $\al F._{\KIn{fin}}\subset\al F._s(\ot H._1)$ of
finite particle vectors (cf.~\cite[Theorem~X.41]{bReedII}).
The creation and annihilation operators on the symmetric Fock space
over the one-particle Hilbert space $\ot H._1$ are defined as usual:
For $\varphi,\varphi_1,\ldots,\varphi_n\in\ot H._1$ we put
\begin{eqnarray*}
  a(\varphi)\Omega &:=& 0\\
  a(\varphi) \al S._n (\varphi_1\otimes\dots\otimes\varphi_n)
   &:=&\sqrt{n}\; \sum_{l=1}^n \langle\varphi,\varphi_l\rangle\,
       \al S._{n-1}(\varphi_1\otimes\dots
       \hat{\varphi_l}\dots\otimes\varphi_n)\\
  a(\varphi)^*\Omega &=& \varphi \\
   a(\varphi)^* \al S._n (\varphi_1\otimes\dots\otimes\varphi_n)
   &=&\frac{1}{\sqrt{n+1}}\,
     \al S._{n+1}( \varphi\otimes\varphi_1\otimes\dots\otimes\varphi_n)\,,
\end{eqnarray*}
where the hat means omission and $\al S._n$ is the symmetrisation operator
$\al S._n(\varphi_1\otimes\dots\otimes\varphi_n)
:=\sum_{\sigma\in\P_n}
  \varphi_{\sigma(1)}\otimes\dots\otimes\varphi_{\sigma(n)}$
on the $n$-tensor product space over $h_+\oplus h_-$.
The previous creation and annihilation operators
are mutually adjoint w.r.t.~$\langle \cdot,\cdot\rangle_s$ and 
satisfy the usual commutation relations: 
for $\varphi,\varphi'\in\ot H._1$ one has
\[
 [a(\varphi) ,a(\varphi')^*]=\langle\varphi,\varphi'\rangle_s\,\1\,,
\]
where $[\cdot,\cdot]$ denotes the commutator.

Similarly as in Theorem~\ref{Wcontinuity} we can show the following
continuity statement for the embedding:
\begin{teo}\label{Fcontinuity}
The embedding ${\got I}_{\KIn{$\got F$}}$ (recall Eq.~(\ref{IF}))
is continuous w.r.t.~the corresponding 
Schwartz and Hilbert space topologies.
\end{teo}

We will show next that the Maxwell quantum field also satisfies the 
Wightman axioms as well as the Maxwell equation in a distributional 
sense.

\begin{teo}\label{WigMaxwell}
The Maxwell quantum field $\phi_{\KIn{$\got F$}}(f)$, 
$f\in \al T._{\KIn{$\got F$}}$, defined on $\al F._s(\ot H._1)$ is 
an unbounded, self-adjoint operator that leaves the dense subspace 
$\al F._{\KIn{fin}}$ invariant. Moreover we have
\begin{itemize}
\item[(i)] {\bf (Maxwell equation)} $\phi_{\KIn{$\got F$}}(\cdot)$ satisfies
 the following equation in the distributional sense:
\[
 \phi_{\KIn{$\got F$}}\Big(\partial^{\KIn{$\;CC'$}} 
                 f_{\KIn{$C$}}^{\KIn{$~B'$}}\Big)=0
  \;,\quad f\in \Test{\z{R}^{4},{\cal H}^{\KIn{$(\frac12,\frac12)$}}}\,.
\]
\item[(ii)] {\bf (Poincar\'e invariance and spectral condition)} 
  $\phi_{\KIn{$\got F$}}(\cdot)$ transforms covariantly under the Poincar\'e
  group: Let $T_{\KIn{$\got F$}}$ and $Q$ be the covariant and the second
  quantisation of the canonical representation 
  $V_{\KIn{${\got F}$}}:=V_1\oplus V_2$ on 
  $\al F._s(\ot H._1)$, respectively (recall Section~\ref{FNet}). Then 
\begin{equation}\label{CovQFM} 
 Q(g)\,\phi_{\KIn{$\got F$}}(f)\,Q(g)^{-1}
          =\phi_{\KIn{$\got F$}}(T_{\KIn{$\got F$}}f)\;,\quad
           f\in \al T._{\KIn{$\got F$}}\;,\;\;g\in \UPgr\,.
\end{equation}
Further, the representation
$V_1\oplus V_2$ satisfies the spectral condition on $\ot H._1$ and
$Q(g)\Omega=\Omega$, $g\in \UPgr$.
\item[(iii)] {\bf (Causality)} For
 $f,k\in\al T._{\KIn{$\got F$}}$ such that 
 $\Wort{supp}\,f$ and $\Wort{supp}\, k$ are space-like separated, the 
 commutator of the corresponding smeard fields vanishes:
\[
 [\phi_{\KIn{$\got F$}}(f),\phi_{\KIn{$\got F$}}(k)]=0\;\quad
 (\mr on~.\al F._\KIn{fin}) \,.
\]
\item[(iv)] {\bf (Regularity)} The map
\[
 \al T._{\KIn{$\got F$}}\ni f\mapsto 
      \langle\psi_1, \phi_{\KIn{$\got F$}}(f) \psi_2\rangle_s\;,
      \quad \psi_1,\psi_2\in\al F._{\KIn{fin}}\,,
\]
is a tempered distribution.
\end{itemize}
\end{teo}
\begin{beweis}
The self-adjointness of the field follows from its definition as generator
of a strongly continuous unitary group and the invariance of 
$\al F._\KIn{fin}$ is a consequence of the remarks after 
Definition~\ref{BQF}.

To show (i) recall that e.g.~the embeddings $\ot I._{1/2}$ used to specify
$\ot I._{\KIn{$\got F$}}$ map into the space  of solutions of 
Maxwell equation (cf.~(\ref{IW})). Reasoning as in the proof of 
Theorem~\ref{WigWeyl} on obtains
$\ot I._\ot F.(\partial^{\KIn{$\;CC'$}} f_{\KIn{$C$}}^{\KIn{$~B'$}})=0$
and again this implies
\[
\phi(\partial^{\KIn{$\;CC'$}} 
                 f_{\KIn{$C$}}^{\KIn{$~B'$}})
        = \frac{1}{\sqrt{2}}\left(
         a\Big((\ot I._\ot F.(\partial^{\KIn{$\;CC'$}} 
                 f_{\KIn{$C$}}^{\KIn{$~B'$}}) )\Big)^*
        +a\Big((\ot I._\ot F.(\partial^{\KIn{$\;CC'$}} 
                 f_{\KIn{$C$}}^{\KIn{$~B'$}}))\Big)\right)=0 \,,
\]
where $f\in
\Test{\z{R}^{4},{\cal H}^{\KIn{$(\frac12,\frac{1}{2})$}}}$.

To prove property (ii) note that the Fock state $\omega_J$
is invariant w.r.t.~the Bogoljubov automorphism $\alpha_g$ generated
by $V_{\KIn{$\got F$}}(g)$, i.e.~$\omega_J\circ\alpha_g=\omega_J$, 
$g\in\UPgr$, hence by Remark~\ref{SpecRem}~(ii) we have
$Q(g)\Omega=\Omega$. Further, $Q(g)$ also leaves 
$\al F._{\KIn{fin}}$ invariant and for $\psi\in\al F._{\KIn{fin}}$
we have 
\[
Q(g)\,\Phi_J (\ot I._{\KIn{$\got F$}}f)\,Q(g)^{-1}\;\psi
          =\Phi_J(\ot I._{\KIn{$\got F$}}T_{\KIn{$\got F$}}f)\;\psi
           \;,\quad g\in\UPgr\,.
\]
Since both sides of the previous equation are essentially self-adjoint
operators we finally obtain the covariance relation: 
\[
Q(g)\,\phi_{\KIn{$\got F$}}(f)\,Q(g)^{-1}
          =\phi_{\KIn{$\got F$}}(T_{\KIn{$\got F$}}f)\;,\quad
           f\in \al T._{\KIn{$\got F$}}\,.
\]

The commutation of the field in (iii) is again a consequence of the 
commutation of the generators $W(\ot I._{\KIn{$\got F$}}f)$ and 
$W(\ot I._{\KIn{$\got F$}}k)$ of the CCR-algebra (cf. Lemma~\ref{Lem.4.4.2} 
and Theorem~\ref{Teo.4.4.3}).

The regularity (iv) of the field follows from the continuity of
the embedding $\ot I._{\KIn{$\got F$}}$ (see Theorem~\ref{Wcontinuity}).
Indeed, let $f_n\to 0$ in the Schwartz topology of 
$\al T._{\KIn{$\got F$}}$. Then by Theorem~\ref{Wcontinuity} we have
$\ot I._{\KIn{$\got F$}}f_n\to 0$ in the Hilbert space topology. Now
for any $k$-th particle vector $\psi\in\al F._{\KIn{fin}}$ we have
\[
 \|\phi_{\KIn{$\got F$}}(f_n)\psi \|
    \leq \sqrt{2}\;\sqrt{k+1}\;\|\ot I._{\KIn{$\got F$}}f_n \| \;\|\psi\|
\]
and therefore $\phi_{\KIn{$\got F$}}(f_n)\to 0$ strongly on 
$\al F._{\KIn{fin}}$. The strong continuity of the field 
implies finally the temperedness of the distribution.
\end{beweis}

We show in the next theorem some additional properties satisfied by the 
Maxwell quantum field.

\begin{teo}\label{MAddition}
The field $\phi_{\KIn{$\got F$}}$ transforms in addition 
covariantly w.r.t.~the (fourfold cover of) conformal group in 
Minkowski space $\g{SU}{2,2}$.
\end{teo}
\begin{beweis}
The extension of the covariance property to the conformal group follows
from the results in \cite[Section~5]{Lledo01} (recall also the isomorphy 
between the massless free nets of Section~\ref{WN} and those constructed
in \cite{Lledo01} given Remark~\ref{GenF}~(i)).
\end{beweis}

\begin{rem}
The present construction can also be generalised to produce massless
bosonic free fields with nontrivial (integer) helicity. Hence replacing
for example $\In{$(0,1)$}$ by $\In{$(0,n)$}$ 
with $n\geq 2$ and even, one can similarly define the massless 
fermionic free quantum field with helicity $\frac{n}{2}$ by
\[
\phi_n(f)
        := \frac{1}{\sqrt{2}}\left(
         a((\ot I._n f))^*
        +a((\ot I._n f))\right) \,,\;\;
  f\in \al T._n
  :=\Test{\z{R}^{4},{\cal H}^{\KIn{$(0,\frac{n}{2})$}}}\,.
\]
These fields also satisfy Wightman axioms and the adapted version of
Theorem~\ref{MAddition}. In particular, it satisfies the corresponding
massless relativistic wave equation in the weak sense:
\[
\phi_n\Big(\partial^{\KIn{$\;CC'$}}
         f_\KIn{$C$}^\KIn{$~~C_1'\dots C_{n-1}'$}\Big)=0
\]
(cf.~Corollary~\ref{General}).
\end{rem}


\section{Conclusions}

In a recent paper Brunetti, Guido and Longo proposed
a construction procedure for a bosonic net of 
von Neumann algebras canonically associated to a
positive energy strongly continuous (anti-) unitary Hilbert space
representation of the proper Poincar\'e group $\cal P_+$
(cf.~\cite{Brunetti02}). They also 
used the suggestive name of free net as in 
\cite[Example~8.3.1]{bBaumgaertel92} (see also \cite{Lledo95,Lledo01}),
since the construction avoids the use of quantum fields
as `coordinates' of the corresponding net. The construction of bosonic
free nets in Section~\ref{AxiomsFreeNets} and the one in
\cite{Brunetti02} are similar in that both use Wigner's cornerstone 
analysis of the unitary irreducible representations of the universal 
cover of the Poincar\'e group, as well as the CCR-algebra. Nevertheless,
in Section~\ref{AxiomsFreeNets} we prefer to work initially with abstract
C*-algebras, while in \cite{Brunetti02} concrete von Neumann algebras 
in a Fock representation are used. 
The crucial difference relies in the choice
of the localisation prescription. We use $\al H.$-valued Schwartz
functions on Minkowski space on which the (algebraically reducible)
covariant representation $T$ of the Poincar\'e group acts 
and, in fact,
we can also canonically construct the corresponding free massless
quantum fields that satisfy Wightman axioms. Brunetti, Guido and
Longo use the relatively
recent notion of modular localisation (see also \cite{Fassarella02})
which does not need test functions on configuration space. There is also
no obvious candidate for covariant representation in this frame. Recall
that the covariance of free nets is expressed at the level of local
reference spaces $h(\al O.)$ of the CAR resp.~CCR-algebras by means of 
the equation
\[
V(g)\,h({\cal O})=h(g{\cal O})\,, \quad g\in\UPgr\;,
\]
where $V$ is the Wigner representation (see Eq.~(\ref{CovRS})).
The proof of the previous equation is based on the 
intertwining equation $V(g)\,\ot I.=\ot I.\,T(g)$, where 
$\ot I.$ is the embedding characterising the free net
(for details see the proof of Theorem~3.6 in \cite{Lledo01}). 
The modular localisation approach uses, instead,  
the Bisognano-Wichmann relations as an essential input to introduce
modular-like objects at the level of the one-particle Hilbert space 
$\ot H.$ and associated to any wedge $W$ in a suitable family 
of wedges $\al W.$. This family is compatible with the action of a
one-parameter group of boosts and a time-reversing reflection
assigned to each $W\in\al W.$. The `Tomita operator' on $\ot H.$ 
naturally selects a family $K_W$, $W\in\al W.$, of $\R$-linear,
closed, standard subspaces of $\ot H.$ that transform covariantly
under the chosen Wigner representation. By means of suitable 
intersections of $K_W$'s one defines a net of subspaces
localised in e.g.~causally complete convex regions
which also transforms covariantly.

A remarkable aspect of the modular localisation approach 
is that one can also naturally
associate a free net $\al O.\to\al M._\mr cont.(\al O.)$
to the `continuous spin' massless representations. These types of 
representations are typically excluded by hand from further considerations.
It is conjectured in \cite[p.~761]{Brunetti02} that this net should not
satisfy the Reeh-Schlieder property for double cones. If so, this would be
conceptually a much more satisfactory explanation of the singular character
that these representations play in nature. 
A natural question that arises in this context is the relation
of the net $\al O.\to\al M._\mr cont.(\al O.)$ with the one associated
to discrete helicity representations. In particular, if it is possible
to describe at the C*-algebraic level in $\al M._\mr cont.$
the choice of nonfaithful representation
of $\al E.(2)$ needed to define discrete helicity. Techniques
of local quantum constraints (see \cite{Grundling85,Lledo00}) 
may possibly be applied to $\al O.\to\al M._\mr cont.(\al O.)$ 
in order to consider this question. (Here, the use of
abstract C*-algebras in a first step can be relevant.) Recall also that
the use of nonfaithful representations of $\al E.(2)$ 
(hence discrete helicity)
is crucial for the extension to the conformal group of the covariance 
of the corresponding net 
(see \cite{Angelopoulos78,Angelopoulos81,Angelopoulos98,Lledo01} 
for further points on this subject).

\paragraph{Acknowledgments}
The present paper is a revised and considerably extended version of 
some parts of the authors PHD at the University of Potsdam. 
It is a pleasure to thank
Hellmut Baumg\"artel for supervision and many useful remarks. 
I would also like to thank Wolfgang Junker for helpful conversations.



\end{document}